# Empowering Precision Medicine: AI-Driven Schizophrenia Diagnosis via EEG Signals: A Comprehensive Review from 2002-2023

Mahboobeh Jafari[1,2,*], Delaram Sadeghi[2,*], Afshin Shoeibi[1,2,**], Hamid Alinejad-Rokny[3,4,5,**], Amin Beheshti[6], David López García[2], Zhaolin Chen[7], U. Rajendra Acharya[8], Juan M. Gorriz[2,9]

[1] Internship in BioMedical Machine Learning Lab, The Graduate School of Biomedical Engineering, UNSW Sydney, Sydney, NSW, 2052, Australia.

[2] Data Science and Computational Intelligence Institute, University of Granada, Spain.

[3] BioMedical Machine Learning Lab, The Graduate School of Biomedical Engineering, UNSW Sydney, Sydney, NSW, 2052, Australia.

[4] UNSW Data Science Hub, The University of New South Wales, Sydney, NSW, 2052, Australia.

[5] Health Data Analytics Program, Centre for Applied Artificial Intelligence, Macquarie University, Sydney, 2109, Australia.

[6] Data Science Lab, School of Computing, Macquarie University, Sydney, NSW 2109, Australia.

[7] Monash Biomedical Imaging (MBI), Monash University, Melbourne, Australia.

[8] School of Mathematics, Physics and Computing, University of Southern Queensland, Springfield, Australia.

[9] Department of Psychiatry, University of Cambridge, UK.

* Equal contributions

** Corresponding authors: A. Shoeibi (shoeibi@correo.ugr.es), H. Alinejad-Rokny (h.alinejad@unsw.edu.au)

## Abstract

Schizophrenia (SZ) is a prevalent mental disorder characterized by cognitive, emotional, and behavioral changes. Symptoms of SZ include hallucinations, illusions, delusions, lack of motivation, and difficulties in concentration. While the exact causes of SZ remain unproven, factors such as brain injuries, stress, and psychotropic drugs have been implicated in its development. SZ can be classified into different types, including paranoid, disorganized, catatonic, undifferentiated, and residual. Diagnosing SZ involves employing various tools, including clinical interviews, physical examinations, psychological evaluations, the Diagnostic and Statistical Manual of Mental Disorders (DSM), and neuroimaging techniques. Electroencephalography (EEG) recording is a significant functional neuroimaging modality that provides valuable insights into brain function during SZ. However, EEG signal analysis poses challenges for neurologists and scientists due to the presence of artifacts, long-term recordings, and the utilization of multiple channels. To address these challenges, researchers have introduced artificial intelligence (AI) techniques, encompassing conventional machine learning (ML) and deep learning (DL) methods, to aid in SZ diagnosis. This study reviews papers focused on SZ diagnosis utilizing EEG signals and AI methods. The introduction section provides a comprehensive explanation of SZ diagnosis methods and intervention techniques. Subsequently, review papers in this field are discussed, followed by an introduction to the AI methods employed for SZ diagnosis and a summary of relevant papers presented in tabular form. Additionally, this study reports on the most significant challenges encountered in SZ diagnosis, as identified through a review of papers in this field. Future directions to overcome these challenges are also addressed. The discussion section examines the specific details of each paper, culminating in the presentation of conclusions and findings.

## 1. Introduction

Schizophrenia (SZ) is a mental disorder that has detrimental effects on an individual's cognitive processes, emotional state, and behavioral patterns [1-2]. SZ is characterized by symptoms that typically manifest at a young age. These symptoms can be broadly categorized into three types: positive [3], negative [4], and cognitive [5]. Positive symptoms of SZ involve hallucinations, delusions and thought disorders [3]. Negative symptoms, on the other hand, include a lack of motivation and reduced emotional expression [4]; Individuals with SZ in this category face serious challenges in self-care, such as cleaning [4]. Additionally, cognitive symptoms, such as memory problems, attention difficulties, and impaired decision-making, are among the most prominent [5]. The presence of positive, negative and cognitive symptoms profoundly impacts on the daily lives of individuals with SZ. While the precise causes of SZ remain unknown, certain clinical studies suggest that a combination of genetic and environmental factors may contribute to its onset [6-7].

Various methods for treating SZ have been provided by medical professionals, some of the most important of which include psychotherapy [8], social support [9], rehabilitation applications [10], and hospitalization [11]. On the other hand, interventional methods for treating this disorder include medication [17], electroconvulsive therapy (ECT) [12], transcranial magnetic stimulation (TMS) [13], transcranial direct current stimulation (tDCS) [14], deep brain stimulation (DBS) [15], and cognitive-behavioral therapy (CBT) [16]. Antipsychotic medications are used to reduce some positive symptoms such as hallucinations and delusions. In ECT, physicians pass an electrical current through the brain to control and treat SZ [12]. TMS is a non-invasive medical procedure that utilizes magnetic fields to stimulate nerve cells in the brain [13]. In the transcranial direct current stimulation (tDCS) method, physicians non-invasively apply electrical currents to the scalp, varying in amplitude [14]. Research has shown the efficacy of these methods in treating SZ. DBS involves the surgical implantation of electrodes in specific brain regions, which are then utilized for targeted stimulation [15]. Lastly, CBT is another interventional approach employed in the treatment of SZ, aiming to modify the negative thought and behavior patterns . Numerous studies have demonstrated the effectiveness of CBT in the treatment of SZ [16].

Various methods developed for diagnosing SZ, including clinical interviews [18], physical examination [19], psychological evaluation [20], diagnostic and statistical manual of mental disorders (DSM) [21], and utilization of neuroimaging modalities [22]. During a clinical interview, a specialist doctor engages in an information-gathering process and evaluates individuals with SZ, marking the initial step in the diagnostic procedure [18]. Physical examinations are conducted as another method to rule out any underlying medical conditions that might contribute to the individual's symptoms [19]. Psychological evaluations are also employed to assess cognitive abilities, memory and attention, aiding in the identification of brain regions associated with SZ [20]. DSM is a comprehensive reference book that provides diagnostic information for various brain disorders, including SZ [21]. Notably, several versions of the DSM have been publishedsuch as DSM-I (1952) [23], DSM-II (1968) [24], DSM-III (1980) [25], DSM-III-R (1987) [26], DSM-IV (1994) [27], DSM- IV-TR (2000) [28], and DSM-5 (2013) [21].

Neuroimaging modalities play a significant role as diagnostic methods in SZ [29]. These methods have gained popularity among specialist doctors due to their ability to provide crucial information about the structure and function of the brain during SZ [30]. The advantages of neuroimaging modalities in diagnosing brain disorders include their non- invasiveness, objectivity, early detection capability, precision and longitudinal monitoring [29-31]. Generally, neuroimaging modalities are divided into functional and structural methods [29]. Functional neuroimaging modalities assess brain function during the occurrence of SZ [30]. Notable examples of functional neuroimaging modalities are functional magnetic resonance imaging (fMRI) [32], positron emission tomography (PET) [33], single-photon emission computerized tomography (SPECT) [34], electroencephalography (EEG) [35], and magnetoencephalography (MEG) [36]. On the other hand, structural neuroimaging modalities focus on the brain's structure during SZ and provide important information to physicians [29]. Structural neuroimaging modalities encompass structural MRI (sMRI) [29], computed tomography (CT) [37], diffusion tensor imaging (DTI) [38], and magnetic resonance spectroscopy (MRS) [39].EEG recording is one of the functional neuroimaging modalities used in diagnosing brain disorders, including SZ[35][40]. During EEG recording, electrodes are placed on the scalp to capture the brain's electrical activity p [40]. In recent times, EEG signals have emerged as a powerful diagnostic tool for brain disorders such as SZ [40]. EEG provides specialist doctors with real-time information about the brain and when combined with other neuroimaging modalities, it enhances the understanding of brain activities for neurologists [41]. Some of the key advantages of EEG signals in diagnosing SZ include non- invasiveness, real-time monitoring, high temporal resolution and low cost [42-43]. However, there are some challenges associated with EEG signals that pose difficulties in SZ diagnosis for clinicians. Limited spatial resolution, susceptibility to artifacts, interpretation challenges and the limited ability to detect structural abnormalities are among the most significant challenges of EEG signals in SZ diagnosis [44-45].

To address the challenges associated with diagnosing SZ using EEG signals, researchers have proposed the idea of utilizing a computer-aided diagnosis system (CADS) based on AI techniques [46]. Extensive research is underway on CADS to diagnose various brain disorders including SZ from EEG signals [108-111]. CADS offers the capability tto automatically remove noise from EEG signals, thereby enhancing the accuracy of SZ diagnosis. Furthermore, the incorporatio of AI techniques in CADS implementation can improve the diagnostic accuracy of SZ diagnosis [120-122]. Researchers have already employed AI techniques, including ML and DL methods s in diagnosing SZ from EEG signals and have achieved promising outcomes. The primary objective behind the presentation of diverse AI techniques in CADS implementation is to aid in the early diagnosis of SZ and alleviate the workload of specialist doctors. It is anticipated that AI-based CADS can be implemented in hospitals and medical clinics to assist patients with SZ in the near future.

This study presents a comprehensive review of papers published on SZ diagnosis from EEG signals using ML and DL techniques. This review paper focuses on providing insights into future research directions in this field, examining SZ diagnosis research in detail, including its challenges and future prospects. The second section explores review papers on the diagnosis and prediction of SZ using AI techniques with neuroimaging modalities. The third section provides delves into the details of AI-based CADS for diagnosing SZ from EEG signals; Initially, it describes the key ML and DL techniques employed in SZ diagnosis.. Additionally, a summary of papers in this field based on ML and DL techniques is presented in Tables (3) and (4), respectively. The fifth section discusses the most important challenges associated with diagnosing SZ from EEG signals. In another section, the future directions are introduced along with pertinent details. The sixth section includes a discussion on the reviewed methods. Finally, we conclude our findings in the conclusion section.

## 2. Comparison of review papers in diagnosis of Schizophrenia

Extensive research is currently being conducted for the early diagnosis of SZ using AI methods in neuroimaging modalities. In this section, a comprehensive discussion of the review papers on SZ diagnosis using AI techniques is provided. In the following, first, review papers on the diagnosis and prediction of SZ disorder using different ML and DL techniques are examined and summarized in Table (1). Sadeghi et al [29] conducted a review of SZ diagnosis papers based on structural and functional MRI modalities using ML and DL techniques. They also identified the most significant challenges and future directions in this field. Likewise, authors in [53] reviewed SZ diagnosis papers using MRI neuroimaging modalities and ML methods. Steardo et al. [52] discussed SZ diagnosis papers that utilized SVM methods based on ML from functional MRI modalities, to investigate the efficacy of SVM techniques in enhancing the accuracy of SZ diagnosis. Other researchers have also reviewed SZ diagnosis papers using neuroimaging modalities and AI methods. For instance, Lai et al [51] conducted a review of papers on the diagnosis of SZ from EEG and MRI modalities. The authors' primary objective in this paper is to explore the significance of ML techniques in the diagnosis of SZ from EEG and MRI modalities. Cortes-Briones et al [48] reviewed the papers on the diagnosis and prediction of SZ using DL techniques, with a particular emphasis on studies that employed MRI and EEG modalities for SZ diagnosis. In another study, the authors presented a review paper on the diagnosis of SZ from EEG and MRI modalities using AI techniques [47]. The authors also introduced the most important future directions for SZ diagnosis using EEG and MRI modalities. Finally, review papers on SZ diagnosis from EEG signals were discussed in [49-50]. Luján et al. [50] conducted a comprehensive review of research applications of neurofeedback in SZ, with a particular focus on EEG signals and ML techniques. Barros et al. [49] presented a comprehensive review of SZ prediction papers that utilized EEG signals and AI techniques.

Table 1. Summary of review papers published on the diagnosis of SZ using AI methods.

| Ref | Year | Publisher | Propose | Modalities | AI Methods | Citations |
|---|---|---|---|---|---|---|
| [47] | 2023 | ArXiv | SZ detection using AI methods, Future works | MRI – EEG | ML – DL | 2 |
| [48] | 2022 | Elsevier | SZ detection and Prediction using DL models | MRI – EEG | DL | 31 |
| [29] | 2022 | Elsevier | SZ detection from MRI modalities using DL models | MRI Modalities | ML – DL | 70 |
| [49] | 2021 | Elsevier | SZ prediction using AI models | EEG | ML – DL | 36 |
| [50] | 2021 | MDPI | Applications of neurofeedback in SZ | EEG | ML | 18 |
| [51] | 2021 | MDPI | SZ detection using ML classifier methods | MRI – EEG | ML | 17 |
| [52] | 2020 | Frontiers | SZ detection using SVM methods | Functional MRI | ML | 50 |
| [53] | 2019 | Taylor | SZ detection using MRI neuroimaging modalities | MRI Modalities | ML | 63 |

Table 2. Exclusion and inclusion criteria used during the selection of papers.

| Inclusion | Exclusion |
|---|---|
| 1. EEG Signals<br>3. Different types of SZ.<br>3. SZ detection<br>4. ML methods<br>4. DL networks<br>6. Rehabilitation (IoT, Cloud computing, Hardware, etc.) | 1. Treatment of SZ<br>2. Clinical methods for SZ treatment<br>3. Rehabilitation systems (Without AI techniques) |

## 3. Search strategy

In this study, the search strategy for SZ diagnosis is carried out based on preferred reporting items for systematic reviews and meta-analyses (PRISMA) [54] guidelines at three levels. In our study, papers published from 2002 to 2023 focusing on SZ diagnosis from valid citation databases including Science Direct, Frontiers, IEEE, Nature, Springer and Wiley are chosen. To identify such papers, we used keywords including "schizophrenia", "EEG", "Electroencephalography", "machine learning", "feature extraction", and "deep learning".

Next, the literature review of SZ diagnosis using the proposed PRISMA [54] guidelines is discussed. Figure (1) displays the block diagram of the proposed PRISMA guideline, which consists of three levels of decomposition. According to the diagram, first, the authors downloaded 219 papers related to SZ diagnosis. The first stage of PRISMA is dedicated to removing out-of-scope 48 papers. In the second stage, 33 papers non-EEG signals related papers were removed. The third phase of PRISMA focused on investigating AI techniques for SZ diagnosis using EEG signals. At this stage, 12 papers were filtered for not using AI techniques. Finally, 126 papers were selected for this study.
Inclusion and exclusion criteria used in this SZ diagnosis research based on EEG signals and AI techniques are presented in Table (2). According to Table (2), the most important inclusion criteria in this field include SZ, EEG signals, ML methods and DL models. SZ disorder is the first inclusion criterion, as many individuals worldwide suffer from SZ brain disorder and their health is threatened with serious risks. Therefore, reviewing and summarizing the papers in this field can lead to rapid diagnosis of SZ using AI techniques. EEG recording is the second inclusion criterion. As mentioned earlier, EEG recording is one of the most important diagnostic methods for SZ due to its ability to provide important information about brain function and accessibility Consequently, much research has been conducted in this area using EEG modalities, and it is of a great significance to review the papers in this field. The main focus of this review paper is on SZ diagnosis using AI methods, which leads to ML and DL methods being the third and fourth inclusion criteria, respectively. A summary of SZ diagnosis papers using ML and DL techniques is reported in Tables (3) and (4), respectively. On the other hand, the exclusion criteria also include the treatment of brain diseases and clinical methods, which are not discussed in this work.

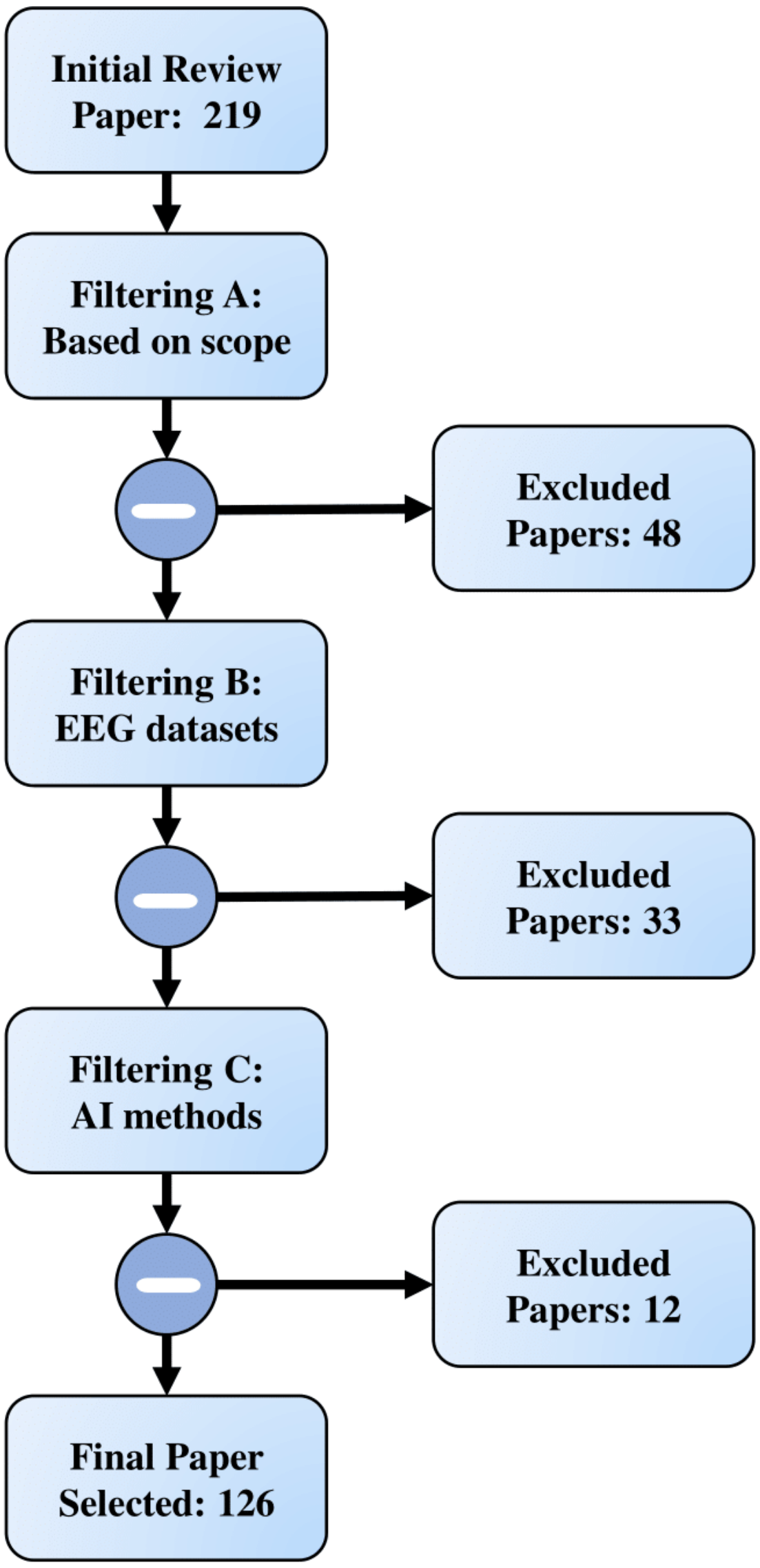

**Fig. 1.** Papers selection process employed using PRISMA guidelines for automated SZ detection.

## 4. Artificial intelligence in diagnosis of Schizophrenia from EEG signals

Today, the early diagnosis of various brain disorders including SZ using neuroimaging modalities has become a serious challenge for specialist doctors [29]. Neuroimaging modalities such as EEG present multiple challenges, which include the presence of various artifacts, long recording time, diverse signals for different subjects, complex interpretation of patterns, limited spatial and temporal resolutions [45-46], as well as overlapping symptoms of some brain disorders at the time of diagnosis, making it difficult for specialist doctors to differentiate between them. To address the challenges raised, so far extensive research has been carried out in the field of developing CADS for SZ detection and prediction from EEG signals using AI techniques [49]. Section 4 covers the reporting of review papers that explore the application of AI techniques for diagnosing and predicting SZ. It was observed that researchers have utilized both ML and DL techniques in their research. Dataset, preprocessing, feature extraction, dimensionality reduction, and classification are the most important components of CADS for SZ diagnosis [29]. ML-based CADS in SZ diagnosis offers numerous advantages, including explainability, training with limited input data, low computational power and reduced likelihood of overfitting [55-58]. On the other hand, the implementation of ML techniques in CADS for SZ diagnosis also involves several challenges, such as the need for trial and error in selecting feature extraction algorithms to achieve high performance, lack of efficiency in real-world applications, and inadequate performance when handling large input data [29][59-60]. In recent years, researchers have employed DL techniques in SZ diagnosis to address these challenges [48]. Compared to ML methods, DL techniques offer the benefits of automated feature extraction, better performance, scalability, generalizability, robustness to noise, and the ability to learn complex relationships [61-62]. Figure (2) displays the steps involved in CADS for SZ diagnosis using both ML and DL methods. In the rest of this section, ML methods in CADS diagnosis of SZ are described first, and then the relevant papers are summarized in Table (3). Finally, the popular DL architectures in SZ diagnosis are introduced, and then a summary of papers is outlined in Table (4).

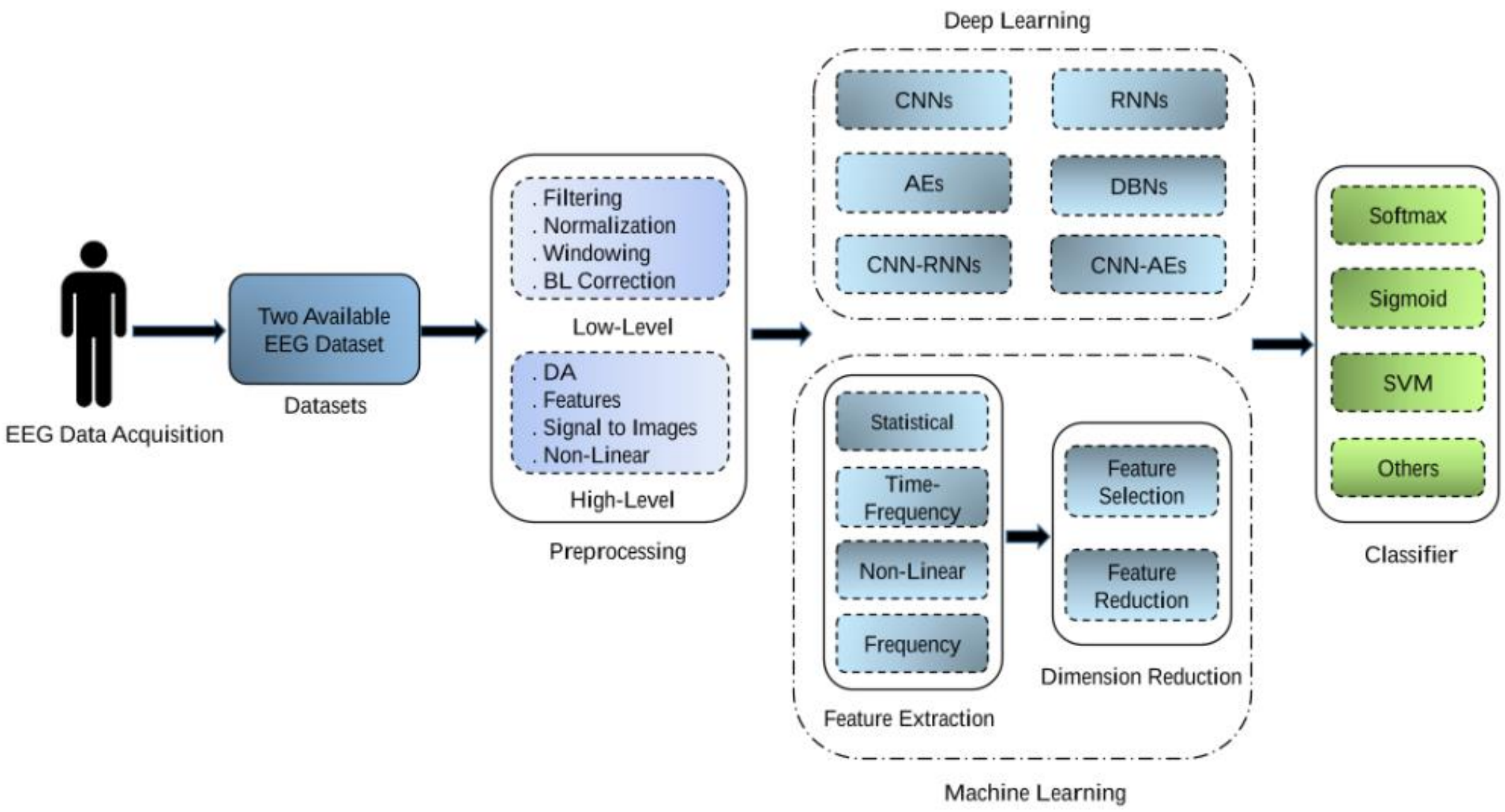

**Fig. 2.** A CADS block diagram for diagnosis of SZ using AI methods.

### 4.1. Available EEG datasets

EEG recording is one of the most important neuroimaging modalities used for diagnosing neurological disorders, such as epilepsy [63], SZ [49], and Parkinson's disease (PD) [64]. EEG is a non-invasive recording method and therefore poses no adverse effects for individuals with SZ. Nowadays, this method is cost-effective and is available in all hospital centers. One of the most significant advantages of EEG recording is its high temporal resolution, which enables doctors to detect abnormal changes in brain activity in real-time [63-65]. Moreover, this method has an objective feature, that is, it performs an objective measurement of brain activity without relying on the subjective report of the patient [63-65]. These advantages have made EEG recording popular among clinicians for diagnosing various brain disorders [63-65]. In this section, the available EEG datasets including RepOD [66] and Kaggle [67] used for SZ detection are introduced. Unfortunately, limited datasets have been provided for research in the field of SZ diagnosis, posing a significant challenge. Addressing this challenge could play a crucial role in developing AI algorithms to aid in the quick diagnosis of SZ. In the following, the details of the RepOD and Kaggle datasets are provided.

#### 4.1.1. RepOD dataset

This dataset was recorded by Olejarczyk et al. in 2017 for graph-based analysis of brain connectivity in SZ disease [66]. The dataset comprises EEG signals obtained from 14 patients aged between 27.9 to 28.3 years and 14 normal individuals of the same age and gender. Data was collected from individuals at the Institute of Psychiatry and Neurology in Warsaw, Poland [66]. All patients and subjects met the ICD-10 International Classification of Diseases criteria for SZ disorder. EEG signals were recorded using the standard 10-20 and a sampling frequency of 256 Hz. EEG recordings of all subjects were taken for 15 minutes with eyes closed and at a resting state. Additionally, EEG recording for all subjects with 19 channels including Fp1, Fp2, F7, F3, Fz, F4, F8, T3, C3, Cz, C4, T4, T5, P3, Pz, P4, T6, O1, O2 [66].

#### 4.1.2. Kaggle dataset

This section introduces the Kaggle dataset for SZ diagnosis, which comprises EEG signals from 23 individuals with SZ and 22 HC subjects [67]. The EEG signals were acquired with a BioSemi ActiveTwo system featuring 64 channels and a sampling frequency of 1 KHz from the scalp [67]. A band-pass filter ranging from 0.5 to 15 Hz was then applied. Various preprocessing techniques performed on this dataset include filtering, interpolation of outlier channels, segmentation, baseline correction, canonical correlation analysis, rejection of outlier single trials, removal of outlier components, and channel selection [67]. After the channel selection process, 9 channels, namely Fz, FCz, Cz, FC3, FC4, C3, C4, CP3, and CP4, were chosen for further processing. More details of this dataset are provided in reference [67].

### 4.2. Preprocessing techniques

EEG preprocessing aims to enhance signal quality and eliminate artifacts that hinder the accurate diagnosis of brain disorders [68]. EEG signals are prone to various types of noise, such as muscle artifacts, movements, and environmental interference, which can hide brain activity and make it difficult for clinicians to interpret the data [69]. Preprocessing methods help to remove various noises and improve the signal-to-noise ratio (SNR) of EEG signals. This process facilitates the detection of important patterns in EEG signals, enabling the extraction of meaningful information from this data for the early diagnosis of brain disorders such as SZ. In SZ diagnosis, preprocessing techniques play a critical role in improving the efficiency of AI-based CADS. In general, preprocessing methods include a variety of low-level and high-level techniques, each with distinct advantages. Low-level preprocessing methods primarily focus on noise removal and normalization of EEG signals. On the other hand, high-level preprocessing usually involves signal-to-image conversion methods, data augmentation (DA) [70], and feature extraction methods (used in DL architectures) that improve the accuracy of CADS for early SZ diagnosis. In the following, further details on low and high-level preprocessing methods are presented in the papers on SZ diagnosis using EEG signals

#### 4.2.1. Low-level preprocessing for EEG signals

This section introduces low-level preprocessing techniques used in studies focused on diagnosing SZ using EEG signals. According to Tables (3) and (4), filtering, segmentation, normalization, and artifact simulation are among the most important low-level preprocessing methods for EEG signals. EEG signals are often contaminated by various sources of noise, such as electrical interference, muscle activity, and movement artifacts [69]. These noises are often removed by different filters such as band-pass, low-pass, and high-pass in the low-level preprocessing stage [70-71]. Windowing is another low-level preprocessing step in EEG signals. During this stage, the EEG signals are segmented into smaller time intervals to enable more precise processing, thereby enhancing the resolution of information for SZ diagnosis using AI techniques [85-86]. The amplitude of EEG signals usually differs between multiple recording sessions or individual subjects. To overcome this challenge, normalization techniques such as z-score or base line correction are used [72-73]. Additionally, artifact simulation is an optional step used to evaluate the efficiency and performance of preprocessing algorithms, utilizing tools such as EEGLab [74]. In Tables (3) and (4), it is shown that researchers have employed various low-level preprocessing techniques in SZ diagnosis papers, with filtering, normalization, and segmentation is the most commonly used methods.

#### 4.2.2. High-level preprocessing for EEG signals

High-level preprocessing techniques play a crucial role in enhancing the efficiency of AI-based CADS for SZ diagnosis. This section outlines the commonly used high-level preprocessing methods used in SZ diagnosis, including artifact removal, spatial filtering, data augmentation (DA), signal-to-image conversion methods, and feature extraction methods (in DL applications). In the following, the details of each high-level preprocessing method are discussed. EEG signals can be affected by some artifacts such as blinks, and eye movements, as well as non-brain-related noise such as electromyogram (EMG). Some researchers have used methods such as independent component analysis (ICA) [80], principal component analysis (PCA) [112], empirical mode decomposition (EMD) [144], discrete wavelet transform (DWT) [88], and tunable Q-factor wavelet transform (TQWT) [114] for artifact removal in SZ diagnosis studies, achieving significant success results. In DL applications, few researchers have converted EEG signals into 2D images during the pre-processing stage [155]. As most DL architectures, including convolutional neural networks (CNNs), are designed for 2D data, the utilization of signal-to-image conversion methods can enhance SZ diagnosis accuracy. To this end, some researchers have utilized techniques like fast Fourier transform (FFT) [173], short-time Fourier transforms (STFT) [134], continuous wavelet transform (CWT) [152], and connectivity [165-166] methods in the pre-processing step to convert EEG signals into 2D images and achieved successful results in diagnosing SZ. Moreover, feature extraction methods are another category of high-level preprocessing techniques used in DL models. For example, Shoeibi et al [196] extracted a variety of fuzzy synchronization likelihood (FSL) features from EEG signals in the preprocessing step. The DL and FSL features were then fused using a concatenated layer.

### 4.3. Diagnosis of SZ using conventional machine learning methods

The application of ML methods in diagnosing various diseases, including brain disorders, has witnessed substantial growth in recent years [55-60]. High accuracy, interpretability, data efficiency, low computational power and speed are among the most significant advantages of ML techniques in various applications [55-58]. In some works, ML techniques have been employed to diagnose brain disorders such as SZ [49], AD [75], PD [76], and epilepsy [77]. Compared to DL architectures, ML methods are often more interpretable, which can aid in diagnosing brain disorders like SZ. ML methods require less training data, which is advantageous when dealing with ML-based CADS that have limited data. Furthermore, the training and evaluation of ML techniques are fast and require less computational power, which is important in clinical environments where the time of disease diagnosis is vital.

Given these advantages, ML methods are widely utilized in the diagnosis of SZ from EEG signals, and researchers hope to develop practical software for use in hospital centers in the future. ML-based CADS for SZ diagnosis generally involves dataset, pre-processing, feature extraction, feature selection, and classification [29]. ML-based and DL-based CAD systems differ in feature extraction and selection sections, while the remaining steps are similar [29]. In ML-based CADS, pre-processing is first carried out on EEG signals, followed by the application of feature extraction and selection methods to extract relevant patterns from EEG signals. The resulting features are then applied to training classification techniques [30]. In the following subsections, the most frequently used feature extraction and feature selection methods in SZ diagnosis studies are presented.

#### 4.3.1. Feature extraction methods

Feature extraction refers to the process of transforming input data into feature vectors, which typically have lower dimensions than the input data and represent its most important features [198-199]. This technique can reduce model complexity and training time by decreasing the input data's dimensions and preventing overfitting [198-199]. In ML-based CADS, feature extraction techniques are applied to identify appropriate patterns and characteristics in EEG signals. Feature extraction methods for EEG signals are classified into four categories: time [200], frequency [201], time-frequency [202] domains, and nonlinear [203]. In Table (3), papers on SZ diagnosis using ML methods are presented. Additionally, according to Table (3), nonlinear feature extraction methods are widely used by authors to diagnose SZ from EEG signals. The reason behind the popularity of feature extraction methods in EEG signals is these methods can capture the subtle changes in the EEG data. Nonlinear feature extraction methods used in SZ diagnosis using EEG signals include entropies [86], fractal dimensions (FDs) [143], graph-based methods [91], etc. These techniques can reveal hidden features in EEG signals, thereby improving the classification accuracy of SZ diagnosis.

**a) Times domain and statistical features**

In this section, time-domain feature extraction techniques used in SZ diagnosis studies from EEG signals are introduced. EEG signals carry valuable information about brain function when brain disorders occur, and the use of appropriate feature extraction methods can aid specialist doctors in quickly diagnosing SZ from EEG signals [81]. The time domain feature extraction methods are directly extracted from the time dimension of EEG signals and include features such as amplitude, variance, mean, skewness, kurtosis and waveform morphology [200]. These features are easy to calculate and interpret, and they can provide useful information about the level of overall brain activity. In part of Table (3), handcrafted feature extraction methods from EEG signals used in SZ diagnosis papers are presented. Based on Table (3), time-domain feature extraction methods have been utilized in numerous researches SZ diagnosis. For instance, some researchers in reference [81] have effectively employed time domain features in SZ diagnosis and have yielded promising results.

**b) Frequency domain features**

Frequency domain features are another set of techniques used to extract features from EEG signals. These techniques involve mapping EEG signals from the time domain to the frequency domain using techniques like FFT [119], after which essential features are extracted from them. Power spectral density (PSD) [100], coherence [204], and phase synchronization (PS) [205] are important techniques for extracting features of the frequency domain. So far, several researches have utilized frequency domain feature extraction methods in SZ diagnosis, as summarized in Table (3). In references [100], the authors used PSD in conjunction with other feature extraction methods to diagnose SZ. Other studies used linear predictive coding (LPC) [96] and frequency bands [106] methods for SZ diagnosis. It has been demonstrated that frequency features contain more critical information for diagnosing brain disorders compared to time-domain features. The section also notes that frequency domain techniques are used as high-level preprocessing in DL research. Table (4) illustrates that frequency domain techniques have been employed as high-level preprocessing.

**c) Time-frequency domain features**

Time-frequency domain feature extraction methods are introduced to address the challenges posed by time and frequency domain methods [202]. The features of this approach offer several advantages, including capturing dynamic changes, improved resolution, better identification of event-related potentials (ERPs) and enhanced classification accuracy [206-208]. As discussed, EEG signals exhibit highly chaotic behavior and change over time. Time-frequency domain features are capable of simultaneously extracting these changes in both time and frequency domains [202]. Furthermore, these features display better resolution compared to features extracted solely in the time or frequency domain, as they enable the analysis of EEG signals at different frequencies and time points [202]. Small changes in EEG signals contain valuable information, which is called ERPs. ERPs typically occur in response to specific stimuli or events and therefore their analysis is of great significance [209-210]. References [210-211] have indicated that time-frequency domain features play a critical role in identifying ERPs. The most important feature extraction methods in this domain include DWTs [164], CWTs [152], STFTs [134], EMDs [144], and TQWTs [114]. Table (3) shows that some studies on SZ detection have employed time-frequency domain methods as a feature extraction step. For instance, researchers utilized EMD [144], Stockwell [95], and TQWT coefficients [131] to diagnose SZ from EEG signals and achieved good outcomes. Additionally, Table (4) reports papers on SZ diagnosis that employed DL techniques, where time-frequency domain methods have been exploited as a high-level preprocessing stage in these studies.

**d) Nonlinear features**

Nonlinear features are the most important feature extraction methods in EEG signals and are widely employed for brain disorders diagnosis [203]. EEG signals exhibit non-linear behavior over time, so nonlinear methods have the ability to capture complex patterns in these signals that cannot be identified with other methods [212-213]. Nonlinear methods provide a range of advantages, such as Improved accuracy, increased sensitivity, and robustness as well as a better understanding of brain function and the potential for personalized medicine [212-213]. Nonlinear techniques exhibit a greater degree of sensitivity toward changes in EEG signals, as compared to other techniques. This advantage significantly contributes to improving the accuracy and effectiveness of diagnosing brain disorders from EEG signals. Furthermore, these methods are robustness against artifacts when compared to linear feature extraction techniques [212-213]. Additionally, non-linear feature extraction methods possess a unique capability to display synchronization and desynchronization patterns between brain regions, which cannot be detected by alternative feature extraction techniques [212-213]. Hence, the advantages provided by nonlinear feature extraction methods can be important in the development of diagnostic software for neurological disorders based on EEG signals. Nonlinear features such as FDs [143], entropies [86], graphs [91], and synchronization likelihood methods [196] are significant in the analysis of EEG signals. Table (3) presents the feature extraction algorithms utilized for SZ diagnosis. According to the Table, researchers predominantly utilized non-linear features, such as entropies [86], graph analysis [91], FDs, and connectivity methods [98] for the diagnosis of SZ.

### 4.3.2. Dimension reduction techniques

Dimension reduction is a technique used in ML to reduce the number of features or variables in a dataset [212-213]. It is often used to simplify the complexity of a dataset, to make it easier to analyze or visualize, or to improve the performance of a ML model [212-213]. These methods are divided into two categories: feature reduction [214-215] and feature selection methods [216-217]. Feature reduction is an important technique for reducing the dimensionality of data and improving the performance of ML models [214-215]. The choice of feature reduction technique depends on the nature of the data and the problem at hand. Linear techniques, such as principal component analysis (PCA) [107] and linear discriminant analysis (LDA) [97], are often used when the data is linearly separable or when the goal is to reduce the dimensionality of the data. Non-linear techniques, such as t-distributed stochastic neighbor embedding (t-SNE) [218], are often used when the data is non-linear or when the goal is to visualize the data in a low-dimensional space. Feature selection involves selecting a subset of the original features that are most relevant to the problem at hand [216-217]. This can be done using various techniques, such as correlation analysis [219], mutual information [220], Wrapper [221], recursive feature elimination (RFE) [222], and regularization [223]. The goal is to retain only the most informative features while discarding irrelevant or redundant ones. The dimension reduction methods are displayed in part of Table (3).

Table 3. A summary of ML research for diagnosis of SZ from EEG signals.

| Works | Dataset | Number of Cases | Preprocessing | | Feature Extraction | Dimension Reduction | Classifier | Performance (%) |
|---|---|---|---|---|---|---|---|---|
| | | | Low Level | High-Level | Linear | | | |
| [78] | Clinical | 48 | Filtering | -- | SpEn, InEn, ShEn, HFD, KOL, ApEn | ANOVA | SVM | Acc = 88.5 |
| [79] | Kaggle | 49 SZ, 32 HC | Filtering | -- | 10 IMF Components | KW | EBT | Acc = 89.59<br>Sens = 89.76<br>Spec = 89.32 |
| [80] | RepOD | 14 SZ, 14 HC | -- | ICA | DFA, HE, RQA, FD, KOL, LZC, LLE | BH | SVM | Acc = 92.17 |
| [81] | Clinical | 26 SZ, 22 HC | -- | EWT | Statistical Features | KW | SVM | Acc = 88.7<br>Sens = 91.13<br>Spec = 89.29 |
| [82] | RepOD | 14 SZ, 14 HC | Normalization | ICA, Signal Decomposition | EEG Decomposition Coefficients | NA | RF | Acc = 71.43<br>Sens = 100<br>Spec = 60.0 |
| [83] | Clinical | 14 SZ, 23 HC | -- | -- | Autoregressive parameters (AR) | NA | ANN | -- |
| [84] | Clinical | 65 SZ, 40 HC | Segmentation, Filtering | -- | Different Features | KW | DT | Acc = 71.76 |
| [85] | Clinical | 54 SZ, 54 HC | Filtering, Normalization | -- | Auditory P300 + Visual P300 + MMN | NA | FR | Sens = 84.4<br>Spec = 85.0 |
| [86] | Kaggle dataset | 49 SZ, 32 HC | Filtering | -- | Kolmogorov Complexity and Sample Entropy | NA | MLP | Acc = 91.25<br>Sens = 90.8<br>Spec = 93.2<br>Prec = 96.0 |
| [87] | Clinical | 5 SZ, 5 HC | Segmentation, Filtering | | Statistical Features, Sample Entropy | GA | SVM | Acc = 88.24<br>Sens = 89.48<br>Spec = 87.0 |
| [88] | Clinical | 63 SZ, 70 HC | Filtering | DWT, ICA | Symbolic Transfer Entropy (STE) | NA | SVM, KNN | Acc = 96.15<br>Sens = 100<br>Spec = 92.86 |
| [89] | RepOD | 14 SZ, 14 HC | Segmentation, Filtering | DWT | AVLSAC, ShEn, SpEn, ApEn | NA | ANFIS | Acc = 99.92 |
| [90] | RepOD | 14 SZ, 14 HC | -- | ICA | EM-PCA, PLS-NLR | Flower Pollination | Adaboost | Acc = 98.77 |
| [91] | RepOD | 14 SZ, 14 HC | Filtering | Signal Decomposition, ICA | Graph Features Extracted from phase lag index (PLI) | ANOVA | Logistic regressio n | Acc = 97.0<br>Sens = 95.0<br>Spec = 98.0 |
| [92] | MHRC | 45 SZ, 39 HC | -- | -- | SLBP | Correlation | LogitBoo st | Acc = 91.66<br>Sens = 93.33<br>Spec = 89.74 |
| [93] | RepOD | 14 SZ, 14 HC | Segmentation, Filtering | DWT | $\ell 1$ norm | NA | KNN | Acc = 99.21<br>Sens = 99.42<br>Spec = 99.05 |
| [94] | MHRC | 45 SZ, 39 HC | Segmentation, Filtering | ICA | Dynamic Connectivity Features | K-S test | SVM | Acc = 94.05<br>Sens = 95.56<br>Spec = 92.31 |
| [95] | Clinical | 7 SZ, 7 HC | Segmentation, Filtering, Normalization | ICA | Time/Frequency Representation via Stockwell Transform | NA | KNN | Acc = 88.7<br>Sens = 77.4<br>Spec = 100 |
| [96] | RepOD | 14 SZ, 14 HC | -- | STFT | LPC Coefficients | t-test | SVM, KNN, DT | Acc = 97.2<br>Sens = 96.81<br>Spec = 97.67 |
| [97] | Clinical | 19 SZ,<br>23 HC | Filtering, Segmentation, Baseline Correction | -- | SPN Features | NA | LDA, SVM | Acc= 90.48<br>Sen= 89.47<br>Spe= 91.30 |
| [98] | Lomonosov Moscow State University | 45 SZ,<br>39 HC | Filtering | -- | Connectivity Features | NA | RF | Acc= 82.36 |
| [99] | COBRE | 86 Subjects | -- | -- | | Recursive Feature Elimination | SVM,RF, NB | Acc= 90.697 |
| | | | | | | | ANN | Acc= 88.37 |
| [100] | Clinical | 52 SZ,<br>29 HC | Filtering | -- | Welch Power Spectral Density (PSD) | t-test | SVM | Acc= 88<br>Sen= 91<br>Spe= 86 |
| | | | | | | | BPN | Acc= 88<br>Sen= 90<br>Spe= 86 |
| [101] | Clinical | 31 SZ,<br>31 HC | Filtering | DWT | Permutation Entropy, Kolmogorov Entropy, Correlation Dimension, Spectral Entropy | Fisher | KNN, SVM, **MLP** | Acc= 86.1 |
| [102] | Clinical | 5 SZ,<br>5 HC | Filtering | DWT | Statistical Features, Sample Entropy | GA | SVM | Acc= 88.24<br>Sen= 89.48<br>Spe= 87 |
| [103] | Online Resources | 49 SZ,<br>32 HC | Baseline Correction, Filtering | -- | Different Features | NA | RF | Acc= 81.1 |
| [104] | Clinical | 119 SZ,<br>119 HC | Filtering | -- | Brain Network Features (Global and Local Clustering Coefficients, and Global Path Length) | SFS | LDA | Acc= 80.66<br>Sen= 78.83<br>Spe= 82.48 |
| [105] | Clinical | 34 SZ,<br>34 HC | Filtering | -- | Source Level and Sensor Level Features | Fisher | SVM | Acc= 78.24 |
| [106] | Clinical | 40 SZ,<br>12 HC | Filtering, Segmentation | | Scalp locations, Information Processing Stages, Frequency Bands | Wrapper method | SVM | Multiple results |
| [107] | Clinical | 49 SZ,<br>50 HC | Filtering | Signal Decomposition | Extract Multiple Topological Attributes, Constructing Brain Function Network | RFE, PCA, ANOVA | SVM, RF, LDA, LR, KNN | Acc= 91.7<br>AUC= 96.5<br>Sen= 91.7<br>Spe= 91.7 |
| [108] | Clinical | 45 SZ, 30 HC | Filtering, ERPs Analysis | ICA | Functional Connectivity, Graph Features | Fisher score, T-test | SVM | Acc= 84.48 |
| [109] | Clinical | 25 SZ, 25 HC | Filtering, Normalization | ICA | Connectivity Features | Fisher | NA | Acc= 93.8<br>Spe= 100<br>Sen= 87.6 |
| [110] | Clinical | 16 SZ, 31 HC | Filtering, Segmentation | ICA | Different Time and Frequency Features | Different Methods | MLP, SVM | Spe= 96.73,<br>Sen= 87.27 |
| [111] | RepOD | 14 Paranoid SZ, 14 HC | -- | -- | Complexity, HFD, LLE | NA | PNN | Acc= 100<br>Spe= 100<br>Sen= 100 |
| [112] | Clinical | 17 Moderately Ill, 17 Markedly Ill, 10 HC | Filtering | PCA | Entropy Features | NA | SVM | Different Results |
| [113] | Clinical | 15 SZ,<br>18 HC | Filtering | Visual Inspections | Statistical Features, Two Spectral Features, Two Non-Linear Features | T-test | KNN | Acc= 94 |
| [114] | Kaggle | 49 SZ, 32 HC | Baseline Correction, Filtering | TQWT | Statistical Features | KW | F-LSSVM | Acc= 91.39 Sen= 92.65 Spe= 93.22 |
| [115] | Clinical | 312 SZ, 320 HC | Filtering | -- | Fuzzy Features | NA | RFB | Acc=93 |
| [116] | Clinical | 54 SZ, 54 HC | Filtering, Baseline Correction | -- | Peak Related Features, Peak to Peak Related Features, and Signal Related Features | Boruta feature selection | Multiple Kernel Learning | Acc=86 |
| [117] | Kaggle | 49 SZ, 32 HC | -- | RVMD, ICA | Statistical Features | KW | OELM | Acc=92.93 Sen=97.15 Pre=93.94 |
| [118] | RepOD | 14 SZ, 14 HC | -- | -- | Entropy Features | ANOVA | Naive Bayes | Acc=100 |
| [119] | RepOD | 14 SZ, 14 HC | Segmentation | FFT | Spectral Features | NA | RF | Acc=74.99 |
| [120] | RepOD | 14 SZ, 14 HC | -- | MDWT | Statistical Features | NA | Ada-Boost | Acc=85.71 |
| [121] | Clinical | 11 SZ, 20 HC | ICA | | Different Linear Features, Different Non-linear Features | PCA | SVM | Acc=89 |
| [122] | Clinical | 2677 SZ | Segmentation | -- | Multi-class Spatial Pattern of the Network (MSPN) Features | ANOVA | SVM | Acc=71.58 |
| [123] | LMSU | 45 SZ, 39 HC | -- | DWT | Time Domain Features, Time-Frequency Features | KW | ANN | Acc=100 |
| [124] | Clinical | 35 SZ, 30 Ultra-High-Risk (UHR) | Segmentation | -- | Dynamic Features | KW | ??? | Acc=71.76 |
| [125] | Moscow State University, RepOD | D1: 45 SZ, 39 HC<br>D2: 14 SZ, 14 HC | -- | -- | HLV Features, SLBP Features | CBFS | Ada-Boost | Acc=99.36 |
| [126] | Moscow State University, RepOD | D1: 45 SZ, 39 HC<br>D2: 14 SZ, 14 HC | -- | ICA, Optimization Methods (WBA, GSO, and BA) | Time Domain and Statistical Features, Spectral Features | ReliefF | ANFIS | Acc=99.54 |
| [127] | Clinical | 70 SZ, 75 HC | Down-sampling, Segmentation | ICA | Statistical Features | ANOVA | SVM | Acc=82.7 Sen=83.5 Spe=85.3 |
| [128] | RepOD | 14 SZ, 14 HC | Re-referencing, Segmentation | -- | Microstate Features | NA | SVM | Acc=90.93 Sen=91.3 Spe=90.48 |
| [129] | RepOD | 14 SZ, 14 HC | Segmentation | Signal Decomposition | Cyclic Group of Prime Order Pattern (CGP17Pat) | INCA | KNN | Acc=99.91 |
| [130] | RepOD | 14 SZ, 14 HC | Segmentation | -- | Non-linear Features | T-test | SVM | Acc=92.91 |
| [131] | RepOD | 14 SZ, 14 HC | Segmentation | TQWT | TQWT Sib-bands Coefficients | ReliefF | KNN | Acc=99.12 |
| [132] | Clinical | 40 SZ, 7 Schizoaffective Disorder | -- | Visual Inspection, ICA | Transfer Entropy (TE) | RSFS | RF | Acc=85.2 Sen=88.7 Spe=81.8 |
| [133] | RepOD | 14 SZ, 14 HC | Re-referencing, Segmentation | -- | Microstate and Conventional Features (Time - Statistical) | RFE | SVM | Acc=75.64 |
| [134] | RepOD | 14 SZ, 14 HC | Segmentation | STFT, EWT | Graphical Features | FSA | KNN | Acc=94.80 Sen=94.3 Spe=95.2 |
| [135] | Two public schizophrenia corpora (DB1 and DB2) | DB1: 626 SZ, 516 HC<br>DB2: 2778 SZ, 1834 HC | -- | Signal Decomposition | Different Features | INCA | KNN | Acc=99.47 |
| [136] | Clinical | 62 SZ, 70 HC | -- | Wavelet-enhanced Independent Component | Symbolic Transfer Entropy (STE) matrix | NA | KNN | Acc=96.92 Sen=95 Spe=98.57 Pre=98.33 |
| | | | | Analysis (wICA) | | | | |
| [137] | RepOD | 14 SZ, 14 HC | Segmentation | Signal Decomposition | Hjorth Parameters | *t*-test | SVM | Acc=98.9<br>Sen=99<br>Spe=98.8 |
| [138] | National Institute of Mental Health | 49 SZ, 32 HC | Filtering | PCA | Hjorth Parameters, Entropy Features | NA | ANN | Acc=93.9 |
| [139] | Clinical | 36 SZ, 36 HC | Segmentation | FFT | Linear Features, Non-linear Features | Feature Ranking | SVM | Acc=99.31 |
| [140] | RepOD | 14 SZ, 14 HC | Segmentation | MEMD | Entropy Features | RFE | SVM | Acc=93 |
| [141] | Clinical | 3002 SZ, 3931 Have Other Mental Disorders | Normalization | -- | Different Features | Different Methods | RF | Acc=72.7 |
| [142] | Clinical | 40 Clinically High-risk Individuals (CHR), 40 FES<br>40 HC | -- | ICA | Neumann Entropy | ANOVA | SVM | Acc=90 |
| [143] | Clinical | 13 SZ, 18 HC | Segmentation | | Autoregressive (AR), Band Power, Fractal Dimension | NA | BDLDA | Acc=87.5 |
| [144] | Kaagle | 49 SZ, 32 HC | Baseline Correction, Canonical Correlation Analysis | EMD | Statistical Features | KW | EBT | Acc=89.59 |
| [145] | RepOD, Kaagle | D1: 14 SZ, 14 HC<br>D2: 49 SZ, 32 HC | -- | MAP Decomposition | Collatz Pattern | INCA | KNN | Acc=99.47 |
| [146] | RepOD | 14 SZ, 14 HC | -- | Decomposition, MSPCA | Graphical Features | FSA | KNN + GRNN | Acc=94.8<br>Sen=94.3<br>Spe=95.2 |
| [147] | Clinical | 40 CHR, 40 FES, 40 HC | -- | ICA, FFT | LES Test | NA | SVM | Acc=79.16 |
| [148] | Clinical | 11 SZ, 9 HC | Re-referencing, Baseline Correction | ICA | Statistical Features | NA | LDA | Acc=71 |
| [149] | Clinical | 12 HR, 14 HC, 19 UHR, 20 FESZ | Segmentation, Re-referencing | -- | Duration, Occurrence and Time Coverage | NA | RF | Acc=92 |
| [150] | RepOD | 14 SZ, 14 HC | Segmentation | -- | Highest Slope of Autoregressive Coefficients (AVLSAC), Shannon, Spectral, and Approximate Entropy | NA | ANFIS | Acc=100 |
| [348] | RepOD, Kaagle | D1: 14 SZ, 14 HC<br>D2: 49 SZ, 32 HC | Segmentation | DFT | Statistical Features, Look Ahead Pattern (LAP) Features | KW | Boosted Trees Classifier | Acc=99.24 |
| [349] | Kaagle | 49 SZ, 32 HC | Segmentation | TQWT | ITQWT TQWT Sub-bands Coefficients | INCA | KNN | Acc=99.2 |
| [350] | RepOD | 14 SZ, 14 HC | Segmentation | Wavelet Scattering Transform (WST), CWT, DWT | Statistical Features | -- | Decision Trees+ LR+ RF | Acc=97.98<br>Sen=98.2 |
| [351] | Clinical | 68 SZ, 132 HC | Segmentation | ICA | Global Average, Max, and Min Features | Sequential Feature Selection | SVM | Sen=91<br>Spe=90.8 |

### 4.4. Diagnosis of SZ using deep learning architectures

Deep learning (DL) has emerged as a sub-field of artificial intelligence, demonstrating remarkable growth in various fields in recent years [224-227]. DL architectures offer a range of features that make them highly desirable, such as automated feature extraction, better performance, scalability, generalizability, robustness to noise, and the capacity to learn complex relationships [228-230]. DL models have proven to be highly successful in diagnosing a wide range of diseases using medical data [231-235]. These networks are capable of analyzing a large amount of medical data and identifying patterns that may not be recognizable to the human eye [231-233]. In particular, DL models have become increasingly popular in diagnosing brain disorders using neuroimaging modalities [236-240]. These models can extract complex patterns and relationships from neuroimaging modalities such as EEG signals, which are challenging to identify using traditional ML methods [239-240]. In the field of SZ diagnosis, considerable research has been conducted utilizing a variety of DL techniques. This review paper focuses on the application of DL techniques for SZ diagnosis using EEG signals. In this section, the most important DL models including CNNs, PreTrained, recurrent neural networks (RNNs), and autoencoders (AEs) are discussed to diagnose SZ from EEG signals.

#### 4.4.1. CNNs

CNN architectures are an important class of DL models that are widely applied in various applications, including signal and medical image classification [244-245]. These models comprise several crucial layers, including convolution, pooling, and fully connected (FC) layers. Of these layers, convolution layers are the most significant part of the architecture of CNNs [241-243]. They apply a set of filters with varying dimensions to the input data, and the output of each convolution layer is a collection of feature maps generated for (from) the input data [241-242]. Additionally, pooling layers are employed to reduce the spatial dimensions of the data while preserving their essential information [241-243]. Usually, this layer is inserted after each convolution layer [241]. Furthermore, FC layers are utilized to classify the input data [241]. In recent years, CNN models have been leveraged for diagnosing SZ disorder from EEG signals, and researchers have achieved promising outcomes. Table (4) presents SZ diagnosis papers based on DL techniques. Researchers have utilized various CNN models, such as 1D-CNNs, 2D-CNNs, and PreTrained models, to diagnose SZ from EEG signals.

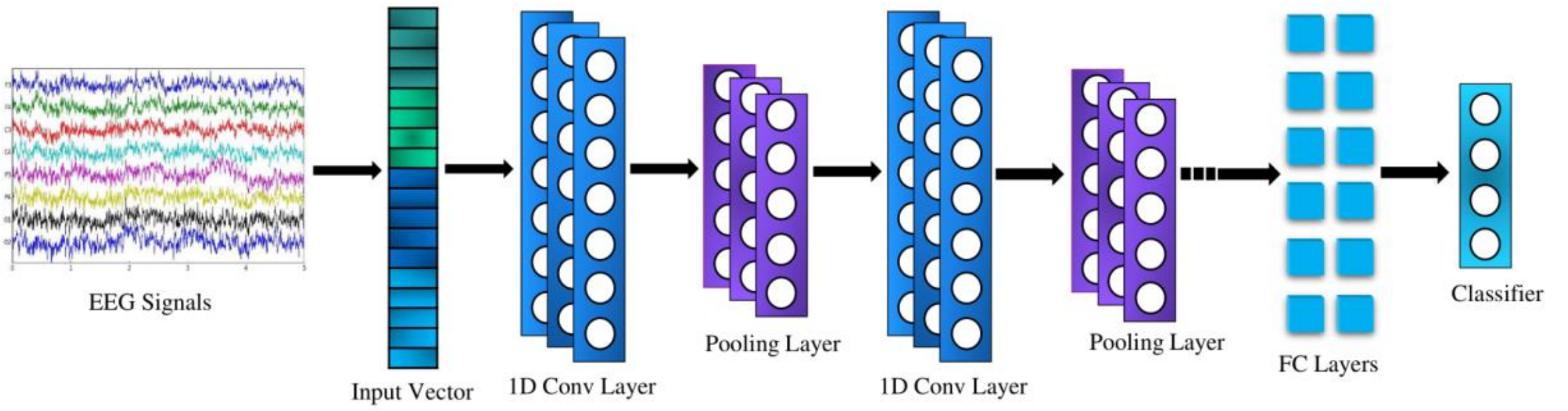

**Fig. 3.** A typical 1D-CNN model used for SZ detection using EEG signals.

#### a) 1D-CNNs

1D-CNNs are a sub-branch of CNNs architectures that are specifically designed for processing time-series such as medical signals [246-247]. Artificial neural networks (ANNs) typically rely on look-up tables for time-series analysis, whereas 1D-CNN models are capable of effectively processing various types of time-series data, including EEG signals [248]. The most crucial component of 1D-CNN models is the 1D convolution layer, which comprises several learnable filters that are applied to time-series data to extract feature maps [241-243]. These maps capture local patterns and relationships among adjacent data points, enabling the network to learn meaningful representations of the input data [241-243]. In addition to the 1D convolution layer, 1D-CNN models also incorporate other layers, such as pooling and FC layers [241-242]. However, 1D-CNN models require more training data and longer training times compared to traditional neural networks [246]. Table (4) provides a summary of SZ diagnosis papers utilizing DL techniques. The Table indicates that numerous researchers have employed different models of 1D-CNNs to diagnose SZ. Moreover, Figure (3) illustrates the general block diagram of the 1D-CNN model for SZ diagnosis.

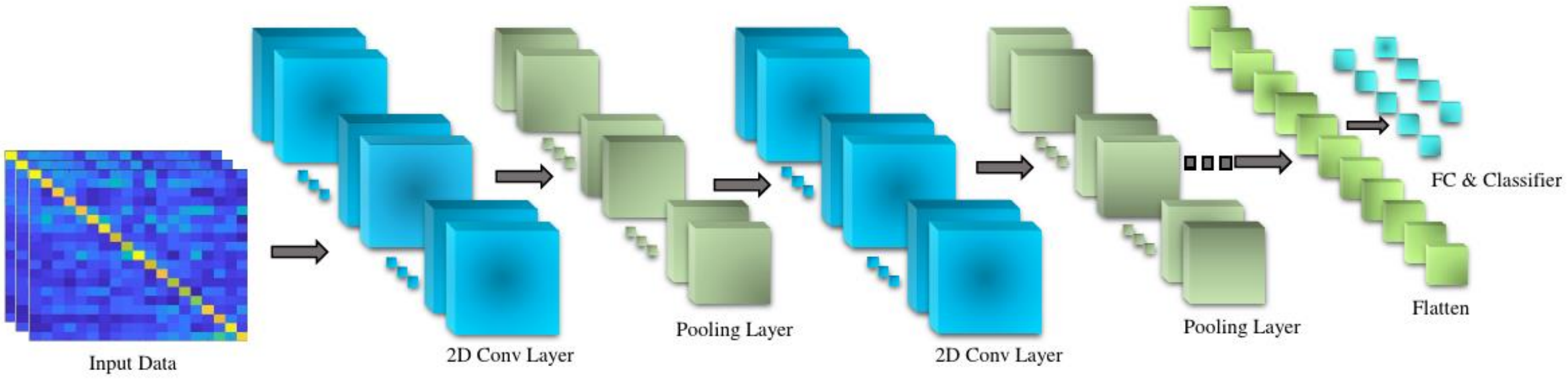

**Fig. 4.** A typical 2D-CNN model used for SZ detection using EEG signals.

**b) 2D-CNNs**

2D-CNNs have gained more popularity than other CNN architectures and are now widely used in a diverse range of applications such as image classification [241-243]. These networks are composed of convolutional layers, pooling layers, and FC layers, similar to 1D-CNN architecture [241-243]. Convolution layers are designed to detect specific patterns such as edges and textures and can capture local information in the image [241-243]. Following convolution layers, pooling layers are commonly applied to preserve important features [241-243]. The final layer of the network, known as the FC layer, is responsible for classification, and its output is typically the Softmax function [241]. The remarkable performance of 2D-CNNs has led to their use in brain disorder diagnosis applications [249-250]. According to Table (4), some researchers have employed 2D-CNNs to improve the accuracy of SZ diagnosis. Since EEG signals are 1D, it is imperative to first convert them into 2D images using high-level preprocessing techniques, such as STFT, to apply 2D-CNN models to extract important frequency features. Despite the advantages of 2D-CNNs, they pose challenges such as high training time and computational complexity. The general block diagram of the 2D-CNN model for SZ detection is displayed in Figure (4).

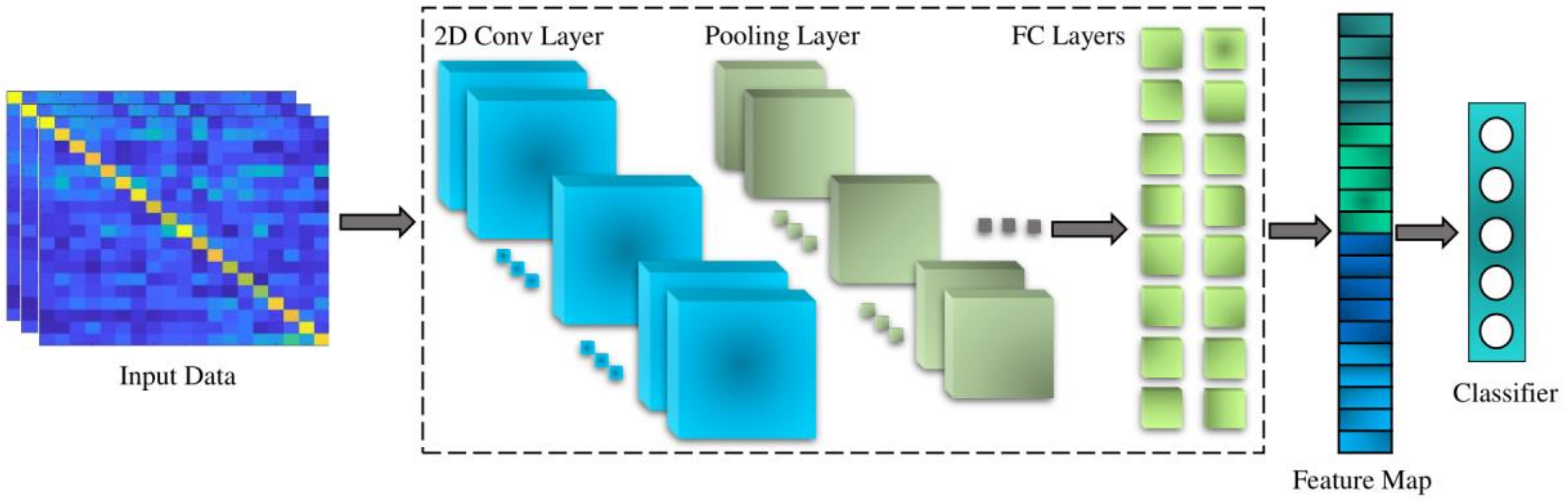

**Fig. 5.** A typical pre-trained model used for SZ detection using EEG signals.

**c) Pre-trained**

Pre-trained models, which are a sub-branch of CNNs, have been pre-trained for various applications like image classification and speech recognition [241-243]. Among the most popular Pre-trained architectures are GoogleNet, VGG, AlexNet, and ResNet [241-243]. Pre-trained networks use supervised learning during their training step, where they are first trained on the large ImageNet dataset, and then their weights are fine-tuned through error backpropagation [241]. The resulting weights are then stored and can be used for diverse applications, including the classification of small datasets [241]. Since Pre-trained models are trained on extensive amounts of data, they can effectively learn complex patterns and relationships in the data [241]. In medical research, the lack of access to large datasets is one of the most significant challenges, and pre-trained models have largely been able to overcome this challenge for medical classification problems [249-250]. To date, many researchers have utilized Pre-trained architectures to diagnose brain disorders from neuroimaging modalities [251-252]. For instance, as shown in Table (4), some researchers have applied Pre-trained models to diagnose SZ disorder. The general block diagram of a pre-trained model for SZ diagnosis is shown in Figure (5).

### 4.4.2. RNNs

RNN models are a class of DL architectures that are widely utilized in time series forecasting applications [241-243]. These networks rely on unsupervised learning during the training stage and possess feedback connections that enable them to preserve a state or internal memory and retrieve information from previous time steps [241-243]. As a result, this memory enables RNNs to model temporal dependencies in data and perform time series prediction. There are different types of RNN models and they include simple RNN, long-short-term memory (LSTM), and Gated recurrent unit (GRU) [241]. prominent features of RNN architectures include their ability to process sequential data, memory, flexibility and feature representation [241]. However, RNNs have some limitations, including vanishing gradient, computationally expensive, difficulty capturing long-term dependencies and sensitivity to hyperparameters [241-242]. In recent years, the use of RNNs techniques in the diagnosis and prediction of brain disorders from EEG signals has grown significantly [253-254]. Table (4) presents papers on SZ diagnosis from EEG signals using DL techniques. some researchers have reported satisfactory results in SZ diagnosis using RNN models, as seen in references [151][173]. The general block diagram of an RNN model for SZ detection is shown in Figure (6).

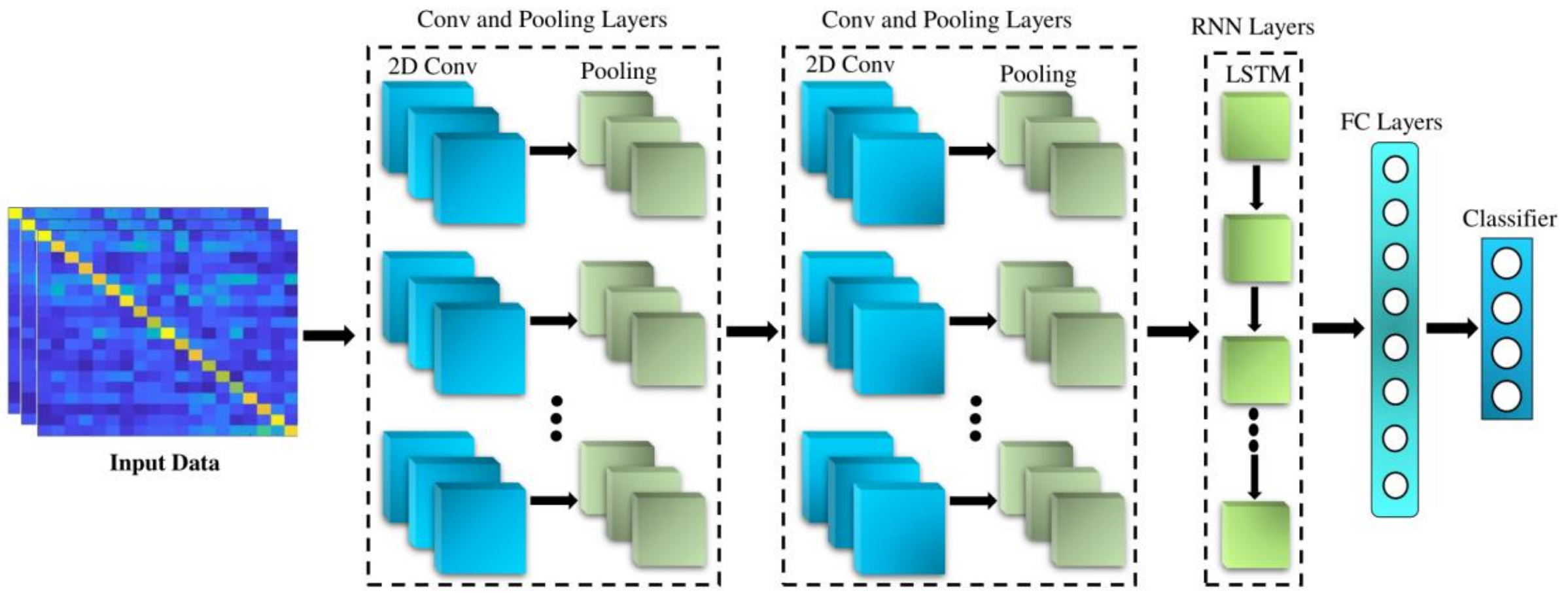

**Fig. 6.** A typical CNN-RNN model used for SZ detection using EEG signals.

### 4.4.3. AEs

Autoencoder (AE) networks are a group of DL networks commonly employed for compression, data reconstruction, or feature extraction applications [241-242]. These networks consist of two main components, namely the encoder and decoder [241-242]. An encoder takes input data and maps it to a hidden space, which is a lower-dimensional representation of the data [241]. The decoder then takes the representation of the hidden space and returns it to the original input space [241]. During training, the autoencoder learns to minimize the reconstruction error between the input and output data of the decoder. The most important AE architectures include basic AEs, Sparse AEs, denoising AEs, Stacked AEs, and convolutional AE (CAE) [241]. AEs architectures offer several advantages, such as dimensionality reduction, feature extraction, noise reduction, generative modeling and unsupervised learning [241-242]. However, they also have some limitations, including Overfitting, difficulty in handling large datasets, limited interpretability, and difficulty in handling high-dimensional data [241-242]. In medical research, AEs models are frequently employed to extract features from medical data [255-256]. AEs models have been utilized in research to diagnose brain disorders based on neuroimaging modalities [257-258]. In references [157][180], researchers have achieved satisfactory results in diagnosing SZ from EEG signals using AEs models. In Figure (7), the general block diagram of an AE architecture for SZ detection is shown.

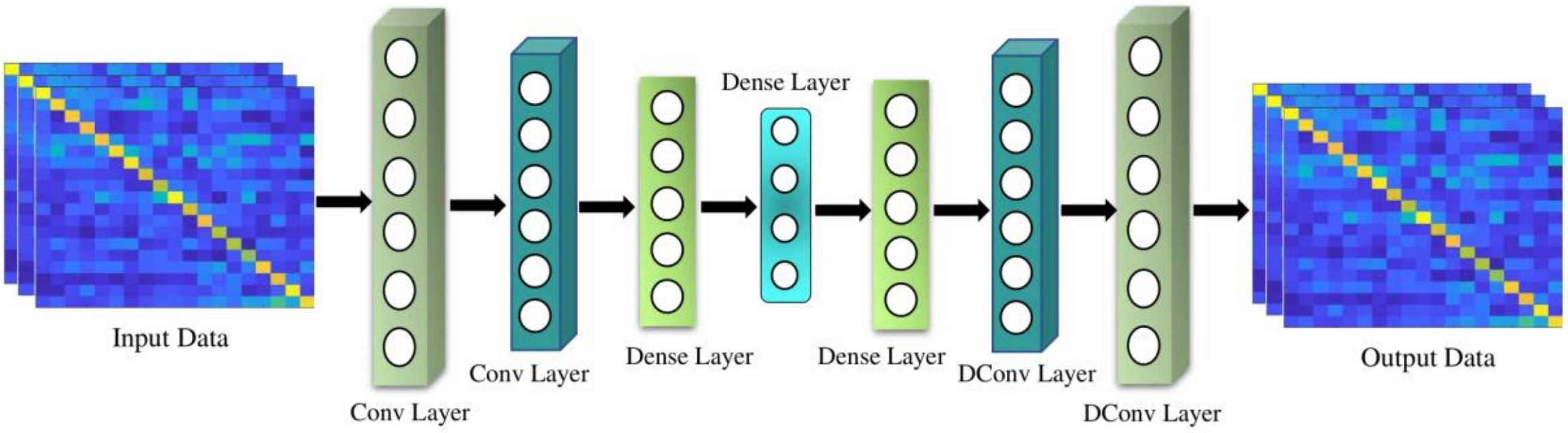

**Fig. 7.** A typical CNN-AE model used for SZ detection using EEG signals.

Table 4. A summary of DL models for diagnosis of SZ from EEG signals.

| Works | Dataset | Number of Cases | Preprocessing | | Deep Learning Model | Classifier | Performance (%) |
|---|---|---|---|---|---|---|---|
| | | | Low Level | High Level | | | |
| [151] | RepOD | 14 SZ, 14 HC | -- | Handcrafted Feature Extraction | LSTM | FC | Acc = 99.0<br>Prec = 99.2<br>Rec = 98.9 |
| [152] | RepOD | 14 SZ, 14 HC | Filtering | CWT | AlexNet, VGG-19, ResNet-18, Inception-v3 | SVM | Acc = 98.6<br>Sens = 99.65<br>Spec = 96.92 |
| [153] | Clinical | 187 RSZ, 127 TD | Filtering | -- | 2D-CNN-LSTM | Sigmoid | Acc = 89.98 |
| [154] | RepOD | 14 SZ, 14 HC | Segmentation | -- | 1D- CNN | FC | Acc = 98.07<br>Sens = 97.32<br>Spec = 98.17 |
| [155] | MHRC, RepOD | 45 SZ, 39 HC<br>14 SZ, 14 HC | Segmentation | Spectrogram | VGG-16 | Softmax | Acc = 97.4<br>Perc = 98.0<br>Rec = 96.0 |
| [156] | PSYKOSE | 22 SZ, 32 HC | -- | -- | Bi-LSTM + attention mechanism | FC | Acc = 86.6<br>Prec = 87.12<br>Rec = 87.03 |
| [157] | Kaggle dataset | 49 SZ, 32 HC | Filtering, Re-reference Filtering, Segmentation | 21-D markers (Information Theory, Connectivity, and Spectrum Markers) | 2D-CNN-AE | Fully connected | Acc = 92.0 |
| [158] | RepOD, MHRC | 14 SZ, 14 HC<br>45 SZ, 39 HC | Segmentation | Acquisition of Hilbert Spectrum (HS) for the first 4 Intrinsic Mode Functions (IMF) Components | VGG16-CNN, XCeption, DenseNet121, ResNet152, Inception V3-CNN | Softmax | Acc = 98.2<br>Sens = 98.0<br>Prec = 99.0<br>F1-S = 98.0 |
| [159] | Kaggle dataset | 49 SZ, 32 HC | -- | STFT, CWT, and SPWVD, Spectrogram, Scalogram, SPWVD, TFR | AlexNet, VGG-16, ResNet, 2D-CNN | Softmax | Acc = 93.36<br>Sens = 94.25<br>Spec = 92.03<br>F1-S= 0.945<br>Prec = 94.66 |
| [160] | Clinical | 40 CHR, 40 FES, 40 HC | Filtering | Feature Extraction (Delta, Theta, Alpha, Beta, and Gamma, Amplitude of Fourier Transform) | 1D-CNN, RNN | RF | Acc = 96.7 |
| [161] | Clinical | 45 SZ, 39 HC | -- | VAR Model Coefficients, PDC, Network Topology-Based Complex Network (CN), Connectivity features (VAR + PDC + CN) | Multi-domain Connectome CNN (MDC-CNN) | Softmax | Acc = 91.69<br>Sens = 91.11<br>Spec = 92.50<br>Prec = 94.14 |
| [162] | Kaggle dataset | 49 SZ, 32 HC | -- | -- | AlexNet | Fully connected | Acc = 76.0<br>Prec = 76.0<br>Rec = 73.5 |
| [163] | MHRC, RepOD | 45 SZ, 39 HC<br>14 SZ, 14 HC | Filtering, Segmentation | FFT, Feature Extraction (Mean Spectral Amplitude (MSA), Spectral Power (Pspectral), Activity, Mobility, Complexity) | 1D-CNN, LSTM | SoftMax | Acc = 98.56 |
| [164] | MHRC | 45 SZ, 39 HC | -- | DWT, Feature Extraction (Concatenating 1D local binary pattern (LBP), data augmentation by ELM-AE) | AlexNet, VGG16, ResNet | Softmax | Acc = 97.7<br>Sens = 97.8<br>Spec = 97.7<br>F1-S = 97.6 |
| [165] | Lomonosov Moscow State University | 45 SZ, 39 HC | Normalization | Connectivity Features | DNN-DBN | Softmax | Acc= 95 |
| [166] | -- | 45 SZ, 39 HC | -- | Connectivity Features | CNN tuned | FC | Acc= 55<br>Pre= 55<br>Rec= 100 |
| [167] | Clinical | 47 SZ, 54 HC | Down-sampling, Up-sampling | -- | 1D-CNN | Softmax | Acc=75.9<br>Sen=73.9<br>Spec=77.9 |
| [168] | RepOD | 14 SZ, 14 HC | Segmentation | -- | Recurrent Auto-encoder (RAE) | FC | Acc=81.81<br>Sen=80.3<br>Spec=83.37 |
| [169] | Moscow State University | 45 SZ, 39 HC | -- | Transformation | CNN–Bi- LSTM | Sigmoid | Acc=72.8 |
| [170] | Kaggle | 49 SZ, 32 HC | -- | -- | SchizoGoogLeNet | SVM | Acc=98.84 |
| [171] | Clinical | 45 SZ, 39 HC | -- | -- | 1D-CNN | Sigmoid | Acc=97 |
| [172] | NIMH | 49 SZ, 32 HC | -- | Transformation | VGGNet | FC | Acc=96.3 |
| [173] | RepOD | 14 SZ, 14 HC | -- | FFT, Wavelet Transform | CNN–LSTM | Sigmoid | Acc=99.04 |
| [174] | Clinical | 45 SZ, 39 HC | -- | -- | 1D-CNN | SVM | Acc=90 |
| [175] | Clinical | 45 SZ, 39 HC | -- | PCA | LSTM | Softmax | Acc=98 |
| [176] | Mental Health Research Center (MHRC), Clinial | MHRC: 45 SZ, 39 HC<br>Clinical: 14 SZ, 14 HC | -- | CWT, DA | VGG16 | FC | Acc=98 |
| [177] | RepOD | 14 SZ, 14 HC | -- | Effective Connectivity (Transfer Entropy) | EfficientNetB0-LSTM | Softmax | Acc=99.93 |
| [178] | RepOD | 14 SZ, 14 HC | Re-referencing | ICA | 1D-CNN | Softmax | Acc=93 |
| [179] | RepOD, Clinical | D1: 14 SZ, 14 HC<br>D2: 45 SZ, 39 HC | Segmentation | ICA | CNN-LSTM | Softmax | Acc=99.9 |
| [180] | RepOD | 14 SZ, 14 HC | -- | -- | SAE | Softmax | Acc=97.95 |
| [181] | Clinical, Combination of two public datasets of SZ and MDD | D1: 100 SZ, 100 DP, 100 HC<br>D2: 10 SZ, 10 DP, 10 HC | Segmentation | ICA | MUCHf-Net | Softmax | Acc=91.12 |
| [182] | RepOD | 14 SZ, 14 HC | -- | MSST-based Time-Frequency Analysis | VGG16+Bi-LSTM | Softmax | Acc=86.9 |
| [183] | Clinical | 41 SZ, 31 HC<br>15 SZ, 16 HC, and 12 First-Degree Relatives | -- | WT | CNN based on Transfer Learning | Softmax | Acc=83.2 |
| [184] | Clinical | 10 SZ, 10 HC | -- | Decomposition Learner (LSDL), CWT | SqueezeNet | Softmax | Acc-98 |
| [185] | Moscow State University | 45 SZ, 39 HC | -- | Dimensionality Reduction using Random Projection (RP) | LSTM | Softmax | Acc=98 |
| [186] | RepOD | 14 SZ, 14 HC | PCA | -- | 1D CNN-LSTM, Squeeze Excitation Network-LSTM-Softmax (SLS) | Softmax | Acc=98.65 |
| [187] | Clinical | 54 SZ, 55 HC | Segmentation | Ocular Correction | CNN+LSTM | Softmax | Acc=99.22 |
| [188] | Public EEG Dataset | 45 SZ, 39 HC | Segmentation | -- | 1D-CNN | Softmax | Acc=90 |
| [189] | Mental Health Research Center (MHRC), RepOD | D1: 45 SZ, 39 HC<br>D2: 14 SZ, 14 HC | Segmentation | -- | VGG-16 | Softmax | Acc=97 |
| [190] | Kaggle | 65 SZ, 63 HC | Filtering, Segmentation | -- | 1D-CNN (SzNet-5) | Softmax | Acc=78 |
| [191] | Clinical | 49 SZ, 32 HC | Segmentation | FFT | AE | Softmax | Acc=92 |
| [192] | RepOD, Clinical | D1: 14 SZ, 14 HC | Segmentation | -- | CNN-LSTM | Softmax | Acc=99.9 |
| | | D2: 45 SZ, 39 HC | | | | | |
| [193] | RepOD | 14 SZ, 14 HC | -- | -- | hybridization of CNN | LR | Acc=98.05 |
| [194] | Clinical, RepOD, Kaggle | D1: 45 SZ, 39 HC<br>D2: 14 SZ, 14 HC<br>D3: 45 SZ, 32 HC | -- | -- | 2D-CNN | Softmax | Acc=99.74 |
| [195] | RepOD | 14 SZ, 14 HC | Segmentation | -- | CNN-LSTM | Sigmoid | Acc=99.43 |
| [196] | RepOD | 14 SZ, 14 HC | Segmentation, Normalization | Fuzzy Features | 1D-CNN-LSTM | Sigmoid | Acc=99.25 |
| [197] | Clinical | 312 SZ, 320 HC | -- | -- | Radial Basis Function (RBF) Combined with Fuzzy C-Means (FCM) | -- | Acc=93.4 |
| [352] | Kaggle | 49 SZ, 32 HC | Average Filtering Method | -- | ResNet | SVM | Acc=99.23 |
| [353] | RepOD | 14 SZ, 14 HC | Filtering, Normalization | Spatial-temporal Features | LightwVision Transformer model LeViT | Softmax | Acc=98.99 |
| [354] | RepOD | 14 SZ, 14 HC | Filtering, Segmentation | Brain Connectivity Features | Schizo-Net | Softmax | Acc=99.84 |
| [355] | RepOD | 14 SZ, 14 HC | Filtering, Segmentation, MSPCA | Multitaper method (Different Features) | 1D-CNN | Softmax | Acc=98.76<br>Sen=99.1<br>Spe=98.3 |

## 5. Challenges

This section introduces the challenges associated with using ML and DL techniques to diagnose SZ from EEG signals. Tables (3) and (4) provide detailed information on papers focused on ML and DL techniques for SZ diagnosis, respectively. The challenges of diagnosing SZ based on the review of papers in this field are presented in this study. The primary challenges in SZ diagnosis from EEG signals encompass datasets, ML techniques, DL models, explainability and hardware resources. In the following, each of the raised challenges will be thoroughly examined. Overcoming the existing challenges has the potential to facilitate the development of practical AI-based tools to aid physicians in the early diagnosis of SZ.

### 5.1. Challenges in datasets

In this section, the challenges associated with using EEG datasets for the diagnosis of SZ are discussed. As shown in Section 4.1, only two EEG datasets with a limited number of subjects have been presented in this research field so far. Meanwhile, early diagnosis of SZ is of particular significance for clinicians. The unwillingness of subjects with SZ to undergo EEG recording, ethical considerations, and physicians' lack of trust in CAD systems make it difficult to provide EEG datasets available to a large number of subjects. In Tables (3) and (4), EEG datasets used for SZ diagnosis papers are reported. According to these Tables, some researchers have utilized clinical datasets that were recorded using various methods. Consequently, these datasets may have different recording standards, making it difficult to compare studies in this field. Additionally, the lack of available multimodality datasets with a large number of subjects poses another challenge. If researchers can obtain access to such datasets, can develop state-of-the-art AI methods that will assist specialist doctors in the timely diagnosis of SZ. The challenges mentioned above, including the limited availability of large datasets, lack of standardized EEG protocols, and multimodality, will be discussed in further detail.

**A) limited availability of large datasets**

Section 4.1 provides an overview of available EEG datasets for diagnosing SZ. However, only two EEG datasets with a limited number of subjects are currently available. Given the significant increase in SZ cases worldwide in recent years, researchers require access to different EEG datasets to assist clinicians in the prompt diagnosis of SZ. On the other hand, AI techniques including DL architectures are rapidly advancing, and their effectiveness is greatly enhanced by training them on large datasets. However, the limited availability of datasets for SZ diagnosis poses a challenge to the use of advanced DL models. In part of Tables (3) and (4), the EEG datasets used in SZ diagnosis studies are indicated. It can be seen that many researchers used clinical datasets to diagnose SZ and achieved satisfactory results. They have made efforts to apply DL techniques in the early diagnosis of SZ. However, these datasets are not accessible to other researchers, posing a significant challenge. Researchers' access to available EEG datasets with a large number of subjects can potentially yield valuable research in this field.

**B) Lack of standardized EEG protocols**

One of the major challenges in utilizing datasets for SZ diagnosis is the lack of standardized EEG protocols. Specialist doctors may use different electrode placement methods, such as a 10-20 system, to record EEG signals for each subject [259]. As a result, EEG signals in datasets may not be recorded with the same protocols by different physicians. Additionally, some physicians may record EEG datasets using multiple devices and varying sampling frequencies. The lack of standardized EEG protocols makes it difficult for researchers to compare the findings across different studies based on various datasets. The datasets used in the SZ diagnosis articles are presented in Tables (3) and (4). Based on these tables, it can be seen that researchers have employed different datasets for SZ detection, resulting in difficulties when comparing their findings. On the other hand, certain researchers have utilized different clinical datasets to diagnose SZ, but there is a lack of comprehensive information regarding how they recorded EEG signals for the participants, posing a significant challenge in comparing papers within this field. Establishing a uniform protocol for recording EEG signals would greatly aid researchers in properly comparing the results of different studies.

**C) Multimodality**

Specialist doctors face difficulties in diagnosing SZ due to the complexity and variety of symptoms, as well as underlying pathologies. For this purpose, clinicians utilize multimodality neuroimaging, in which information from two or more structural and functional modalities are integrated [260-262]. For instance, structural modalities can reveal the changes in brain volume such as working memory (WM) in SZ patients, while functional modalities provide a detailed representation of brain neural networks for SZ patients [263-264]. Clinical settings commonly employ fusion neuroimaging modalities, such as MRI-PET [265], EEG-fMRI [266], EEG-MEG [267], and MRI-fMRI [268], to diagnose a range of brain disorders, including Alzheimer's disease (AD), PD, multiple sclerosis (MS), and SZ. Clinical studies have demonstrated that multimodality neuroimaging provides crucial information about brain function and structure for diagnosing SZ [269-270]. To date, multimodality neuroimaging datasets, such as EEG-fMRI and EEG-MEG, have not been accessible to researchers for early SZ diagnosis using AI techniques. Access to multimodality neuroimaging datasets can facilitate valuable research in the field of SZ diagnosis using AI techniques.

### 5.2. Challenges in ML methods

As previously discussed, there are differences between ML-based and DL-based CADS in terms of feature extraction and selection. ML methods are widely used in SZ diagnosis due to their advantages, including high accuracy, interpretability, data efficiency, low computational power, and speed. However, ML-based CADS also faces several challenges, such as feature engineering, complexity, scalability, and generalizability. The primary challenge for SZ diagnosis using ML methods lies in feature engineering, which entails manual feature extraction [271]. In ML-based CAD systems, this process is typically performed manually, which is highly time-consuming and demands significant expertise in this field. Additionally, ML methods face numerous limitations in modeling complex relationships in EEG signals, which is a critical challenge in medical diagnostics [272-273]. Limited scalability is another challenge of ML methods, as the performance of ML models can degrade considerably as the size of datasets increases. This can limit their suitability for deployment in most applications of medicine including, SZ diagnosis. Finally, limited generalizability is a significant challenge for ML techniques, as they often fail to generalize to different datasets, resulting in the performance of an ML-based CADS that performs well on one dataset but not on a similar one.

### 5.3. Challenges in DL methods

In the previous section, the most important challenges associated with using ML methods to diagnose SZ from EEG signals were mentioned. To address these difficulties, researchers have proposed and subsequently developed and expanded DL techniques in various medical applications [274-275]. As a result, DL techniques are widely adopted in diagnosing and predicting brain disorders from EEG signals. Researchers are currently exploring the use of DL techniques to develop a real-world tool for the rapid diagnosis of brain disorders, including SZ [164-166]. The advantages of DL techniques include automated feature extraction, better performance, scalability, generalizability, robustness to noise, and the capacity to learn complex relationships [276-277]. Today, these benefits have led to a growing number of researchers utilizing DL techniques over ML methods for diagnosing brain disorders. DL architectures, however, encounter certain challenges, including data requirements, overfitting, and training time [278]. Table (4) provides a summary of studies employing DL techniques for SZ diagnosis. As is apparent, the limited accessibility of EEG datasets with a considerable number of subjects has posed serious challenges in employing advanced DL techniques, such as attention models [279], graphs [280], and mutual learning [281] for SZ diagnosis. DL models face the issue of overfitting when they encounter insufficient input data. This is because these networks require a large amount of data for effective training, and the availability of EEG data is limited in the field of SZ diagnosis. Additionally, DL models for complex applications, such as diagnosing brain disorders from EEG data, generally demand prolonged training periods, posing a challenge for applications that necessitate real-time decision-making or prompt response time.

### 5.4. Challenges in explainability

DL models are considered black boxes due to their complex mathematical operations that are difficult for humans to interpret [282-283]. This issue poses challenges in various applications, including the diagnosis of brain disorders using neuroimaging modalities. With the rapid growth of DL models, their interpretation is becoming increasingly challenging [284]. Researchers are striving to develop techniques to enhance the interpretability of DL models [284]. Despite numerous methods being proposed to interpret DL networks, further efforts are required to attain successful outcomes in practical scenarios. Tables (3) and (4) reported SZ diagnosis papers utilizing ML and DL techniques, respectively. Explainable AI (XAI) techniques have not been exploited in SZ diagnosis studies using EEG signals. However, various XAI methods have been employed in diagnosing brain disorders from medical imaging in various studies [284-285]. This indicates that XAI techniques have not been extensively developed for diagnosing brain disorders from biological signals, which presents another challenge. Addressing these challenges could enhance clinicians' confidence in utilizing AI techniques for diagnosing brain disorders, including SZ.

### 5.5. Challenges in hardware resources

This section focuses on the challenges of hardware resources in implementing DL architecture for SZ diagnosis. The limitations of hardware resources pose one of the major hurdles in training and implementing complex DL models for various applications [286-287]. These networks necessitate a large amount of computational power and memory for data processing as well as complex calculations [288]. While graphics processing units (GPUs) are the most commonly used hardware for training DL architectures, they are expensive and not readily accessible to everyone [286-288]. Furthermore, training DL models using these processors can take several days. The memory capacity of GPUs also poses a challenge in this domain [286-288]. DL models require a huge amount of memory to store the weights obtained during calculations, and if the model is too large for the GPU memory, it cannot be trained on that hardware [286]. While companies like Google and Amazon offer computing servers to researchers, these servers have limitations in terms of memory and time for running DL models. Consequently, these computational servers are not suitable for diagnosing SZ from EEG signals.

### 5.6. Challenges rehabilitation systems

This section discusses the challenges associated with rehabilitation systems in SZ diagnosis and predictions. Researchers are currently investigating the use of AI-based online monitoring systems for all types of patients, including those with heart, brain, and other conditions [287-290]. To this end, extensive research is being conducted to develop rehabilitation systems such as brain-computer interface (BCI) [291], neurofeedback [292], wearable devices [293], etc. to assist patients. While some preliminary work has been done in the field of neurofeedback for SZ patients, suitable solutions to aid these patientsare yet to be found [50]. The challenges related to the datasets available and high-power computing hardware resources are among the most important challenges that have prevented the development of effective rehabilitation systems. Moreover, most BCI hardware and smart wearable devices are designed for diseases like blood pressure, stress, and epilepsy, and have proven beneficial in assisting patients with these conditions [294-297]. However, this hardware has yet to be employed for SZ patients, which presents a challenge. These tools can process and monitor patients' information in real-time, potentially preventing harm and enhancing patients' safety. Thus, developing such hardware for online SZ diagnosis and prediction applications could significantly improve the quality of life for individuals with SZ.

## 6. Discussion

Section 6 presents a comprehensive discussion of the literature related to automated SZ detection using EEG signals. In Section 2, we previously reported the review articles published in the field of SZ diagnosis. In this section, we first compare our work with published review papers on SZ characterization to emphasize the novelty of this work. The subsequent discussions cover critical aspects of SZ detection articles, including the dataset, feature extraction, feature selection, classification, and evaluation parameters. More details about the evaluation parameters are provided in Appendix A.

### 6.1. Comparison

The use of AI techniques for SZ diagnosis has significantly increased in recent years. Section 2 presents the review papers published on SZ diagnosis using AI techniques. Most of the researchers have utilized neuroimaging modalities for diagnosing and predicting SZ. Review papers on the diagnosis and prediction of SZ using EEG and MRI modalities are presented in references [47-48][51]. Additionally, some authors have focused on SZ diagnosis using MRI neuroimaging modalities and AI techniques [29][52-53]. Moreover, SZ diagnosis papers using EEG modality and AI methods were discussed in [49-50]. Authors in [50], reported works related to the application of neurofeedback in SZ using EEG signals and AI techniques. Another study by Barros et al. [49] reviewed the papers in the field of SZ prediction using EEG signals and AI techniques.

In this study, CADS developed to diagnose SZ using ML and DL methods from EEG signals is presented. Subsequently, a summary of related papers published was provided in Tables (3) and (4). In contrast, other works have not comprehensively reviewed the literature on SZ diagnosis from EEG signals utilizing ML and DL techniques. The second novelty of the current study is introducing the most important challenges involved in diagnosing SZ using EEG signals. These challenges include datasets, ML techniques, DL networks, rehabilitation, and hardware resources. However, these challenges have not been addressed in previous review papers. Future works and findings are an additional novelty of this work. Specifically, this review paper provides a detailed description of future work in the areas of datasets, ML methods, DL techniques, XAI, uncertainty quantification (UQ), hardware resources, and rehabilitation systems. Figure (8) shows the comparison of our work with other review papers published in this field and is graphically displayed.

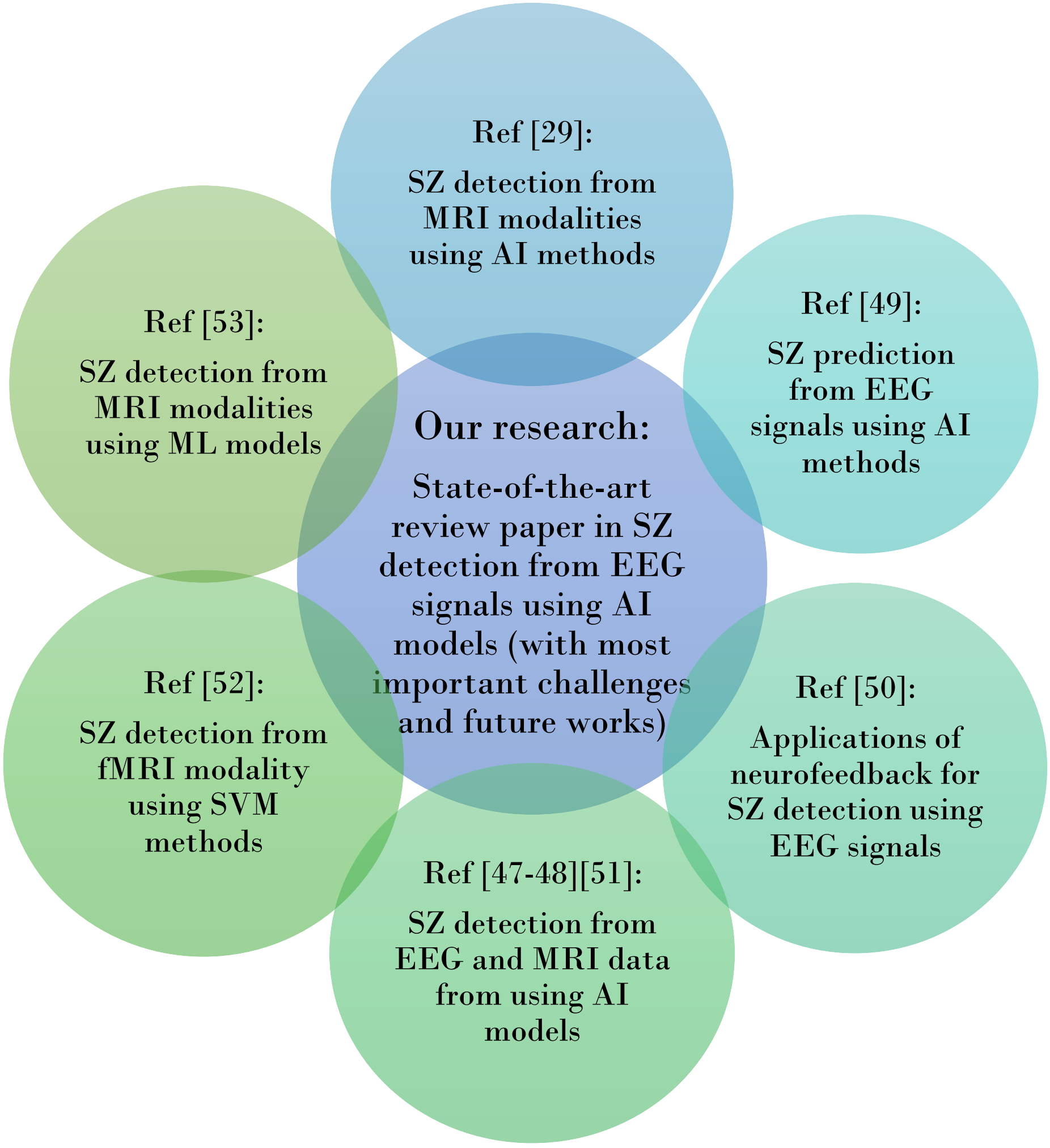

**Fig. 8.** Comparison of our study with other review papers published on automated SZ detection.

### 6.2. Dataset

Section 4.1 of this study introduces available EEG datasets, including RepOD [66] and Kaggle [67], that have been utilized in SZ diagnosis research. The datasets employed in papers utilizing ML and DL techniques in this field are reported in Tables (3) and (4), as well as the number of datasets used in SZ diagnosis papers is graphically displayed in Figure (9). Based on Figure (9), it can be observed that the RepOD dataset was the most frequently used in SZ diagnosis papers. This dataset comprises an equal number of subjects for both HC and SZ classes. First, various pre-processing techniques have been applied to this dataset before being made available to researchers. The mentioned reasons have contributed to the popularity of this dataset in ML and DL research for diagnosing SZ. Additionally, Figure (9) highlights the utilization of clinical datasets in several SZ diagnosis papers, which are not available to other researchers due to limited accessibility. This dataset comprises an equal number of subjects for both healthy control (HC) and SZ classes and has been subjected to various pre-processing steps before being made available to researchers. These features have contributed to the widespread use of this dataset in ML and DL research for diagnosing SZ. Furthermore, Figure (9) illustrates that clinical datasets have been utilized in several papers, but access to these datasets is restricted to researchers.

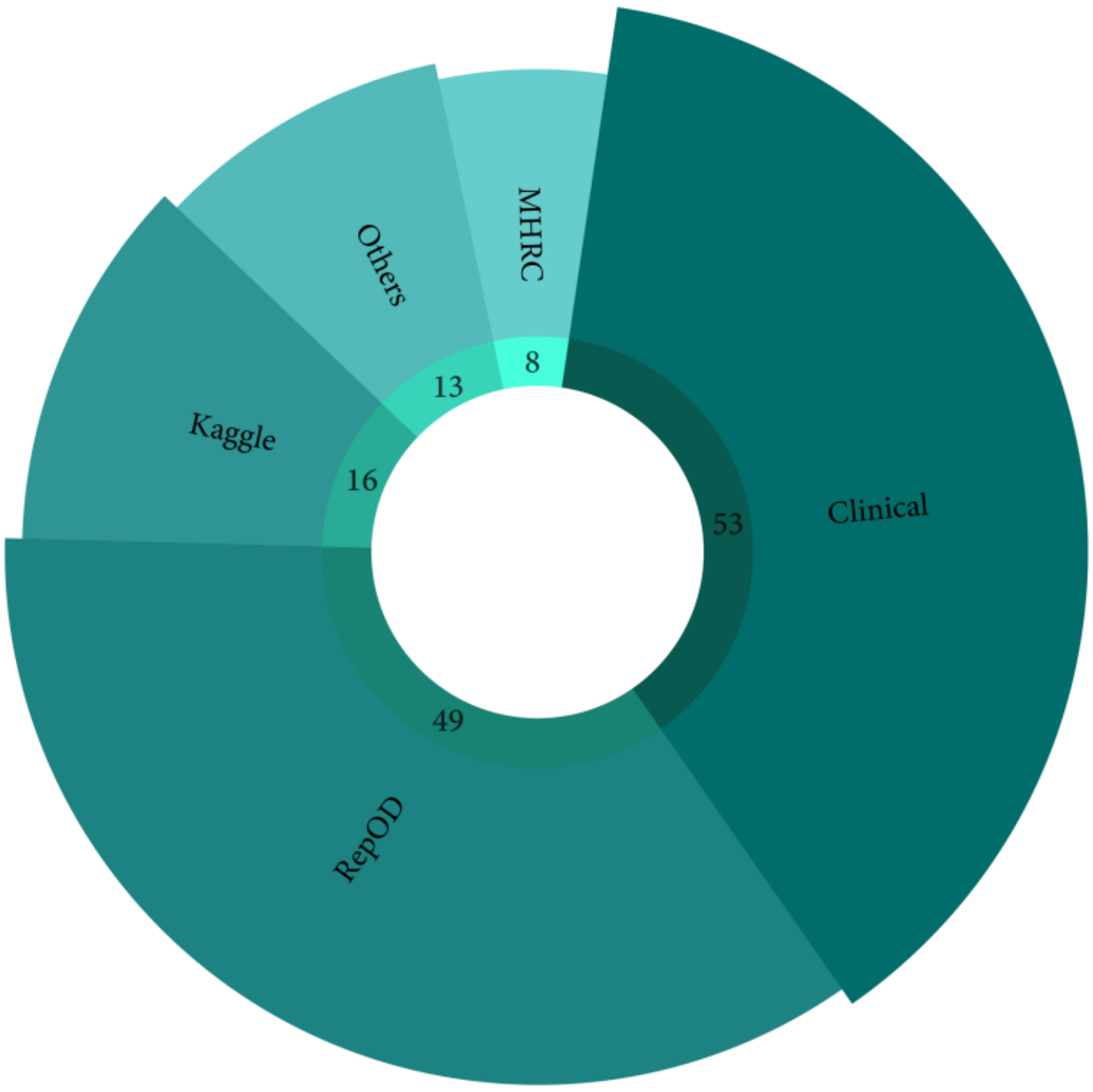

**Fig. 9.** Number of datasets used in the diagnosis of SZ from EEG signals using AI methods

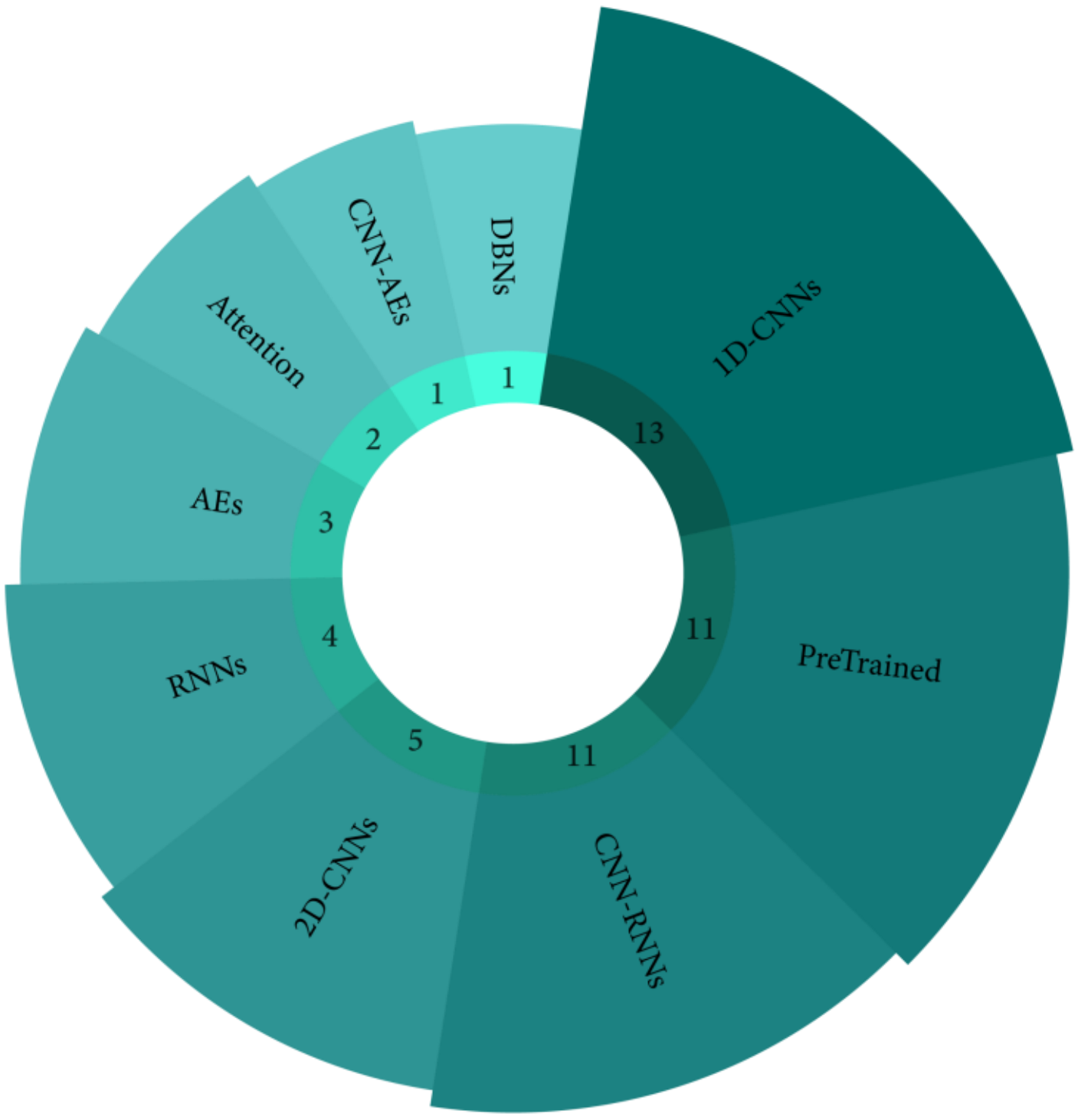

**Fig. 10.** Number of DL methods for diagnosis of SZ from EEG signals.

### 6.3. Compared DL with ML Research in SZ detection

In this section, SZ diagnosis papers using ML and DL methods are compared. Tables (3) and (4) provide information on SZ diagnosis using ML and DL techniques. These tables reveal that ML methods have been used in a greater number of studies for SZ diagnosis compared to DL methods. The lack of access to EEG datasets containing a large number of subjects has limited the utilization of DL networks in SZ diagnosis research. According to Table (4), researchers have frequently employed standard CNNs architectures to diagnose SZ. However, dataset limitations have hindered the feasibility of utilizing advanced DL models such as graph models and transformers in diagnosing SZ. These reasons have led researchers to utilize ML techniques in a large number of papers to diagnose SZ. The availability of EEG datasets containing a larger number of subjects provides hope to researchers, as it may facilitate the use of advanced DL techniques for the diagnosis and prediction of SZ disorder in the future.

### 6.4. DL

In this section, the primary focus is on the most significant DL networks used in the papers related to SZ diagnosis from EEG signals. A summary of widely employed DL networks used for SZ diagnosis is reported in Section 4.4. As mentioned in the previous subsection, the lack of a huge number of available EEG datasets may restrict the use of advanced DL techniques for diagnosing SZ. As depicted in Table (4), standard CNNs, RNNs, AEs, and CNN-RNNs are frequently employed in research. Figure (10) displays the number of DL networks employed in SZ diagnosis papers utilizing EEG signals, indicating that CNN models have been the most commonly used in SZ diagnosis research from EEG signals.

### 6.5. Classifiers

The final component of AI-based CADS is comprised of classification algorithms. This section focuses on classification methods utilized in SZ diagnosis. The classification techniques employed in SZ diagnosis papers utilizing ML methods are presented in Table (3). Figure (11a) illustrates the number of classification algorithms exploited in SZ diagnosis based on EEG signals, with the SVM classifier [298] which is most popular in SZ diagnosis. In contrast, the classification algorithms used in DL networks are reported in Table (4). In Figure (11b), the number of classification techniques used in SZ diagnosis papers using DL techniques are displayed, with the Softmax algorithm being the most frequently employed in SZ diagnosis. Softmax [299] is an efficient method used in DL models and also used for other diverse applications.

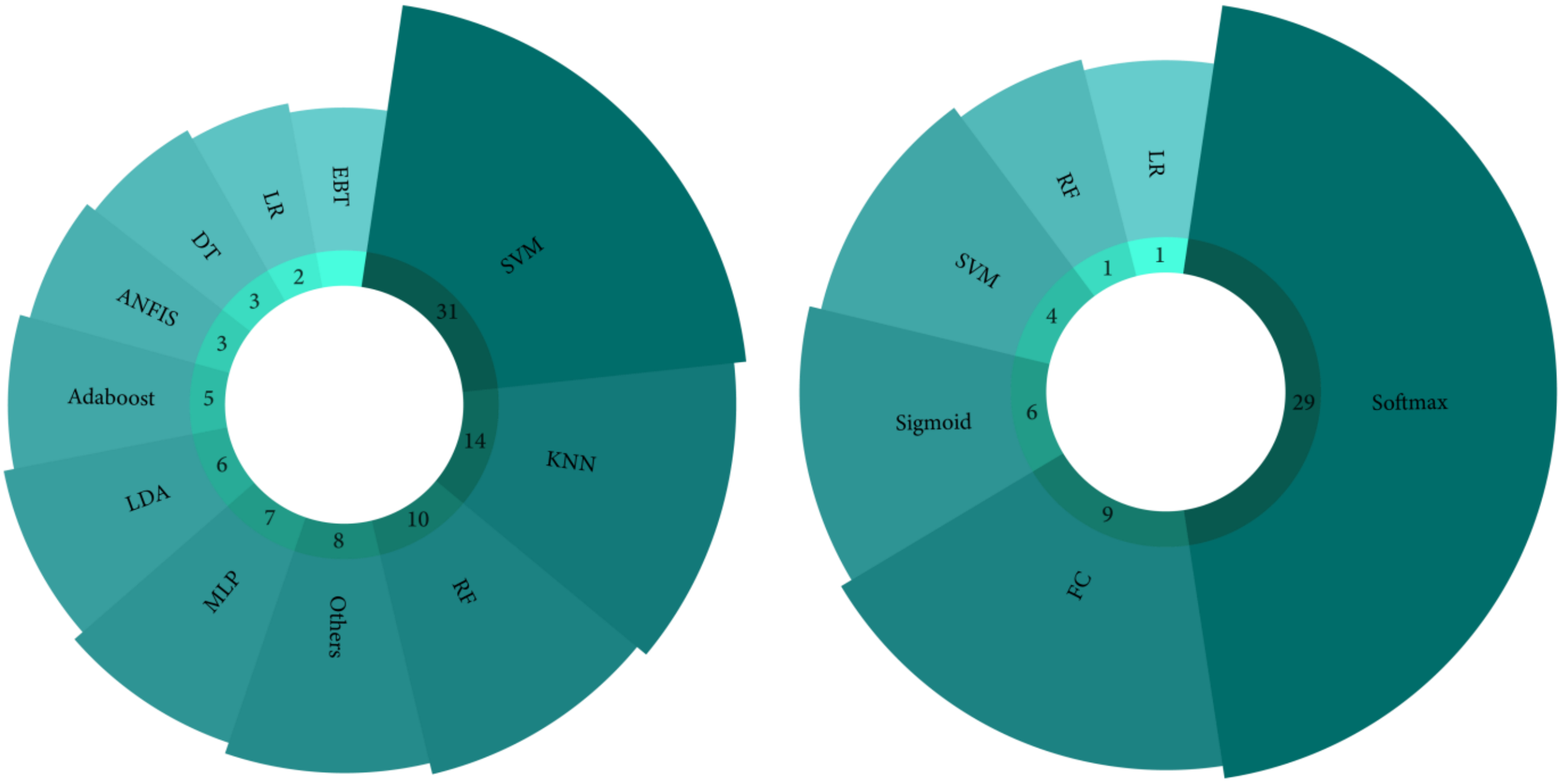

(a) (b)

**Fig. 11.** Illustration of the number of various classifier algorithms used in AI methods for automated detection of SZ. a) ML research and b) DL research.

## 7. Future Directions

This section introduces the most important future direction in SZ diagnosis research using EEG signals. The challenges of this field are discussed in detail in Section 5. Overcoming the current challenges in this field can yield invaluable research findings regarding SZ diagnosis by leveraging both EEG signals and AI techniques. This section outlines future directions, including datasets, ML techniques, DL networks, XAI techniques, UQ, hardware resources and rehabilitation systems. These directions can facilitate researchers in utilizing state-of-the-art AI techniques to enhance the accuracy of SZ diagnosis for future research. The subsequent section discusses each of these future directions in greater detail.

### 7.1. Future works in the dataset

The development of AI-based CAD systems for brain disorders, including SZ, heavily relies on the availability of appropriate datasets. The lack of available EEG datasets with a large number of subjects poses a significant challenge in SZ diagnosis research, which has significantly limited research in this field. In the future, the provision of readily available EEG datasets with more subjects has the potential to facilitate practical research in this field. It is worth noting that SZ disorder exhibits varying levels of severity [300], yet thus far, EEG datasets have not been provided to diagnose SZ disorder at different levels. Therefore, providing researchers with access to these datasets can offer new opportunities for research in this field. As mentioned in the introduction section, TMS, tDCS, CBT and DBS are among the most important interventional methods for the treatment of SZ disorder [12-16]. The provision of available EEG datasets based on interventional methods can aid specialist doctors in the rapid treatment of SZ disorder. However, if these interventional methods are not appropriately utilized, they can pose a risk to human health. Therefore, employing AI techniques to analyze these data can minimize the likelihood of treatment errors and ultimately assist SZ patients.

### 7.2. Future works in multimodality datasets

Multimodality neuroimaging data have been instrumental in improving the accuracy of diagnosing brain disorders, such as SZ, by integrating structural and functional information [260-262]. Typically, specialist doctors combine a structural modality with a functional modality to diagnose brain disorders [262]. Clinical research has demonstrated that multimodality neuroimaging data, such as MRI-PET [265], EEG-fMRI [266], EEG-MEG [267], and MRI-fMRI [268], have yielded invaluable results in diagnosing brain disorders, including SZ, PD, and AD. However, researchers currently lack access to multimodality neuroimaging datasets to diagnose SZ accurately. As a potential future direction, providing researchers with access to MRI-EEG and EEG-MEG datasets could lead to improved accuracy in SZ diagnosis using AI techniques. In other words, AI techniques can assist specialist doctors in quickly diagnosing SZ by simultaneously analyzing the structure and function of the brain from these datasets.

### 7.3. Future works in ML methods

In this section, novel ML methods are introduced for the diagnosis of SZ. These techniques are proposed as future directions in feature extraction and classification sections. Handcrafted feature extraction methods are outlined in part of Table (3). In EEG signals, these methods are categorized into four categories: time [200], frequency [201], time-frequency [202] domains, and nonlinear [203]. Additionally, connectivity methods [301-302] are highlighted as a crucial category of feature extraction techniques for neuroimaging modalities, including EEG signals. These methods encompass various types of functional connectivity [303], dynamic connectivity [304], and effective connectivity [305]. As a potential future research direction, the utilization of connectivity feature extraction methods is recommended to enhance the accuracy of SZ diagnosis. Furthermore, novel feature extraction techniques, such as synchronization likelihood (SL) [196], graph-based approaches [305], and entropies techniques [310], are proposed as attractive options for future research works. The classification algorithms employed in previous studies are presented in a separate part of the Table. These papers utilize diverse classifier methods, including support vector machine (SVM) [298], K-nearest neighbors (KNN) [306], random forest (RF) [307], adaptive neuro-fuzzy inference system (ANFIS) [308], and decision tree (DT) [309]. However, type-2 fuzzy classification methods have not been exploited in SZ diagnosis papers. Therefore, type-2 fuzzy classification methods are introduced as one of the future directions in this section. Moreover, classification methods based on graph theory have not been employed in SZ diagnosis papers, and graph-based classification techniques are suggested as another promising direction for future research in this domain.

### 7.4. Future works in DL models

The utilization of DL architectures in various medical domains has experienced significant growth in recent years. The section proposes future research directions in the field of SZ diagnosis using novel DL models. In Table (4), papers on SZ diagnosis from EEG signals using various DL methods are summarized. In these papers, standard DL models including CNNs, RNNs, AEs and CNN-RNNs have been employed to enhance the accuracy of SZ diagnosis. Additionally, potential future works are introduced, including DL architectures that incorporate deep attention mechanisms (DAMs) [279], deep graph models [280], deep mutual learning (DML) [281], deep multi-task learning (DMTL) [311], and federated learning (FL) [312]. In the following, some other promising future research directions in the field of XAI [282], hardware resources [287], and uncertainty [313] are presented.

#### a) Attention mechanism

Deep attention mechanism (DAM) architectures are a new category of DL networks that focus on the most important parts of the input data through the use of attention mechanisms [314-316]. DAM models offer several advantages, including variable-length inputs, capturing long-range dependencies, and providing interpretability [314-316]. The ability to manage variable-length inputs is a crucial advantage of DAM models. In contrast to ANN, attention mechanisms enable DL models to selectively attend to various parts of the input data, irrespective of its length [314]. Moreover, DAM models are capable of capturing long-range dependencies, which is another significant advantage. Therefore, attention mechanisms facilitate the modeling of complex relationships between input data [315-316]. Additionally, DAM models are highly interpretable, making them a popular choice in medical applications [316-317]. As a potential direction for future research, various types of DAM models, including graph attention [318], attention RNNs [319], and attention AEs [320], could be explored for SZ diagnosis from EEG signals.

#### b) Graph models

Graph models are a novel category of DL architectures that employ neural networks to learn representations of graphs [321-323]. These networks are gaining popularity rapidly in medical applications, particularly for the diagnosis of brain disorders [324-325]. Researchers utilized graph models to analyze brain networks, and some recent studies have employed graph models, such as Graph Convolutional Neural Networks (GCNN) to diagnose brain disorders from EEG signals, resulting in significant findings [326]. Notable deep graph models include GCNNs, Graph Attention Networks (GATs) [318], Graph AEs (GAEs) [327], and Graph RNNs (GRNNs) [328]. In future work, the application of deep graph models has the potential to yield valuable insights into the diagnosis of SZ from EEG signals.

#### c) Multi-task learning

Deep multi-task learning (DMTL) is a recent development in DL architectures [329]. In DMTL models, a single network performs multiple tasks concurrently, eliminating the need for training separate models for each task [329]. This design enables the model to learn multiple related tasks simultaneously. Some of the most important advantages of DMTL models include improved performance, reduced computational cost, and better feature extraction [329]. To date, DMTL models have been employed in limited research to diagnose brain diseases from neuroimaging modalities such as sMRI data [330-331]. As a potential future direction, various multi-task learning techniques could be exploited for SZ diagnosis.

**d) Federated learning**

Data sharing in medical research poses a challenge due to privacy concerns. To address this challenge, federated learning (FL) techniques have been proposed [332-333]. These techniques enable multiple data recording devices to collaboratively train a network without sharing their data. In this approach, models are trained locally on each device using its data, and then the weights of the networks are transmitted to a central server [332-334]. The network weights of each dataset are aggregated to create a state-of-the-art model. The adoption of FL techniques holds great promise for collaboration, offering advantages such as privacy, scalability, and accuracy [332-333]. Given that CADS systems require access to large amounts of sensitive data, FL can emerge as an important tool to preserve data privacy [332-334]. Based on Tables (3) and (4), it can be observed that most of the research utilized clinical datasets that are not publicly available. In future research, the use of FL methods can potentially enable the development of a state-of-the-art model for the diagnosis of SZ from EEG signals, without compromising data privacy.

**e) Mutual learning**

Deep mutual learning (DML) is another category of DL techniques that has recently captured the attention of researchers in medical research [335-337]. In DML, multiple networks work together to improve each other's performance. Initially, a set of networks are trained on distinct datasets or tasks, and their outputs are subsequently shared [335-336]. These networks are then trained based on each other's outputs and tune their parameters. As a result, DML models demonstrate improved performance for various applications, including data classification [335-337]. The primary advantage of DML models lies in their ability to enable more efficient use of data and computing resources. By sharing information between networks, the training process can be accelerated and the overall accuracy of the models can be enhanced [335-337]. To date, these techniques have not been exploited in SZ diagnosis from EEG signals. Therefore, future research directions could involve the incorporation of DML techniques in the research of SZ diagnosis from EEG signals.

### 7.5. Explainability

DL networks involve complex mathematical operations that can be challenging for humans to interpret. In recent years, XAI techniques have been developed to overcome the challenges of using DL techniques in various applications [282-284]. Using XAI methods can provide critical information on the effectiveness of DL-based CADS for disease diagnosis, which can subsequently increase the trust of specialist doctors in these systems [282-285]. The advent of XAI techniques has significantly enhanced the trust of specialist doctors in the disease diagnosis results obtained through DL models [283]. Additionally, XAI features such as debugging can be applied to a range of DL methods. When a CAD system produces unexpected results in diagnosing a disease from medical data, XAI methods can help identify and fix the problem in various parts of this system [285]. As a future direction, the utilization of XAI methods could result in a notable improvement in the efficiency of DL-based CADS for SZ diagnosis.

### 7.6. Uncertainty

In recent years, researchers have directed their attention toward uncertainty quantification (UQ) in DL models. In references [338-340], the importance of utilizing uncertainty quantification methods in DL models is emphasized and valuable results are achieved. DL architectures are highly capable of analyzing complex data in applications such as forecasting [338-340]. However, one of the challenges in this field is that the efficiency of these models decreases when presented with different data, and UQ techniques are employed to investigate this issue. UQ methods typically encompass model uncertainty, data uncertainty, and parameter uncertainty [338-340]. Despite this, UQ methods have yet to be employed in SZ diagnosis. Therefore, in future works, UQ methods can be utilized to assess the effectiveness of DL techniques in diagnosing SZ.

### 7.7. Hardware resources

The challenges due to limited hardware resources for DL models were mentioned in the previous section. As discussed, DL architectures require extensive computational power and memory for training. To address these challenges, researchers have proposed implementing DL models on chips such as field programmable gate arrays (FPGA) and application-specific integrated circuits (ASIC), which are more efficient for certain types of computations compared to GPUs and have the potential to reduce training time for DL models [341-343]. Therefore, as a future direction, utilizing FPGA and ASIC chips could be an effective solution to implement DL models [341-343]. Additionally, some researchers have proposed compression methods, such as quantization, pruning, and knowledge distillation, to reduce training time in DL models [344-345]. As another future direction, incorporating compression methods could significantly help to address the challenges of limited hardware resources. Furthermore, recent advancements have introduced deep compact-size CNN architectures, such as TinyNet and MobileNet networks, that can be implemented on simple hardware [346-347]. These architectures hold promise for future SZ detection research.

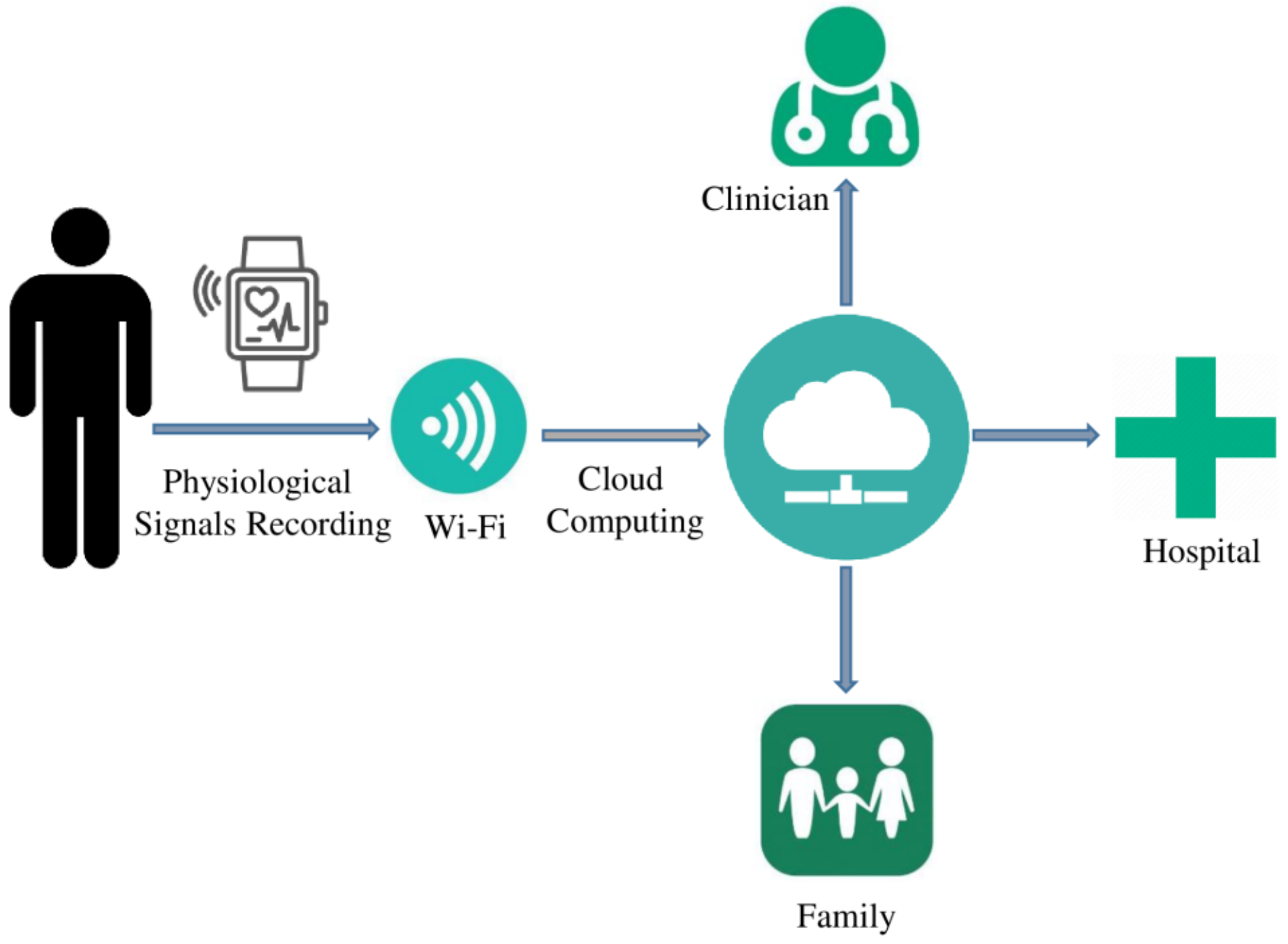

**Fig. 12.** A block diagram of the proposed rehabilitation system.

### 7.8. Future works in rehabilitation systems

The global incidence of brain disorders, such as SZ, has significantly increased in recent years [49]. As a result, numerous studies have been conducted to develop AI-based tools for the rapid diagnosis of brain disorders using neuroimaging modalities. Rehabilitation systems, including BCI [291], neurofeedback [292], and wearable devices [293], are crucial in aiding individuals with brain disorders like SZ. Implementing BCI systems based on EEG signals holds significant potential for future developments in the field of SZ treatment. Recent research has explored the potential of neurofeedback based on EEG signals for SZ patients [50]. In the future, utilizing the latest AI techniques in neurofeedback applications could result in improved treatment outcomes for SZ patients. Various companies have produced smart wearables capable of recording and monitoring biological signals in real-time [294-296]. Also, these systems can be utilized for the online monitoring of SZ patients. The process involves recording the biological signals online, transmitting the data to a cloud server for processing, and providing feedback from the data processing to the treating physician or the patient's family in real-time. The diagram in Figure 12 depicts a block diagram that outlines the general structure of a rehabilitation system used in the diagnosis of SZ.

**8. Conclusion and finding**

Schizophrenia (SZ) is a highly significant mental disorder characterized by a wide range of symptoms that can have a negative impact on various aspects of patient abilities, including thinking, feeling, and behavior [1-2]. Some of the most common symptoms of SZ include hallucinations, delusions, disordered thinking, and lack of motivation. While the precise causes of SZ remain unclear, specialist physicians believe that its development is influenced by a combination of genetic, environmental, and neurological factors [4-6]. Neuroimaging techniques, which encompass a variety of functional and structural methods, play a crucial role in diagnosing SZ and are widely utilized by specialists [29]. Structural neuroimaging methods such as sMRI, DTI, MRS, and CT scans enable the examination of the brain's structure and detect any abnormalities [29]. On the other hand, functional neuroimaging methods including fMRI, PET, SPECT, EEG, MEG, and fNIRS, assess the brain's during SZ episodes [29-30]. Among these, EEG recording is considered one of the most effective methods for SZ diagnosis. AI-based CAD systems have shown promising results in the detection of SZ from EEG signals [49]. However, it is important to note that these systems are still in the developmental stage and require further validation before they their implementation in clinical settings. Numerous studies have been conducted on AI techniques to diagnose SZ using EEG signals [49-50]. in this review paper we examined the papers published on SZ detection from EEG signals using ML and DL techniques for early detection of SZ. While the primary focus of this work is to review the existing literature, it also sheds light on the most important challenges, future research directions, and key findings in.The introduction section of the paper provides a general overview of SZ disorder and its underlying causes. . Furthermore, it introduces the most commonly employed treatment methods for SZ.. Subsequently, the section delves into a discussion on SZ detection methods, providing a comparative analysis of their effectiveness. The importance of utilizing neuroimaging techniques for SZ diagnosis is underscored, elucidating the advantages and limitations associated with each method.. Notably, EEG signals are identified as a particularly promising diagnostic tool due to their non-invasive nature, real-time monitoring capabilities, high temporal resolution, and cost-effectiveness [42-43]. The section concludes by emphasizing the significance of employing AI techniques in SZ detection from EEG signals.

The third section of this study presents a comprehensive review of papers on SZ diagnosis using neuroimaging modalities and AI techniques from 2016 to 2023. These reviews encompass a wide range of topics, including the application of ML and DL techniques for diagnosing SZ from MRI modalities [52-53], as well as the implementation of AI techniques for SZ diagnosis using EEG signals [49-50]. A summary of the findings from these review papers is presented in Table (1). The objective of this section is to compare our research with other review papers in this field. the subsequent discussion section, a detailed analysis of the novel contributions of our study is presented in relation to previous research conducted in this domain.

In a separate section, this work presents the PRISMA guidelines [54] used to search for relevant research papers on SZ diagnosis. The proposed PRISMA framework consists of three levels of analysis: out-of-scope, EEG signals, and AI methods. First, all papers on SZ diagnosis using EEG signals published in reputed journals between 2002 and 2023 were collected. The collected papers were then reviewed using the proposed PRISMA framework with three levels of analysis. A detailed description of the PRISMA framework is presented in Figure (1). This section also includes a brief overview of the inclusion and exclusion criteria used to select papers, along with a comprehensive explanation of each criterion.

In another section, this work introduces the various steps involved in AI-based CADS, including dataset preparation, preprocessing, feature extraction, feature selection, and classification. The section begins with a detailed presentation of ML-based CADS and a summary of SZ detection papers that have employed these methods (Table (3)). ML methods are known for their high accuracy, interpretability, data efficiency, low computational power, and speed, making them valuable tool for SZ diagnosis. The section then goes on to discuss the most important DL architectures used in SZ detection research using EEG signals. DL techniques offer automated feature extraction, superior performance, scalability, generalizability, robustness to noise, and the ability to learn complex relationships. A summary of SZ diagnosis papers using DL techniques is presented in Table (4) at the end of this section.

In another section, this work highlights the significant challenges associated with using AI techniques to diagnose SZ from EEG signals. The challenges include datasets, ML algorithms, DL models, explainability, rehabilitation systems, and hardware resources. In this section, each of these challenges is discussed in detail. For instance, we first introduce the dataset-related challenges, such as the limited availability of large datasets, the absence of standardized EEG protocols, and multimodality, before providing a more detailed explanation of each challenge. Previous research suggests that the lack of access to EEG datasets with a sufficient number of subjects is the most critical challenge in this field (as discussed in Section 5.1). Addressing these dataset challenges could potentially facilitate the development of advanced DL models and rehabilitation systems for SZ diagnosis applications.

In Section 6 of this paper, we provide a comprehensive discussion of the various subsections of SZ diagnosis research. The section builds on the summaries of SZ detection papers using ML and DL techniques reported in Tables (3) and (4), respectively. These Tables contain critical information such as the EEG datasets used, preprocessing techniques, feature extraction, and classification methods employed in each study. In the discussion section, we first compare our work with other review papers in this field. We then provide a thorough analysis of the dataset challenges, a comparison of ML versus DL techniques, DL models, and classification methods. This section aims to provide useful insights for researchers to select the most effective techniques for each CADS section of SZ diagnosis in future research.

In Section 7, we provide an overview ofpotential directions for future work in SZ diagnosis research, encompassing various aspects such as datasets, multimodality, ML methods, DL models, XAI, hardware resources, and rehabilitation systems. The availability of EEG datasets with a sufficient number of subjects is critical for future work in the field of SZ diagnosis. Additionally, researchers can leverage advanced DL models, such as DATMs [318], graph models [321], DMLs [325], DMTLs [329], and FLs [333], to improve SZ diagnosis accuracy. Additionally, the development of rehabilitation systems based on AI techniques, such as BCI [291], neurofeedback [292], and wearable devices [293], could significantly benefit patients with SZ and represents a promising avenue for future investigation.

**- Author Contributions:**

| Author | Contributions |
|---|---|
| Mahboobeh Jafari | Data curation; funding acquisition; investigation; writing – original draft; writing – review and editing. |
| Delaram Sadeghi | formal analysis; investigation; visualization; writing – original draft; writing – review and editing. |
| Afshin Shoeibi | Conceptualization; formal analysis; investigation; project administration; resources; supervision; visualization; writing – original draft; writing – review and editing. |
| Hamid Alinejad-Rokny | Formal analysis; supervision; validation; writing – review and editing. |
| Amin Beheshti | Formal analysis; methodology; validation; writing – review and editing. |
| David López García | writing – review and editing. |
| Zhaolin Chen | formal analysis; investigation; validation; writing – review and editing. |
| U. Rajendra Acharya | Conceptualization; formal analysis; investigation; methodology; supervision; validation; writing – original draft; writing – review and editing. |
| Juan Manuel Gorriz | Conceptualization; investigation; project administration; resources; supervision; validation; visualization; writing – original draft; writing – review and editing. |

**- Conflicts of Interest:** The authors declare no conflict of interest.

**- Ethical and informed consent for data used:** This is a review paper and we do not use any data.

**- Data availability and access:** This is a review paper and we do not use any data.

**Appendix A:**

**Accuracy** can be defined as the proportion of correctly predicted observations to the total number of observations [88].

$$Acc = \frac{TP + TN}{FP + FN + TP + TN}$$

**Sensitivity**, also referred to as recall, can be defined as the proportion of correctly predicted positive observations to the total number of cases that have the particular condition of interest [88].

$$Sen = \frac{TP}{FN + TP}$$

**Specificity** can be defined as the proportion of correctly predicted negative observations to the total number of observations that are negative [88].

$$Spec = \frac{TN}{FN + TN}$$

Precision, also known as positive predictive value, represents the proportion of correctly predicted positive observations to the total number of observations that are predicted as positive [88].

$$Prec = \frac{TP}{TP + FP}$$

**Appendix B: Abbreviations**

| |
|---|
| A |
| Absolute value of the highest slope of autoregressive coefficients (AVLSAC) |
| Accuracy (Acc) |
| Adaptive neuro-fuzzy inference system (ANFIS) |
| Alzheimer's disease (AD) |
| Analysis of Variance (ANOVA) |
| Application-specific integrated circuits (ASIC) |
| Approximate Entropy (ApEn) |
| Artificial intelligence (AI) |
| Artificial neural networks (ANNs) |
| Autoencoders (AEs) |
| Autoregressive (AR) |
| B |
| Back propagation network (BPN) |
| Bat optimization (BA) |
| Black Hole (BH) |
| Boosted version of Direct Linear Discriminant Analysis (BDLDA) |
| Brain-computer interface (BCI) |
| C |
| Clinically High-risk (CHR) |
| Cognitive-behavioral therapy (CBT) |
| Complex Network (CN) |
| Computed tomography (CT) |
| Computer-aided diagnosis system (CADS) |
| Continuous wavelet transform (CWT) |
| Convolutional AE (CAE) |
| Convolutional neural networks (CNNs) |
| Correlation-based feature selection (CBFS) |
| Cyclic Group of Prime Order Pattern (CGP17Pat) |
| D |
| Data augmentation (DA) |
| Decision tree (DT) |
| Deep attention mechanisms (DAMs) |
| Deep brain stimulation (DBS) |
| Deep learning (DL) |
| Deep multi-task learning (DMTL) |
| Deep mutual learning (DML) |
| Detrend Fluctuation Analysis (DFA) |
| Diagnostic and statistical manual of mental disorders (DSM) |
| Diffusion tensor imaging (DTI) |
| Discrete Fourier transform (DFT) |
| Discrete wavelet transform (DWT) |
| E |
| Electroconvulsive therapy (ECT) |
| Electroencephalography (EEG) |
| Electromyogram (EMG) |
| Empirical mode decomposition (EMD) |
| Ensemble bagged tree (EBT) |
| Event-related potentials (ERPs) |
| Empirical wavelet transform (EWT) |
| Expectation Maximization based Principal Component Analysis (EM-PCA) |
| Explainable AI (XAI) |
| Extreme learning machine (ELM) |
| F |
| Fast Fourier transform (FFT) |
| Feature ranking (FR) |
| Federated learning (FL) |
| Field programmable gate arrays (FPGA) |
| Flexible least square support vector machine (F-LSSVM) |
| Fractal Dimension (FD) |
| Fully connected (FC) |
| Functional magnetic resonance imaging (fMRI) |
| Fuzzy C-Means (FCM) |
| Fuzzy synchronization likelihood (FSL) |
| G |
| Gated recurrent unit (GRU) |
| Genetic Algorithm (GA) |
| Graph AEs (GAEs) |
| Graph RNNs (GRNNs) |
| Graph Attention Networks (GATs) |
| Graph Convolutional Neural Networks (GCNN) |
| Graphics processing units (GPUs) |
| Grey-Wolf optimization (GSO) |
| H |
| Healthy control (HC) |
| Higuchi's Fractal Dimension (HFD) |
| Hilbert Spectrum (HS) |
| Histogram of local variance (HLV) |
| Hurst Exponent (HE) |
| I |
| Independent component analysis (ICA) |
| Information Entropy (InEn) |
| Intrinsic Mode Functions (IMF) |
| Iterative neighborhood component analysis (INCA) |
| Iterative tunable q-factor wavelet transform (ITQWT) |
| J |
| K |
| K-nearest neighbors (KNN) |
| Kolmogorov Complexity (KOL) |
| Kruskal Wallis (KW) |
| L |
| Largest Lyapunov Exponent (LLE) |
| Lempel Ziv Complexity (LZC) |
| Linear discriminant analysis (LDA) |
| Linear predictive coding (LPC) |
| Linear series decomposition learner (LSDL) |
| Local binary pattern (LBP) |
| Logistic regression (LR) |
| Look Ahead Pattern (LAP) |
| long-short-term memory (LSTM) |
| Lyapunov exponents (Lya) |
| M |
| Machine learning (ML) |
| Magnetic resonance spectroscopy (MRS) |
| Magnetoencephalography (MEG) |
| maximum absolute pooling (MAP) |
| Mean Spectral Amplitude (MSA) |
| Mental Health Research Center (MHRC) |
| Multi-Channel Frequency Network (MUCHf-Net) |
| Multi-class Spatial Pattern of the Network (MSPN) |
| Multi-domain Connectome CNN (MDC-CNN) |
| Multi-Layer Perceptron (MLP) |
| Multi-level Discrete Wavelet Transformation (MDWT) |
| Multiple sclerosis (MS) |
| Multiscale principal component analysis (MSPCA) |
| Multisynchrosqueezing transform (MSST) |
| Multi-variate empirical mode decomposition (MEMD) |
| N |
| O |
| Optimized extreme learning machine (OELM) |
| P |
| Parkinson's disease (PD) |
| Partial directed coherence (PDC) |
| Partial Least Squares Non linear Regression (PLS-NLR) |
| Phase lag index (PLI) |
| Phase synchronization (PS) |
| Positron emission tomography (PET) |
| Power spectral density (PSD) |
| Precision (Pre) |
| Preferred reporting items for systematic reviews and meta-analyses (PRISMA) |
| Principal component analysis (PCA) |
| Probabilistic neural network (PNN) |
| Q |
| R |
| Radial Basis Function (RBF) |
| Random Forest (RF) |
| Random Subset Feature Selection (RSFS) |
| Recall (Re) |
| Recurrence Quantification Analysis (RQA) |
| Recurrent Auto-encoder (RAE) |
| Recurrent neural networks (RNNs) |
| Recursive feature elimination (RFE) |
| Robust variational mode decomposition (RVMD) |
| S |
| Schizophrenia (SZ) |
| Sensitivity (Sen) |
| Sequential forward selection (SFS) |
| Shannon entropy (ShEn) |
| Short-time Fourier transforms (STFT) |
| Signal-to-noise ratio (SNR) |
| Single-photon emission computerized tomography (SPECT) |
| Smoothed pseudo-Wigner–Ville distribution (SPWVD) |
| Sparse Autoencoder (SAE) |
| Specificity (Spe) |
| Spectral eEntropy (SpEn) |
| Squeeze Excitation Network-LSTM- Softmax (SLS) |
| Structural magnetic resonance imaging (sMRI) |
| Support Vector Machine (SVM) |
| Symbolic Transfer Entropy (STE) |
| Symmetrically weighted local binary patterns (SLBP) |
| Synchronization likelihood (SL) |
| T |
| t-distributed stochastic neighbor embedding (t-SNE) |
| Time–frequency representation (TFR) |
| Transcranial direct current stimulation (tDCS) |
| Transcranial magnetic stimulation (TMS) |
| Transfer Entropy (TE) |
| Tunable Q-factor wavelet transform (TQWT) |
| U |
| Uncertainty quantification (UQ) |
| V |
| Vector autoregressive (VAR) |
| W |
| Wavelet-enhanced Independent Component Analysis (wICA) |
| Wavelet Scattering Transform (WST) |
| Wavelet transform (WT) |
| Wolf-Bat Algorithm (WBA) |
| X |
| Y |
| Z |

**References**

[1] Insel, T. R. (2010). Rethinking schizophrenia. *Nature*, *468*(7321), 187-193.

[2] McCutcheon, R. A., Marques, T. R., & Howes, O. D. (2020). Schizophrenia—an overview. *JAMA psychiatry*, *77*(2), 201-210.

[3] Fletcher, P. C., & Frith, C. D. (2009). Perceiving is believing: a Bayesian approach to explaining the positive symptoms of schizophrenia. *Nature Reviews Neuroscience*, *10*(1), 48-58.

[4] Andreasen, N. C. (1982). Negative symptoms in schizophrenia: definition and reliability. *Archives of general psychiatry*, *39*(7), 784-788.

[5] Simpson, E. H., Kellendonk, C., & Kandel, E. (2010). A possible role for the striatum in the pathogenesis of the cognitive symptoms of schizophrenia. *Neuron*, *65*(5), 585-596.

[6] Henkel, N. D., Wu, X., O'Donovan, S. M., Devine, E. A., Jiron, J. M., Rowland, L. M., ... & McCullumsmith, R. E. (2022). Schizophrenia: A disorder of broken brain bioenergetics. *Molecular Psychiatry*, *27*(5), 2393-2404.

[7] Batiuk, M. Y., Tyler, T., Dragicevic, K., Mei, S., Rydbirk, R., Petukhov, V., ... & Khodosevich, K. (2022). Upper cortical layer–driven network impairment in schizophrenia. *Science Advances*, *8*(41), eabn8367.

[8] Dickerson, F. B., & Lehman, A. F. (2006). Evidence-based psychotherapy for schizophrenia. *The Journal of nervous and mental disease*, *194*(1), 3-9.

[9] Smolak, A., Gearing, R. E., Alonzo, D., Baldwin, S., Harmon, S., & McHugh, K. (2013). Social support and religion: mental health service use and treatment of schizophrenia. *Community mental health journal*, *49*, 444-450.

[10] Spaulding, W. D., Fleming, S. K., Reed, D., Sullivan, M., Storzbach, D., & Lam, M. (1999). Cognitive functioning in schizophrenia: implications for psychiatric rehabilitation. *Schizophrenia bulletin*, *25*(2), 275-289.

[11] Sun, S. X., Liu, G. G., Christensen, D. B., & Fu, A. Z. (2007). Review and analysis of hospitalization costs associated with antipsychotic nonadherence in the treatment of schizophrenia in the United States. *Current medical research and opinion*, *23*(10), 2305-2312.

[12] Tharyan, P., & Adams, C. E. (2005). Electroconvulsive therapy for schizophrenia. *Cochrane Database of Systematic Reviews*, (2).

[13] Agarwal, S. M., Shivakumar, V., Bose, A., Subramaniam, A., Nawani, H., Chhabra, H., ... & Venkatasubramanian, G. (2013). Transcranial direct current stimulation in schizophrenia. *Clinical Psychopharmacology and Neuroscience*, *11*(3), 118.

[14] Lindenmayer, J. P., Kulsa, M. K. C., Sultana, T., Kaur, A., Yang, R., Ljuri, I., ... & Khan, A. (2019). Transcranial direct-current stimulation in ultra-treatment-resistant schizophrenia. *Brain stimulation*, *12*(1), 54-61.

[15] Corripio, I., Roldán, A., McKenna, P., Sarró, S., Alonso-Solis, A., Salgado, L., ... & Portella, M. (2022). Target selection for deep brain stimulation in treatment resistant schizophrenia. *Progress in Neuro-Psychopharmacology and Biological Psychiatry*, *112*, 110436.

[16] Turkington, D., Dudley, R., Warman, D. M., & Beck, A. T. (2006). Cognitive-behavioral therapy for schizophrenia: a review. *Focus*, *10*(2), 5-233.

[17] Zygmunt, A., Olfson, M., Boyer, C. A., & Mechanic, D. (2002). Interventions to improve medication adherence in schizophrenia. *American Journal of Psychiatry*, *159*(10), 1653-1664.

[18] Bermanzohn, P. C., Porto, L., Arlow, P. B., Pollack, S., Stronger, R., & Siris, S. G. (2000). At issue: hierarchical diagnosis in chronic schizophrenia: a clinical study of co-occurring syndromes. *Schizophrenia Bulletin*, *26*(3), 517-525.

[19] Martin, C. T. (2016). The value of physical examination in mental health nursing. *Nurse Education in Practice*, *17*, 91-96.

[20] Tsuang, M. T., Stone, W. S., & Faraone, S. V. (2000). Toward reformulating the diagnosis of schizophrenia. *American Journal of Psychiatry*, *157*(7), 1041-1050.

[21] American Psychiatric Association, D., & American Psychiatric Association. (2013). *Diagnostic and statistical manual of mental disorders: DSM-5* (Vol. 5, No. 5). Washington, DC: American psychiatric association.

[22] Kraguljac, N. V., McDonald, W. M., Widge, A. S., Rodriguez, C. I., Tohen, M., & Nemeroff, C. B. (2021). Neuroimaging biomarkers in schizophrenia. *American Journal of Psychiatry*, *178*(6), 509-521.

[23] Cooper, R., & Blashfield, R. K. (2016). Re-evaluating DSM-I. *Psychological Medicine*, *46*(3), 449-456.

[24] Horwitz, A. V. (2014). DSM-I and DSM-II. *The encyclopedia of clinical psychology*, 1-6.

[25] Spitzer, R. L., Williams, J. B., & Skodol, A. E. (1980). DSM-III: the major achievements and an overview. *The American Journal of Psychiatry*.

[26] Krueger, R. F., Caspi, A., Moffitt, T. E., & Silva, P. A. (1998). The structure and stability of common mental disorders (DSM-III-R): a longitudinal-epidemiological study. *Journal of abnormal psychology*, *107*(2), 216.

[27] American Psychiatric Association, A. P., & American Psychiatric Association. (1994). *Diagnostic and statistical manual of mental disorders: DSM-IV* (Vol. 4). Washington, DC: American psychiatric association.

[28] Segal, D. L. (2010). Diagnostic and statistical manual of mental disorders (DSM-IV-TR). *The corsini encyclopedia of psychology*, 1-3.

[29] Sadeghi, D., Shoeibi, A., Ghassemi, N., Moridian, P., Khadem, A., Alizadehsani, R., ... & Acharya, U. R. (2022). An overview of artificial intelligence techniques for diagnosis of Schizophrenia based on magnetic resonance imaging modalities: Methods, challenges, and future works. *Computers in Biology and Medicine*, 105554.

[30] Shoeibi, A., Ghassemi, N., Khodatars, M., Moridian, P., Khosravi, A., Zare, A., ... & Rajendra Acharya, U. (2022). Automatic diagnosis of schizophrenia and attention deficit hyperactivity disorder in rs-fMRI modality using convolutional autoencoder model and interval type-2 fuzzy regression. *Cognitive Neurodynamics*, 1-23.

[31] Arbabshirani, M. R., Plis, S., Sui, J., & Calhoun, V. D. (2017). Single subject prediction of brain disorders in neuroimaging: Promises and pitfalls. *Neuroimage*, *145*, 137-165.

[32] Gur, R. E., & Gur, R. C. (2022). Functional magnetic resonance imaging in schizophrenia. *Dialogues in clinical neuroscience*.

[33] Patel, N. H., Vyas, N. S., Puri, B. K., Nijran, K. S., & Al-Nahhas, A. (2010). Positron emission tomography in schizophrenia: a new perspective. *Journal of Nuclear Medicine*, *51*(4), 511-520.

[34] Laruelle, M., Abi-Dargham, A., Van Dyck, C. H., Gil, R., D'Souza, C. D., Erdos, J., ... & Innis, R. (1996). Single photon emission computerized tomography imaging of amphetamine-induced dopamine release in drug-free schizophrenic subjects. *Proceedings of the National Academy of Sciences*, *93*(17), 9235-9240.

[35] Shoeibi, A., Rezaei, M., Ghassemi, N., Namadchian, Z., Zare, A., & Gorriz, J. M. (2022, May). Automatic diagnosis of schizophrenia in EEG signals using functional connectivity features and CNN-LSTM model. In *Artificial Intelligence in Neuroscience: Affective Analysis and Health Applications: 9th International Work-Conference on the Interplay Between Natural and Artificial Computation, IWINAC 2022, Puerto de la Cruz, Tenerife, Spain, May 31–June 3, 2022, Proceedings, Part I* (pp. 63-73). Cham: Springer International Publishing.

[36] Edgar, J. C., Guha, A., & Miller, G. A. (2020). Magnetoencephalography for schizophrenia. *Neuroimaging Clinics*, *30*(2), 205-216.

[37] Srivastava, N. K., Khanra, S., Chail, V., & Khess, C. R. (2015). Clinical correlates of enlarged cavum septum pellucidum in schizophrenia: A revisit through computed tomography. *Asian Journal of Psychiatry*, *15*, 21-24.

[38] Samartzis, L., Dima, D., Fusar-Poli, P., & Kyriakopoulos, M. (2014). White matter alterations in early stages of schizophrenia: a systematic review of diffusion tensor imaging studies. *Journal of neuroimaging*, *24*(2), 101-110.

[39] Rowland, L. M., Pradhan, S., Korenic, S., Wijtenburg, S. A., Hong, L. E., Edden, R. A., & Barker, P. B. (2016). Elevated brain lactate in schizophrenia: a 7 T magnetic resonance spectroscopy study. *Translational psychiatry*, *6*(11), e967-e967.

[40] Shoeibi, A., Sadeghi, D., Moridian, P., Ghassemi, N., Heras, J., Alizadehsani, R., ... & Gorriz, J. M. (2021). Automatic diagnosis of schizophrenia in EEG signals using CNN-LSTM models. *Frontiers in neuroinformatics*, 58.

[41] Shoeibi, A., Moridian, P., Khodatars, M., Ghassemi, N., Jafari, M., Alizadehsani, R., ... & Acharya, U. R. (2022). An overview of deep learning techniques for epileptic seizures detection and prediction based on neuroimaging modalities: Methods, challenges, and future works. *Computers in Biology and Medicine*, 106053.

[42] Merlin Praveena, D., Angelin Sarah, D., & Thomas George, S. (2022). Deep learning techniques for EEG signal applications–a review. *IETE Journal of Research*, *68*(4), 3030-3037.

[43] Shoeibi, A., Ghassemi, N., Khodatars, M., Moridian, P., Alizadehsani, R., Zare, A., ... & Gorriz, J. M. (2022). Detection of epileptic seizures on EEG signals using ANFIS classifier, autoencoders and fuzzy entropies. *Biomedical Signal Processing and Control*, *73*, 103417.
[44] He, C., Chen, Y. Y., Phang, C. R., Stevenson, C., Chen, I. P., Jung, T. P., & Ko, L. W. (2023). Diversity and Suitability of the State-of-the-Art Wearable and Wireless EEG Systems Review. *IEEE Journal of Biomedical and Health Informatics*.
[45] Cherian, R., & Kanaga, E. G. (2022). Theoretical and methodological analysis of EEG based seizure detection and prediction: An exhaustive review. *Journal of Neuroscience Methods*, 109483.
[46] Acharya, U. R., Sree, S. V., Swapna, G., Martis, R. J., & Suri, J. S. (2013). Automated EEG analysis of epilepsy: a review. *Knowledge-Based Systems*, *45*, 147-165.
[47] Verma, S., Goel, T., Tanveer, M., Ding, W., Sharma, R., & Murugan, R. (2023). Machine learning techniques for the Schizophrenia diagnosis: A comprehensive review and future research directions. *arXiv preprint arXiv:2301.07496*.
[48] Cortes-Briones, J. A., Tapia-Rivas, N. I., D'Souza, D. C., & Estevez, P. A. (2022). Going deep into schizophrenia with artificial intelligence. *Schizophrenia Research*, *245*, 122-140.
[49] Barros, C., Silva, C. A., & Pinheiro, A. P. (2021). Advanced EEG-based learning approaches to predict schizophrenia: Promises and pitfalls. *Artificial intelligence in medicine*, *114*, 102039.
[50] Luján, M. Á., Jimeno, M. V., Mateo Sotos, J., Ricarte, J. J., & Borja, A. L. (2021). A survey on eeg signal processing techniques and machine learning: Applications to the neurofeedback of autobiographical memory deficits in schizophrenia. *Electronics*, *10*(23), 3037.
[51] Lai, J. W., Ang, C. K. E., Acharya, U. R., & Cheong, K. H. (2021). Schizophrenia: a survey of artificial intelligence techniques applied to detection and classification. *International journal of environmental research and public health*, *18*(11), 6099.
[52] Steardo Jr, L., Carbone, E. A., De Filippis, R., Pisanu, C., Segura-Garcia, C., Squassina, A., ... & Steardo, L. (2020). Application of support vector machine on fMRI data as biomarkers in schizophrenia diagnosis: a systematic review. *Frontiers in psychiatry*, *11*, 588.
[53] de Filippis, R., Carbone, E. A., Gaetano, R., Bruni, A., Pugliese, V., Segura-Garcia, C., & De Fazio, P. (2019). Machine learning techniques in a structural and functional MRI diagnostic approach in schizophrenia: a systematic review. *Neuropsychiatric disease and treatment*, 1605-1627.
[54] Moher, D., Liberati, A., Tetzlaff, J., Altman, D. G., & PRISMA Group*, T. (2009). Preferred reporting items for systematic reviews and meta-analyses: the PRISMA statement. *Annals of internal medicine*, *151*(4), 264-269.
[55] Khan, P., Kader, M. F., Islam, S. R., Rahman, A. B., Kamal, M. S., Toha, M. U., & Kwak, K. S. (2021). Machine learning and deep learning approaches for brain disease diagnosis: principles and recent advances. *IEEE Access*, *9*, 37622-37655.
[56] Cho, G., Yim, J., Choi, Y., Ko, J., & Lee, S. H. (2019). Review of machine learning algorithms for diagnosing mental illness. *Psychiatry investigation*, *16*(4), 262.
[57] Zhang, L., Wang, M., Liu, M., & Zhang, D. (2020). A survey on deep learning for neuroimaging-based brain disorder analysis. *Frontiers in neuroscience*, *14*, 779.
[58] Shatte, A. B., Hutchinson, D. M., & Teague, S. J. (2019). Machine learning in mental health: a scoping review of methods and applications. *Psychological medicine*, *49*(9), 1426-1448.
[59] Tzimourta, K. D., Christou, V., Tzallas, A. T., Giannakeas, N., Astrakas, L. G., Angelidis, P., ... & Tsipouras, M. G. (2021). Machine learning algorithms and statistical approaches for Alzheimer's disease analysis based on resting-state EEG recordings: A systematic review. *International journal of neural systems*, *31*(05), 2130002.
[60] Rasheed, K., Qayyum, A., Qadir, J., Sivathamboo, S., Kwan, P., Kuhlmann, L., ... & Razi, A. (2020). Machine learning for predicting epileptic seizures using EEG signals: A review. *IEEE Reviews in Biomedical Engineering*, *14*, 139-155.
[61] Shoeibi, A., Khodatars, M., Jafari, M., Ghassemi, N., Moridian, P., Alizadesani, R., ... & Gorriz, J. M. (2022). Diagnosis of brain diseases in fusion of neuroimaging modalities using deep learning: A review. *Information Fusion*.
[62] Pathak, D., Kashyap, R., & Rahamatkar, S. (2022). A study of deep learning approach for the classification of electroencephalogram (EEG) brain signals. In *Artificial Intelligence and Machine Learning for EDGE Computing* (pp. 133-144). Academic Press.
[63] Shoeibi, A., Ghassemi, N., Alizadehsani, R., Rouhani, M., Hosseini-Nejad, H., Khosravi, A., ... & Nahavandi, S. (2021). A comprehensive comparison of handcrafted features and convolutional autoencoders for epileptic seizures detection in EEG signals. *Expert Systems with Applications*, *163*, 113788.
[64] Oh, S. L., Hagiwara, Y., Raghavendra, U., Yuvaraj, R., Arunkumar, N., Murugappan, M., & Acharya, U. R. (2020). A deep learning approach for Parkinson's disease diagnosis from EEG signals. *Neural Computing and Applications*, *32*, 10927-10933.

[65] Yasin, S., Hussain, S. A., Aslan, S., Raza, I., Muzammel, M., & Othmani, A. (2021). EEG based Major Depressive disorder and Bipolar disorder detection using Neural Networks: A review. *Computer Methods and Programs in Biomedicine*, *202*, 106007.
[66] Olejarczyk, E., & Jernajczyk, W. (2017). Graph-based analysis of brain connectivity in schizophrenia. *PloS one*, *12*(11), e0188629.
[67] https://www.kaggle.com/datasets/broach/button-tone-sz
[68] Kim, S. P. (2018). Preprocessing of EEG. *Computational EEG Analysis: Methods and Applications*, 15-33.
[69] Sazgar, M., Young, M. G., Sazgar, M., & Young, M. G. (2019). EEG artifacts. *Absolute Epilepsy and EEG Rotation Review: Essentials for Trainees*, 149-162.
[70] Lakshmi, M. R., Prasad, T. V., & Prakash, D. V. C. (2014). Survey on EEG signal processing methods. *International journal of advanced research in computer science and software engineering*, *4*(1).
[71] Thakor, N. V., & Sherman, D. L. (2012). EEG signal processing: Theory and applications. In *Neural Engineering* (pp. 259-303). Boston, MA: Springer US.
[72] Babiloni, C., Pennica, A., Del Percio, C., Noce, G., Cordone, S., Lopez, S., ... & Andreoni, M. (2016). Antiretroviral therapy affects the z-score index of deviant cortical EEG rhythms in naïve HIV individuals. *NeuroImage: Clinical*, *12*, 144-156.
[73] Kwak, Y., Kong, K., Song, W. J., & Kim, S. E. (2023). Subject-Invariant Deep Neural Networks based on Baseline Correction for EEG Motor Imagery BCI. *IEEE Journal of Biomedical and Health Informatics*.
[74] Delorme, A., & Makeig, S. (2004). EEGLAB: an open source toolbox for analysis of single-trial EEG dynamics including independent component analysis. *Journal of neuroscience methods*, *134*(1), 9-21.
[75] AlSharabi, K., Salamah, Y. B., Abdurraqeeb, A. M., Aljalal, M., & Alturki, F. A. (2022). EEG signal processing for Alzheimer's disorders using discrete wavelet transform and machine learning approaches. *IEEE Access*, *10*, 89781-89797.
[76] Xu, S., Wang, Z., Sun, J., Zhang, Z., Wu, Z., Yang, T., ... & Cheng, C. (2020). Using a deep recurrent neural network with EEG signal to detect Parkinson's disease. *Annals of translational medicine*, *8*(14).
[77] Shoeibi, A., Khodatars, M., Ghassemi, N., Jafari, M., Moridian, P., Alizadehsani, R., ... & Acharya, U. R. (2021). Epileptic seizures detection using deep learning techniques: A review. *International Journal of Environmental Research and Public Health*, *18*(11), 5780.
[78] Thilakvathi, B., Devi, S. S., Bhanu, K., & Malaippan, M. (2017). EEG signal complexity analysis for schizophrenia during rest and mental activity. *Biomedical Research-India*, *28*(1), 1-9.
[79] Siuly, S., Khare, S. K., Bajaj, V., Wang, H., & Zhang, Y. (2020). A computerized method for automatic detection of schizophrenia using EEG signals. *IEEE Transactions on Neural Systems and Rehabilitation Engineering*, *28*(11), 2390-2400.
[80] Prabhakar, S. K., Rajaguru, H., & Kim, S. H. (2020). Schizophrenia EEG signal classification based on swarm intelligence computing. *Computational Intelligence and Neuroscience*, *2020*.
[81] Khare, S. K., Bajaj, V., Siuly, S., & Sinha, G. R. (2020). Classification of schizophrenia patients through empirical wavelet transformation using electroencephalogram signals. In *Modelling and Analysis of Active Biopotential Signals in Healthcare, Volume 1*. IOP Publishing.
[82] Buettner, R., Hirschmiller, M., Schlosser, K., Rössle, M., Fernandes, M., & Timm, I. J. (2019, October). High-performance exclusion of schizophrenia using a novel machine learning method on EEG data. In *2019 IEEE International Conference on E-Health Networking, Application & Services (HealthCom)* (pp. 1-6). IEEE.
[83] Vasios, C., Papageorgiou, C., Matsopoulos, G. K., Nikita, K. S., & Uzunoglu, N. (2002). A decision support system of evoked potentials for the classification of patients with first-episode schizophrenia. *German Journal of Psychiatry*, *5*, 78-84.
[84] Chang, Q., Li, C., Zhang, J., & Wang, C. (2022). Dynamic brain functional network based on EEG microstate during sensory gating in schizophrenia. *Journal of Neural Engineering*, *19*(2), 026007.
[85] Laton, J., Van Schependom, J., Gielen, J., Decoster, J., Moons, T., De Keyser, J., ... & Nagels, G. (2014). Single-subject classification of schizophrenia patients based on a combination of oddball and mismatch evoked potential paradigms. *Journal of the neurological sciences*, *347*(1-2), 262-267.
[86] Prabhu, S., & Martis, R. J. (2020, July). Diagnosis of schizophrenia using Kolmogorov complexity and sample entropy. In *2020 IEEE International Conference on Electronics, Computing and Communication Technologies (CONECCT)* (pp. 1-4). IEEE.
[87] Khare, S. K., & Bajaj, V. (2021). A self-learned decomposition and classification model for schizophrenia diagnosis. *Computer Methods and Programs in Biomedicine*, *211*, 106450.
[88] Masychev, K., Ciprian, C., & Ravan, M. (2020, December). Machine Learning Approach to Diagnose Schizophrenia Based on Effective Connectivity of Resting EEG Data. In *2020 IEEE Signal Processing in Medicine and Biology Symposium (SPMB)* (pp. 1-6). IEEE.

[89] Najafzadeh, H., Esmaeili, M., Farhang, S., Sarbaz, Y., & Rasta, S. H. (2021). Automatic classification of schizophrenia patients using resting-state EEG signals. *Physical and Engineering Sciences in Medicine*, *44*(3), 855-870.
[90] Prabhakar, S. K., Rajaguru, H., & Lee, S. W. (2020). A framework for schizophrenia EEG signal classification with nature inspired optimization algorithms. *IEEE Access*, *8*, 39875-39897.
[91] Azizi, S., Hier, D. B., & Wunsch, D. C. (2021, November). Schizophrenia classification using resting state EEG functional connectivity: source level outperforms sensor level. In *2021 43rd Annual International Conference of the IEEE Engineering in Medicine & Biology Society (EMBC)* (pp. 1770-1773). IEEE.
[92] Rajesh, K. N., & Kumar, T. S. (2021, November). Schizophrenia Detection in Adolescents from EEG Signals using Symmetrically weighted Local Binary Patterns. In *2021 43rd Annual International Conference of the IEEE Engineering in Medicine & Biology Society (EMBC)* (pp. 963-966). IEEE.
[93] Sharma, M., & Acharya, U. R. (2021). Automated detection of schizophrenia using optimal wavelet-based l 1 norm features extracted from single-channel EEG. *Cognitive Neurodynamics*, *15*(4), 661-674.
[94] Xin, J., Zhou, K., Wang, Z., Wang, Z., Chen, J., Wang, X., & Chen, Q. (2022). Hybrid High-order Brain Functional Networks for Schizophrenia-Aided Diagnosis. *Cognitive Computation*, *14*(4), 1303-1315.
[95] Dvey-Aharon, Z., Fogelson, N., Peled, A., & Intrator, N. (2015). Schizophrenia detection and classification by advanced analysis of EEG recordings using a single electrode approach. *PloS one*, *10*(4), e0123033.
[96] WeiKoh, J. E., Rajinikanth, V., Vicnesh, J., Pham, T. H., Oh, S. L., Yeong, C. H., ... & Cheong, K. H. (2022). Application of local configuration pattern for automated detection of schizophrenia with electroencephalogram signals. *Expert Systems*, e12957.
[97] Li, F., Wang, J., Liao, Y., Yi, C., Jiang, Y., Si, Y., ... & Xu, P. (2019). Differentiation of schizophrenia by combining the spatial EEG brain network patterns of rest and task P300. *IEEE Transactions on Neural Systems and Rehabilitation Engineering*, *27*(4), 594-602.
[98] Bougou, V., Mporas, I., Schirmer, P., & Ganchev, T. (2019, November). Evaluation of EEG Connectivity Network Measures based Features in Schizophrenia Classification. In *2019 International Conference on Biomedical Innovations and Applications (BIA)* (pp. 1-4). IEEE.
[99] Almutairi, M. M., Alhamad, N., Alyami, A., Alshobbar, Z., Alfayez, H., Al-Akkas, N., ... & Olatunji, S. O. (2019, May). Preemptive diagnosis of schizophrenia disease using computational intelligence techniques. In *2019 2nd International Conference on Computer Applications & Information Security (ICCAIS)* (pp. 1-6). IEEE.
[100] Thilakavathi, B., Shenbaga Devi, S., Malaiappan, M., & Bhanu, K. (2019). EEG power spectrum analysis for schizophrenia during mental activity. *Australasian Physical & Engineering Sciences in Medicine*, *42*(3), 887-897.
[101] Zhao, Q., Hu, B., Li, Y., Peng, H., Li, L., Liu, Q., ... & Feng, J. (2013, November). An Alpha resting EEG study on nonlinear dynamic analysis for schizophrenia. In *2013 6th International IEEE/EMBS Conference on Neural Engineering (NER)* (pp. 484-488). IEEE.
[102] Hiesh, M. H., Andy, Y. Y. L., Shen, C. P., Chen, W., Lin, F. S., Sung, H. Y., ... & Lai, F. (2013, July). Classification of schizophrenia using genetic algorithm-support vector machine (ga-svm). In *2013 35th Annual International Conference of the IEEE Engineering in Medicine and Biology Society (EMBC)* (pp. 6047-6050). IEEE.
[103] Zhang, L. (2019, July). EEG signals classification using machine learning for the identification and diagnosis of schizophrenia. In *2019 41st Annual International Conference of the IEEE Engineering in Medicine and Biology Society (EMBC)* (pp. 4521-4524). IEEE.
[104] Kim, J. Y., Lee, H. S., & Lee, S. H. (2020). EEG source network for the diagnosis of schizophrenia and the identification of subtypes based on symptom severity—A machine learning approach. *Journal of Clinical Medicine*, *9*(12), 3934.
[105] Shim, M., Hwang, H. J., Kim, D. W., Lee, S. H., & Im, C. H. (2016). Machine-learning-based diagnosis of schizophrenia using combined sensor-level and source-level EEG features. *Schizophrenia research*, *176*(2-3), 314-319.
[106] Johannesen, J. K., Bi, J., Jiang, R., Kenney, J. G., & Chen, C. M. A. (2016). Machine learning identification of EEG features predicting working memory performance in schizophrenia and healthy adults. *Neuropsychiatric electrophysiology*, *2*(1), 1-21.
[107] Ke, P. F., Xiong, D. S., Li, J. H., Pan, Z. L., Zhou, J., Li, S. J., ... & Wu, K. (2021). An integrated machine learning framework for a discriminative analysis of schizophrenia using multi-biological data. *Scientific reports*, *11*(1), 1-11.

[108] Zhao, Z., Wang, C., Yuan, Q., Zhao, J., Ren, Q., Xu, Y., ... & Yu, Y. (2020). Dynamic changes of brain networks during feedback-related processing of reinforcement learning in schizophrenia. *Brain Research*, *1746*, 146979.
[109] Dvey-Aharon, Z., Fogelson, N., Peled, A., & Intrator, N. (2017). Connectivity maps based analysis of EEG for the advanced diagnosis of schizophrenia attributes. *PloS one*, *12*(10), e0185852.
[110] Santos-Mayo, L., San-José-Revuelta, L. M., & Arribas, J. I. (2016). A computer-aided diagnosis system with EEG based on the P3b wave during an auditory odd-ball task in schizophrenia. *IEEE transactions on biomedical engineering*, *64*(2), 395-407.
[111] Goshvarpour, A., & Goshvarpour, A. (2020). Schizophrenia diagnosis using innovative EEG feature-level fusion schemes. *Physical and Engineering Sciences in Medicine*, *43*(1), 227-238.
[112] Chu, W. L., Huang, M. W., Jian, B. L., & Cheng, K. S. (2017). Analysis of EEG entropy during visual evocation of emotion in schizophrenia. *Annals of General Psychiatry*, *16*(1), 1-9.
[113] Sabeti, M., Behroozi, R., & Moradi, E. (2016). Analysing complexity, variability and spectral measures of schizophrenic EEG signal. *International Journal of Biomedical Engineering and Technology*, *21*(2), 109-127.
[114] Khare, S. K., & Bajaj, V. (2021). A self-learned decomposition and classification model for schizophrenia diagnosis. *Computer Methods and Programs in Biomedicine*, *211*, 106450.
[115] Luján, M. Á., Sotos, J. M., Santos, J. L., & Borja, A. L. (2022). Accurate Neural Network Classification Model for Schizophrenia Disease Based on Electroencephalogram Data.
[116] Febles, E. S., Ortega, M. O., Sosa, M. V., & Sahli, H. (2022). Machine Learning techniques for the diagnosis of Schizophrenia based on Event Related Potentials. *medRxiv*.
[117] Khare, S. K., & Bajaj, V. (2022). A hybrid decision support system for automatic detection of Schizophrenia using EEG signals. *Computers in Biology and Medicine*, *141*, 105028.
[118] Goshvarpour, A., & Goshvarpour, A. (2022). Schizophrenia diagnosis by weighting the entropy measures of the selected EEG channel. *Journal of Medical and Biological Engineering*, *42*(6), 898-908.
[119] Ellis, C. A., Sattiraju, A., Miller, R., & Calhoun, V. (2022, November). Examining Reproducibility of EEG Schizophrenia Biomarkers Across Explainable Machine Learning Models. In *2022 IEEE 22nd International Conference on Bioinformatics and Bioengineering (BIBE)* (pp. 305-308). IEEE.
[120] Sahu, P. K. (2023). Artificial intelligence system for verification of schizophrenia via theta-EEG rhythm. *Biomedical Signal Processing and Control*, *81*, 104485.
[121] de Miras, J. R., Ibáñez-Molina, A. J., Soriano, M. F., & Iglesias-Parro, S. (2023). Schizophrenia classification using machine learning on resting state EEG signal. *Biomedical Signal Processing and Control*, *79*, 104233.
[122] Li, F., Jiang, L., Liao, Y., Li, C., Zhang, Q., Zhang, S., ... & Dai, J. (2022). Recognition of the Multi-class Schizophrenia Based on the Resting-State EEG Network Topology. *Brain Topography*, *35*(4), 495-506.
[123] Sairamya, N. J., Subathra, M. S. P., & George, S. T. (2022). Automatic identification of schizophrenia using EEG signals based on discrete wavelet transform and RLNDiP technique with ANN. *Expert Systems with Applications*, *192*, 116230.
[124] Chang, Q., Li, C., Zhang, J., & Wang, C. (2022). Dynamic brain functional network based on EEG microstate during sensory gating in schizophrenia. *Journal of Neural Engineering*, *19*(2), 026007.
[125] Kumar, T. S., Rajesh, K. N., Maheswari, S., Kanhangad, V., & Acharya, U. R. (2023). Automated Schizophrenia detection using local descriptors with EEG signals. *Engineering Applications of Artificial Intelligence*, *117*, 105602.
[126] Balasubramanian, K., Ramya, K., & Gayathri Devi, K. (2022). Optimized adaptive neuro-fuzzy inference system based on hybrid grey wolf-bat algorithm for schizophrenia recognition from EEG signals. *Cognitive Neurodynamics*, 1-19.
[127] Baradits, M., Bitter, I., & Czobor, P. (2020). Multivariate patterns of EEG microstate parameters and their role in the discrimination of patients with schizophrenia from healthy controls. *Psychiatry research*, *288*, 112938.
[128] Keihani, A., Sajadi, S. S., Hasani, M., & Ferrarelli, F. (2022). Bayesian optimization of machine learning classification of resting-state EEG microstates in schizophrenia: a proof-of-concept preliminary study based on secondary analysis. *Brain Sciences*, *12*(11), 1497.
[129] Aydemir, E., Dogan, S., Baygin, M., Ooi, C. P., Barua, P. D., Tuncer, T., & Acharya, U. R. (2022, March). CGP17Pat: Automated schizophrenia detection based on a cyclic group of prime order patterns using EEG signals. In *Healthcare* (Vol. 10, No. 4, p. 643). MDPI.
[130] Jahmunah, V., Oh, S. L., Rajinikanth, V., Ciaccio, E. J., Cheong, K. H., Arunkumar, N., & Acharya, U. R. (2019). Automated detection of schizophrenia using nonlinear signal processing methods. *Artificial intelligence in medicine*, *100*, 101698.

[131] Baygin, M. (2021). An accurate automated schizophrenia detection using TQWT and statistical moment based feature extraction. *Biomedical Signal Processing and Control*, *68*, 102777.
[132] Min, B., Kim, M., Lee, J., Byun, J. I., Chu, K., Jung, K. Y., ... & Kwon, J. S. (2020). Prediction of individual responses to electroconvulsive therapy in patients with schizophrenia: Machine learning analysis of resting-state electroencephalography. *Schizophrenia research*, *216*, 147-153.
[133] Kim, K., Duc, N. T., Choi, M., & Lee, B. (2021). EEG microstate features for schizophrenia classification. *PloS one*, *16*(5), e0251842.
[134] Akbari, H., Ghofrani, S., Zakalvand, P., & Sadiq, M. T. (2021). Schizophrenia recognition based on the phase space dynamic of EEG signals and graphical features. *Biomedical Signal Processing and Control*, *69*, 102917.
[135] Baygin, M., Yaman, O., Tuncer, T., Dogan, S., Barua, P. D., & Acharya, U. R. (2021). Automated accurate schizophrenia detection system using Collatz pattern technique with EEG signals. *Biomedical Signal Processing and Control*, *70*, 102936.
[136] Ciprian, C., Masychev, K., Ravan, M., Manimaran, A., & Deshmukh, A. (2021). Diagnosing schizophrenia using effective connectivity of resting-state EEG data. *Algorithms*, *14*(5), 139.
[137] Das, K., & Pachori, R. B. (2021). Schizophrenia detection technique using multivariate iterative filtering and multichannel EEG signals. *Biomedical Signal Processing and Control*, *67*, 102525.
[138] Aksöz, A., Akyüz, D., BAYIR, F., YILDIZ, N. C., Orhanbulucu, F., & Latifoğlu, F. Analysis and Classification of Schizophrenia Using Event Related Potential Signals. *Computer Science*, 32-36.
[139] URAL, A. B., & Uğur, E. R. A. Y. AUTOMATED PSYCHIATRIC DATA ANALYSIS from SINGLE CHANNEL EEG with SIGNAL PROCESSING and ARTIFICIAL INTELLIGENCE METHODS. *Journal of Scientific Reports-A*, (050), 106-123.
[140] Krishnan, P. T., Raj, A. N. J., Balasubramanian, P., & Chen, Y. (2020). Schizophrenia detection using MultivariateEmpirical Mode Decomposition and entropy measures from multichannel EEG signal. *Biocybernetics and Biomedical Engineering*, *40*(3), 1124-1139.
[141] Góngora Alonso, S., Marques, G., Agarwal, D., De la Torre Díez, I., & Franco-Martín, M. (2022). Comparison of machine learning algorithms in the prediction of hospitalized patients with schizophrenia. *Sensors*, *22*(7), 2517.
[142] Liu, H., Zhang, T., Ye, Y., Pan, C., Yang, G., Wang, J., & Qiu, R. C. (2017). A data driven approach for resting-state EEG signal classification of schizophrenia with control participants using random matrix theory. *arXiv preprint arXiv:1712.05289*.
[143] Boostani, R., Sadatnezhad, K., & Sabeti, M. (2009). An efficient classifier to diagnose of schizophrenia based on the EEG signals. *Expert Systems with Applications*, *36*(3), 6492-6499.
[144] Siuly, S., Khare, S. K., Bajaj, V., Wang, H., & Zhang, Y. (2020). A computerized method for automatic detection of schizophrenia using EEG signals. *IEEE Transactions on Neural Systems and Rehabilitation Engineering*, *28*(11), 2390-2400.
[145] Baygin, M., Yaman, O., Tuncer, T., Dogan, S., Barua, P. D., & Acharya, U. R. (2021). Automated accurate schizophrenia detection system using Collatz pattern technique with EEG signals. *Biomedical Signal Processing and Control*, *70*, 102936.
[146] Akbari, H., Ghofrani, S., Zakalvand, P., & Sadiq, M. T. (2021). Schizophrenia recognition based on the phase space dynamic of EEG signals and graphical features. *Biomedical Signal Processing and Control*, *69*, 102917.
[147] Liu, H., Zhang, T., Ye, Y., Pan, C., Yang, G., Wang, J., & Qiu, R. C. (2017). A data driven approach for resting-state EEG signal classification of schizophrenia with control participants using random matrix theory. *arXiv preprint arXiv:1712.05289*.
[148] Devia, C., Mayol-Troncoso, R., Parrini, J., Orellana, G., Ruiz, A., Maldonado, P. E., & Egaña, J. I. (2019). EEG classification during scene free-viewing for schizophrenia detection. *IEEE Transactions on Neural Systems and Rehabilitation Engineering*, *27*(6), 1193-1199.
[149] Luo, Y., Tian, Q., Wang, C., Zhang, K., Wang, C., & Zhang, J. (2020). Biomarkers for prediction of schizophrenia: Insights from resting-state EEG microstates. *IEEE Access*, *8*, 213078-213093.
[150] Najafzadeh, H., Esmaeili, M., Farhang, S., Sarbaz, Y., & Rasta, S. H. (2021). Automatic classification of schizophrenia patients using resting-state EEG signals. *Physical and Engineering Sciences in Medicine*, *44*(3), 855-870.
[151] Nikhil Chandran, A., Sreekumar, K., & Subha, D. P. (2021). EEG-based automated detection of schizophrenia using long short-term memory (LSTM) network. In *Advances in Machine Learning and Computational Intelligence: Proceedings of ICMLCI 2019* (pp. 229-236). Springer Singapore.

[152] Shalbaf, A., Bagherzadeh, S., & Maghsoudi, A. (2020). Transfer learning with deep convolutional neural network for automated detection of schizophrenia from EEG signals. *Physical and Engineering Sciences in Medicine*, *43*, 1229-1239.
[153] Ahmedt-Aristizabal, D., Fernando, T., Denman, S., Robinson, J. E., Sridharan, S., Johnston, P. J., ... & Fookes, C. (2020). Identification of children at risk of schizophrenia via deep learning and EEG responses. *IEEE Journal of biomedical and health informatics*, *25*(1), 69-76.
[154] Oh, S. L., Vicnesh, J., Ciaccio, E. J., Yuvaraj, R., & Acharya, U. R. (2019). Deep convolutional neural network model for automated diagnosis of schizophrenia using EEG signals. *Applied Sciences*, *9*(14), 2870.
[155] Aslan, Z., & Akin, M. (2020). Automatic Detection of Schizophrenia by Applying Deep Learning over Spectrogram Images of EEG Signals. *Traitement du Signal*, *37*(2).
[156] Bondugula, R. K., Sivangi, K. B., & Udgata, S. K. (2022). Identification of schizophrenic individuals using activity records through visualization of recurrent networks. In *Intelligent Systems: Proceedings of ICMIB 2021* (pp. 653-664). Singapore: Springer Nature Singapore.
[157] Guo, Z., Wu, L., Li, Y., & Li, B. (2021, May). Deep neural network classification of EEG data in schizophrenia. In *2021 IEEE 10th Data Driven Control and Learning Systems Conference (DDCLS)* (pp. 1322-1327). IEEE.
[158] Zülfikar, A., & Mehmet, A. (2022). Empirical mode decomposition and convolutional neural network-based approach for diagnosing psychotic disorders from eeg signals. *Applied Intelligence*, *52*(11), 12103-12115.
[159] Khare, S. K., Bajaj, V., & Acharya, U. R. (2021). SPWVD-CNN for automated detection of schizophrenia patients using EEG signals. *IEEE Transactions on Instrumentation and Measurement*, *70*, 1-9.
[160] Chu, L., Qiu, R., Liu, H., Ling, Z., Zhang, T., & Wang, J. (2017). Individual recognition in schizophrenia using deep learning methods with random forest and voting classifiers: Insights from resting state EEG streams. *arXiv preprint arXiv:1707.03467*.
[161] Phang, C. R., Noman, F., Hussain, H., Ting, C. M., & Ombao, H. (2019). A multi-domain connectome convolutional neural network for identifying schizophrenia from EEG connectivity patterns. *IEEE journal of biomedical and health informatics*, *24*(5), 1333-1343.
[162] Laksono, I. K., & Imah, E. M. (2021, August). Schizophrenia Detection Based on Electroencephalogram Using Support Vector Machine. In *2021 International Conference on ICT for Smart Society (ICISS)* (pp. 1-6). IEEE.
[163] Singh, K., Singh, S., & Malhotra, J. (2021). Spectral features based convolutional neural network for accurate and prompt identification of schizophrenic patients. *Proceedings of the Institution of Mechanical Engineers, Part H: Journal of Engineering in Medicine*, *235*(2), 167-184.
[164] Sobahi, N., Ari, B., Cakar, H., Alcin, O. F., & Sengur, A. (2022). A new signal to image mapping procedure and convolutional neural networks for efficient schizophrenia detection in eeg recordings. *IEEE Sensors Journal*, *22*(8), 7913-7919.
[165] Phang, C. R., Ting, C. M., Samdin, S. B., & Ombao, H. (2019, March). Classification of EEG-based effective brain connectivity in schizophrenia using deep neural networks. In *2019 9th International IEEE/EMBS Conference on Neural Engineering (NER)* (pp. 401-406). IEEE.
[166] Alves, C. L., Pineda, A. M., Roster, K., Thielemann, C., & Rodrigues, F. A. (2022). EEG functional connectivity and deep learning for automatic diagnosis of brain disorders: Alzheimer's disease and schizophrenia. *Journal of Physics: Complexity*, *3*(2), 025001.
[167] Ellis, C. A., Sattiraju, A., Miller, R., & Calhoun, V. (2022, November). Examining effects of schizophrenia on EEG with explainable deep learning models. In *2022 IEEE 22nd International Conference on Bioinformatics and Bioengineering (BIBE)* (pp. 301-304). IEEE.
[168] Wu, Y., Xia, M., Wang, X., & Zhang, Y. (2023, April). Schizophrenia detection based on EEG using recurrent auto-encoder framework. In *Neural Information Processing: 29th International Conference, ICONIP 2022, Virtual Event, November 22–26, 2022, Proceedings, Part II* (pp. 62-73). Cham: Springer International Publishing.
[169] Menon, N. G., Bhavana, N. D., & Hema, D. D. Towards better intelligent implementation of Schizophrenia prediction using federated deep learning framework. *framework*, *6*(S3), 5631-5645.
[170] Siuly, S., Li, Y., Wen, P., & Alcin, O. F. (2022). SchizoGoogLeNet: the googlenet-based deep feature extraction design for automatic detection of schizophrenia. *Computational Intelligence and Neuroscience*, *2022*, 1-13.
[171] Alves, C. L., Pineda, A. M., Roster, K., Thielemann, C., & Rodrigues, F. A. (2022). EEG functional connectivity and deep learning for automatic diagnosis of brain disorders: Alzheimer's disease and schizophrenia. *Journal of Physics: complexity*, *3*(2), 025001.
[172] Ko, D. W., & Yang, J. J. (2022). EEG-Based schizophrenia diagnosis through time series image conversion and deep learning. *Electronics*, *11*(14), 2265.
[173] Saeedi, M., Saeedi, A., & Mohammadi, P. (2022). Schizophrenia Diagnosis via FFT and Wavelet Convolutional Neural Networks utilizing EEG signals.

[174] Kose, M., Ahirwal, M. K., & Atulkar, M. (2022). Weighted Ordinal Connection based Functional Network Classification for Schizophrenia Disease Detection using EEG signal.
[175] Supakar, R., Satvaya, P., & Chakrabarti, P. (2022). A deep learning based model using RNN-LSTM for the Detection of Schizophrenia from EEG data. *Computers in Biology and Medicine*, *151*, 106225.
[176] Aslan, Z., & Akin, M. (2022). A deep learning approach in automated detection of schizophrenia using scalogram images of EEG signals. *Physical and Engineering Sciences in Medicine*, *45*(1), 83-96.
[177] Bagherzadeh, S., Shahabi, M. S., & Shalbaf, A. (2022). Detection of schizophrenia using hybrid of deep learning and brain effective connectivity image from electroencephalogram signal. *Computers in Biology and Medicine*, *146*, 105570.
[178] Lillo, E., Mora, M., & Lucero, B. (2022). Automated diagnosis of schizophrenia using EEG microstates and Deep Convolutional Neural Network. *Expert Systems with Applications*, *209*, 118236.
[179] Sharma, G., & Joshi, A. M. (2022). SzHNN: A Novel and Scalable Deep Convolution Hybrid Neural Network Framework for Schizophrenia Detection Using Multichannel EEG. *IEEE Transactions on Instrumentation and Measurement*, *71*, 1-9.
[180] Prabhakar, S. K., & Lee, S. W. (2022). SASDL and RBATQ: Sparse Autoencoder With Swarm Based Deep Learning and Reinforcement Based Q-Learning for EEG Classification. *IEEE Open Journal of Engineering in Medicine and Biology*, *3*, 58-68.
[181] Wang, Z., Feng, J., Jiang, R., Shi, Y., Li, X., Xue, R., ... & Deng, W. (2022). Automated Rest EEG-Based Diagnosis of Depression and Schizophrenia Using a Deep Convolutional Neural Network. *IEEE Access*, *10*, 104472-104485.
[182] Jindal, K., Upadhyay, R., Padhy, P. K., & Longo, L. (2022). Bi-LSTM-deep CNN for schizophrenia detection using MSST-spectral images of EEG signals. In *Artificial Intelligence-Based Brain-Computer Interface* (pp. 145-162). Academic Press.
[183] Korda, A. I., Ventouras, E., Asvestas, P., Toumaian, M., Matsopoulos, G. K., & Smyrnis, N. (2022). Convolutional neural network propagation on electroencephalographic scalograms for detection of schizophrenia. *Clinical Neurophysiology*, *139*, 90-105.
[184] Nsugbe, E., Samuel, O. W., Asogbon, M. G., & Li, G. (2022, June). Intelligence combiner: a combination of deep learning and handcrafted features for an adolescent psychosis prediction using EEG signals. In *2022 IEEE International Workshop on Metrology for Industry 4.0 & IoT (MetroInd4. 0&IoT)* (pp. 92-97). IEEE.
[185] Supakar, R., Satvaya, P., & Chakrabarti, P. (2022). A deep learning based model using RNN-LSTM for the Detection of Schizophrenia from EEG data. *Computers in Biology and Medicine*, *151*, 106225.
[186] Prabhakar, S. K., & Lee, S. W. (2022). Improved sparse representation based robust hybrid feature extraction models with transfer and deep learning for EEG classification. *Expert Systems with Applications*, *198*, 116783.
[187] Sun, J., Cao, R., Zhou, M., Hussain, W., Wang, B., Xue, J., & Xiang, J. (2021). A hybrid deep neural network for classification of schizophrenia using EEG Data. *Scientific Reports*, *11*(1), 1-16.
[188] Naira, C. A. T., & Jos, C. (2019). Classification of people who suffer schizophrenia and healthy people by EEG signals using deep learning. *International Journal of Advanced Computer Science and Applications*, *10*(10).
[189] Aslan, Z., & Akin, M. (2020). Automatic Detection of Schizophrenia by Applying Deep Learning over Spectrogram Images of EEG Signals. *Traitement du Signal*, *37*(2).
[190] Barros, C., Roach, B., Ford, J. M., Pinheiro, A. P., & Silva, C. A. (2022). From sound perception to automatic detection of schizophrenia: an EEG-based deep learning approach. *Frontiers in Psychiatry*, *12*, 2659.
[191] Guo, Z., Wu, L., Li, Y., & Li, B. (2021, May). Deep neural network classification of EEG data in schizophrenia. In *2021 IEEE 10th Data Driven Control and Learning Systems Conference (DDCLS)* (pp. 1322-1327). IEEE.
[192] Sharma, G., & Joshi, A. M. (2021). Novel eeg based schizophrenia detection with iomt framework for smart healthcare. *arXiv preprint arXiv:2111.11298*.
[193] Hassan, F., Hussain, S. F., & Qaisar, S. M. (2023). Fusion of multivariate EEG signals for schizophrenia detection using CNN and machine learning techniques. *Information Fusion*, *92*, 466-478.
[194] Khare, S. K., Bajaj, V., & Acharya, U. R. (2023). SchizoNET: a robust and accurate Margenau-Hill time-frequency distribution based deep neural network model for schizophrenia detection using EEG signals. *Physiological Measurement*.
[195] Shoeibi, A., Rezaei, M., Ghassemi, N., Namadchian, Z., Zare, A., & Gorriz, J. M. (2022, May). Automatic diagnosis of schizophrenia in EEG signals using functional connectivity features and CNN-LSTM model. In *Artificial Intelligence in Neuroscience: Affective Analysis and Health Applications: 9th International Work-Conference on the Interplay Between Natural and Artificial Computation, IWINAC 2022, Puerto de la Cruz, Tenerife, Spain, May 31–June 3, 2022, Proceedings, Part I* (pp. 63-73). Cham: Springer International Publishing.

[196] Shoeibi, A., Sadeghi, D., Moridian, P., Ghassemi, N., Heras, J., Alizadehsani, R., ... & Gorriz, J. M. (2021). Automatic diagnosis of schizophrenia in EEG signals using CNN-LSTM models. *Frontiers in neuroinformatics*, 58.
[197] Luján, M. Á., Mateo Sotos, J., Torres, A., Santos, J. L., Quevedo, O., & Borja, A. L. (2022). Mental Disorder Diagnosis from EEG Signals Employing Automated Leaning Procedures Based on Radial Basis Functions. *Journal of Medical and Biological Engineering*, 1-7.
[198] Mutlag, W. K., Ali, S. K., Aydam, Z. M., & Taher, B. H. (2020, July). Feature extraction methods: a review. In *Journal of Physics: Conference Series* (Vol. 1591, No. 1, p. 012028). IOP Publishing.
[199] Galar, M., Derrac, J., Peralta, D., Triguero, I., Paternain, D., Lopez-Molina, C., ... & Herrera, F. (2015). A survey of fingerprint classification Part I: Taxonomies on feature extraction methods and learning models. *Knowledge-based systems*, *81*, 76-97.
[200] Diykh, M., Li, Y., & Wen, P. (2016). EEG sleep stages classification based on time domain features and structural graph similarity. *IEEE Transactions on Neural Systems and Rehabilitation Engineering*, *24*(11), 1159-1168.
[201] Ramos-Aguilar, R., Olvera-López, J. A., Olmos-Pineda, I., & Sánchez-Urrieta, S. (2020). Feature extraction from EEG spectrograms for epileptic seizure detection. *Pattern Recognition Letters*, *133*, 202-209.
[202] Al-Fahoum, A. S., & Al-Fraihat, A. A. (2014). Methods of EEG signal features extraction using linear analysis in frequency and time-frequency domains. *International Scholarly Research Notices*, *2014*.
[203] Li, M., Chen, W., & Zhang, T. (2017). Automatic epileptic EEG detection using DT-CWT-based non-linear features. *Biomedical Signal Processing and Control*, *34*, 114-125.
[204] Knott, V., Mahoney, C., Kennedy, S., & Evans, K. (2001). EEG power, frequency, asymmetry and coherence in male depression. *Psychiatry Research: Neuroimaging*, *106*(2), 123-140.
[205] Mormann, F., Lehnertz, K., David, P., & Elger, C. E. (2000). Mean phase coherence as a measure for phase synchronization and its application to the EEG of epilepsy patients. *Physica D: Nonlinear Phenomena*, *144*(3-4), 358-369.
[206] Baygin, M., Dogan, S., Tuncer, T., Barua, P. D., Faust, O., Arunkumar, N., ... & Acharya, U. R. (2021). Automated ASD detection using hybrid deep lightweight features extracted from EEG signals. *Computers in Biology and Medicine*, *134*, 104548.
[207] Acharya, U. R., Hagiwara, Y., Deshpande, S. N., Suren, S., Koh, J. E. W., Oh, S. L., ... & Lim, C. M. (2019). Characterization of focal EEG signals: a review. *Future Generation Computer Systems*, *91*, 290-299.
[208] Boonyakitanont, P., Lek-Uthai, A., Chomtho, K., & Songsiri, J. (2020). A review of feature extraction and performance evaluation in epileptic seizure detection using EEG. *Biomedical Signal Processing and Control*, *57*, 101702.
[209] Zang, B., Lin, Y., Liu, Z., & Gao, X. (2021). A deep learning method for single-trial EEG classification in RSVP task based on spatiotemporal features of ERPs. *Journal of Neural Engineering*, *18*(4), 0460c8.
[210] Abibullaev, B., & Zollanvari, A. (2019). Learning discriminative spatiospectral features of ERPs for accurate brain–computer interfaces. *IEEE Journal of Biomedical and Health Informatics*, *23*(5), 2009-2020.
[211] Lemm, S., Curio, G., Hlushchuk, Y., & Muller, K. R. (2006). Enhancing the signal-to-noise ratio of ICA-based extracted ERPs. *IEEE Transactions on Biomedical Engineering*, *53*(4), 601-607.
[212] Ayesha, S., Hanif, M. K., & Talib, R. (2020). Overview and comparative study of dimensionality reduction techniques for high dimensional data. *Information Fusion*, *59*, 44-58.
[213] Reddy, G. T., Reddy, M. P. K., Lakshmanna, K., Kaluri, R., Rajput, D. S., Srivastava, G., & Baker, T. (2020). Analysis of dimensionality reduction techniques on big data. *Ieee Access*, *8*, 54776-54788.
[214] Espadoto, M., Martins, R. M., Kerren, A., Hirata, N. S., & Telea, A. C. (2019). Toward a quantitative survey of dimension reduction techniques. *IEEE transactions on visualization and computer graphics*, *27*(3), 2153-2173.
[215] Palo, H. K., Sahoo, S., & Subudhi, A. K. (2021). Dimensionality reduction techniques: Principles, benefits, and limitations. *Data Analytics in Bioinformatics: A Machine Learning Perspective*, 77-107.
[216] Hussain, A., Kim, C. H., & Mehdi, A. (2021). A comprehensive review of intelligent islanding schemes and feature selection techniques for distributed generation system. *IEEE Access*, *9*, 146603-146624.
[217] Kumar, R. A., Franklin, J. V., & Koppula, N. (2022). A comprehensive survey on metaheuristic algorithm for feature selection techniques. *Materials Today: Proceedings*.
[218] Anowar, F., Sadaoui, S., & Selim, B. (2021). Conceptual and empirical comparison of dimensionality reduction algorithms (pca, kpca, lda, mds, svd, lle, isomap, le, ica, t-sne). *Computer Science Review*, *40*, 100378.
[219] Wang, Y., Cang, S., & Yu, H. (2019). Mutual information inspired feature selection using kernel canonical correlation analysis. *Expert Systems with Applications: X*, *4*, 100014.

[220] Song, X. F., Zhang, Y., Gong, D. W., & Sun, X. Y. (2021). Feature selection using bare-bones particle swarm optimization with mutual information. *Pattern Recognition*, *112*, 107804.
[221] Maldonado, J., Riff, M. C., & Neveu, B. (2022). A review of recent approaches on wrapper feature selection for intrusion detection. *Expert Systems with Applications*, 116822.
[222] Nafis, N. S. M., & Awang, S. (2021). An enhanced hybrid feature selection technique using term frequency-inverse document frequency and support vector machine-recursive feature elimination for sentiment classification. *IEEE Access*, *9*, 52177-52192.
[223] Zhang, H., Wang, J., Sun, Z., Zurada, J. M., & Pal, N. R. (2019). Feature selection for neural networks using group lasso regularization. *IEEE Transactions on Knowledge and Data Engineering*, *32*(4), 659-673.
[224] Aversano, L., Bernardi, M. L., Cimitile, M., & Pecori, R. (2021). A systematic review on Deep Learning approaches for IoT security. *Computer Science Review*, *40*, 100389.
[225] Zeng, C., Gu, L., Liu, Z., & Zhao, S. (2020). Review of deep learning approaches for the segmentation of multiple sclerosis lesions on brain MRI. *Frontiers in Neuroinformatics*, *14*, 610967.
[226] Jafari, M., Shoeibi, A., Khodatars, M., Ghassemi, N., Moridian, P., Alizadehsani, R., ... & Acharya, U. R. (2023). Automated diagnosis of cardiovascular diseases from cardiac magnetic resonance imaging using deep learning models: A review. *Computers in Biology and Medicine*, 106998.
[227] Moridian, P., Shoeibi, A., Khodatars, M., Jafari, M., Pachori, R. B., Khadem, A., ... & Ling, S. H. (2022). Automatic diagnosis of sleep apnea from biomedical signals using artificial intelligence techniques: Methods, challenges, and future works. *Wiley Interdisciplinary Reviews: Data Mining and Knowledge Discovery*, *12*(6), e1478.
[228] Moridian, P., Ghassemi, N., Jafari, M., Salloum-Asfar, S., Sadeghi, D., Khodatars, M., ... & Acharya, U. R. (2022). Automatic autism spectrum disorder detection using artificial intelligence methods with MRI neuroimaging: A review. *arXiv preprint arXiv:2206.11233*.
[229] Huang, S. C., Pareek, A., Seyyedi, S., Banerjee, I., & Lungren, M. P. (2020). Fusion of medical imaging and electronic health records using deep learning: a systematic review and implementation guidelines. *NPJ digital medicine*, *3*(1), 136.
[230] Zhou, T., Ruan, S., & Canu, S. (2019). A review: Deep learning for medical image segmentation using multi-modality fusion. *Array*, *3*, 100004.
[231] Aggarwal, R., Sounderajah, V., Martin, G., Ting, D. S., Karthikesalingam, A., King, D., ... & Darzi, A. (2021). Diagnostic accuracy of deep learning in medical imaging: a systematic review and meta-analysis. *NPJ digital medicine*, *4*(1), 65.
[232] Chen, X., Wang, X., Zhang, K., Fung, K. M., Thai, T. C., Moore, K., ... & Qiu, Y. (2022). Recent advances and clinical applications of deep learning in medical image analysis. *Medical Image Analysis*, 102444.
[233] Van der Laak, J., Litjens, G., & Ciompi, F. (2021). Deep learning in histopathology: the path to the clinic. *Nature medicine*, *27*(5), 775-784.
[234] Budd, S., Robinson, E. C., & Kainz, B. (2021). A survey on active learning and human-in-the-loop deep learning for medical image analysis. *Medical Image Analysis*, *71*, 102062.
[235] Liu, X., Faes, L., Kale, A. U., Wagner, S. K., Fu, D. J., Bruynseels, A., ... & Denniston, A. K. (2019). A comparison of deep learning performance against health-care professionals in detecting diseases from medical imaging: a systematic review and meta-analysis. *The lancet digital health*, *1*(6), e271-e297.
[236] Yin, W., Li, L., & Wu, F. X. (2022). Deep learning for brain disorder diagnosis based on fMRI images. *Neurocomputing*, *469*, 332-345.
[237] Noor, M. B. T., Zenia, N. Z., Kaiser, M. S., Mamun, S. A., & Mahmud, M. (2020). Application of deep learning in detecting neurological disorders from magnetic resonance images: a survey on the detection of Alzheimer's disease, Parkinson's disease and schizophrenia. *Brain informatics*, *7*, 1-21.
[238] Ebrahimighahnavieh, M. A., Luo, S., & Chiong, R. (2020). Deep learning to detect Alzheimer's disease from neuroimaging: A systematic literature review. *Computer methods and programs in biomedicine*, *187*, 105242.
[239] Craik, A., He, Y., & Contreras-Vidal, J. L. (2019). Deep learning for electroencephalogram (EEG) classification tasks: a review. *Journal of neural engineering*, *16*(3), 031001.
[240] Kora, P., Meenakshi, K., Swaraja, K., Rajani, A., & Raju, M. S. (2021). EEG based interpretation of human brain activity during yoga and meditation using machine learning: A systematic review. *Complementary therapies in clinical practice*, *43*, 101329.
[241] Goodfellow, I., Bengio, Y., & Courville, A. (2016). *Deep learning*. MIT press.
[242] Bengio, Y., Goodfellow, I., & Courville, A. (2017). *Deep learning* (Vol. 1). Cambridge, MA, USA: MIT press.

[243] Gulli, A., & Pal, S. (2017). *Deep learning with Keras*. Packt Publishing Ltd.
[244] Anwar, S. M., Majid, M., Qayyum, A., Awais, M., Alnowami, M., & Khan, M. K. (2018). Medical image analysis using convolutional neural networks: a review. *Journal of medical systems*, *42*, 1-13.
[245] Yamashita, R., Nishio, M., Do, R. K. G., & Togashi, K. (2018). Convolutional neural networks: an overview and application in radiology. *Insights into imaging*, *9*, 611-629.
[246] Ismail Fawaz, H., Forestier, G., Weber, J., Idoumghar, L., & Muller, P. A. (2019). Deep learning for time series classification: a review. *Data mining and knowledge discovery*, *33*(4), 917-963.
[247] Längkvist, M., Karlsson, L., & Loutfi, A. (2014). A review of unsupervised feature learning and deep learning for time-series modeling. *Pattern recognition letters*, *42*, 11-24.
[248] Wang, X., Wang, X., Liu, W., Chang, Z., Kärkkäinen, T., & Cong, F. (2021). One dimensional convolutional neural networks for seizure onset detection using long-term scalp and intracranial EEG. *Neurocomputing*, *459*, 212-222.
[249] Tan, C., Sun, F., Kong, T., Zhang, W., Yang, C., & Liu, C. (2018). A survey on deep transfer learning. In *Artificial Neural Networks and Machine Learning–ICANN 2018: 27th International Conference on Artificial Neural Networks, Rhodes, Greece, October 4-7, 2018, Proceedings, Part III 27* (pp. 270-279). Springer International Publishing.
[250] Zhuang, F., Qi, Z., Duan, K., Xi, D., Zhu, Y., Zhu, H., ... & He, Q. (2020). A comprehensive survey on transfer learning. *Proceedings of the IEEE*, *109*(1), 43-76.
[251] Morid, M. A., Borjali, A., & Del Fiol, G. (2021). A scoping review of transfer learning research on medical image analysis using ImageNet. *Computers in biology and medicine*, *128*, 104115.
[252] Cheplygina, V., de Bruijne, M., & Pluim, J. P. (2019). Not-so-supervised: a survey of semi-supervised, multi-instance, and transfer learning in medical image analysis. *Medical image analysis*, *54*, 280-296.
[253] Boveiri, H. R., Khayami, R., Javidan, R., & Mehdizadeh, A. (2020). Medical image registration using deep neural networks: a comprehensive review. *Computers & Electrical Engineering*, *87*, 106767.
[254] Liu, W., Wang, Z., Liu, X., Zeng, N., Liu, Y., & Alsaadi, F. E. (2017). A survey of deep neural network architectures and their applications. *Neurocomputing*, *234*, 11-26.
[255] Baur, C., Denner, S., Wiestler, B., Navab, N., & Albarqouni, S. (2021). Autoencoders for unsupervised anomaly segmentation in brain MR images: a comparative study. *Medical Image Analysis*, *69*, 101952.
[256] Wei, R., & Mahmood, A. (2020). Recent advances in variational autoencoders with representation learning for biomedical informatics: A survey. *Ieee Access*, *9*, 4939-4956.
[257] Nayak, D. R., Padhy, N., Mallick, P. K., & Singh, A. (2022). A deep autoencoder approach for detection of brain tumor images. *Computers and Electrical Engineering*, *102*, 108238.
[258] Mallick, P. K., Ryu, S. H., Satapathy, S. K., Mishra, S., Nguyen, G. N., & Tiwari, P. (2019). Brain MRI image classification for cancer detection using deep wavelet autoencoder-based deep neural network. *IEEE Access*, *7*, 46278-46287.
[259] Tekgul, H., Bourgeois, B. F., Gauvreau, K., & Bergin, A. M. (2005). Electroencephalography in neonatal seizures: comparison of a reduced and a full 10/20 montage. *Pediatric neurology*, *32*(3), 155-161.
[260] Uludağ, K., & Roebroeck, A. (2014). General overview on the merits of multimodal neuroimaging data fusion. *Neuroimage*, *102*, 3-10.
[261] Biessmann, F., Plis, S., Meinecke, F. C., Eichele, T., & Muller, K. R. (2011). Analysis of multimodal neuroimaging data. *IEEE reviews in biomedical engineering*, *4*, 26-58.
[262] Ewers, M., Frisoni, G. B., Teipel, S. J., Grinberg, L. T., Amaro Jr, E., Heinsen, H., ... & Hampel, H. (2011). Staging Alzheimer's disease progression with multimodality neuroimaging. *Progress in neurobiology*, *95*(4), 535-546.
[263] Aine, C. J., Bockholt, H. J., Bustillo, J. R., Cañive, J. M., Caprihan, A., Gasparovic, C., ... & Calhoun, V. D. (2017). Multimodal neuroimaging in schizophrenia: description and dissemination. *Neuroinformatics*, *15*, 343-364.
[264] Sui, J., Pearlson, G. D., Du, Y., Yu, Q., Jones, T. R., Chen, J., ... & Calhoun, V. D. (2015). In search of multimodal neuroimaging biomarkers of cognitive deficits in schizophrenia. *Biological psychiatry*, *78*(11), 794-804.
[265] Panigrahy, C., Seal, A., Gonzalo-Martín, C., Pathak, P., & Jalal, A. S. (2023). Parameter adaptive unit-linking pulse coupled neural network based MRI–PET/SPECT image fusion. *Biomedical Signal Processing and Control*, *83*, 104659.
[266] Cichy, R. M., & Oliva, A. (2020). AM/EEG-fMRI fusion primer: resolving human brain responses in space and time. *Neuron*, *107*(5), 772-781.

[267] Lecaignard, F., Bertrand, O., Caclin, A., & Mattout, J. (2021). Empirical Bayes evaluation of fused EEG-MEG source reconstruction: Application to auditory mismatch evoked responses. *Neuroimage*, *226*, 117468.
[268] Liu, R., Huang, Z. A., Hu, Y., Zhu, Z., Wong, K. C., & Tan, K. C. (2022). Attention-Like Multimodality Fusion With Data Augmentation for Diagnosis of Mental Disorders Using MRI. *IEEE Transactions on Neural Networks and Learning Systems*.
[269] Du, Y., He, X., Kochunov, P., Pearlson, G., Hong, L. E., van Erp, T. G., ... & Calhoun, V. D. (2022). A new multimodality fusion classification approach to explore the uniqueness of schizophrenia and autism spectrum disorder. *Human brain mapping*, *43*(12), 3887-3903.
[270] Tikàsz, A., Dumais, A., Lipp, O., Stip, E., Lalonde, P., Laurelli, M., ... & Potvin, S. (2019). Reward-related decision-making in schizophrenia: A multimodal neuroimaging study. *Psychiatry Research: Neuroimaging*, *286*, 45-52.
[271] Lemley, J., Bazrafkan, S., & Corcoran, P. (2017). Deep Learning for Consumer Devices and Services: Pushing the limits for machine learning, artificial intelligence, and computer vision. *IEEE Consumer Electronics Magazine*, *6*(2), 48-56.
[272] Pramod, A., Naicker, H. S., & Tyagi, A. K. (2021). Machine learning and deep learning: Open issues and future research directions for the next 10 years. *Computational analysis and deep learning for medical care: Principles, methods, and applications*, 463-490.
[273] Janiesch, C., Zschech, P., & Heinrich, K. (2021). Machine learning and deep learning. *Electronic Markets*, *31*(3), 685-695.
[274] Esteva, A., Robicquet, A., Ramsundar, B., Kuleshov, V., DePristo, M., Chou, K., ... & Dean, J. (2019). A guide to deep learning in healthcare. *Nature medicine*, *25*(1), 24-29.
[275] Faust, O., Hagiwara, Y., Hong, T. J., Lih, O. S., & Acharya, U. R. (2018). Deep learning for healthcare applications based on physiological signals: A review. *Computer methods and programs in biomedicine*, *161*, 1-13.
[276] Pang, G., Shen, C., Cao, L., & Hengel, A. V. D. (2021). Deep learning for anomaly detection: A review. *ACM computing surveys (CSUR)*, *54*(2), 1-38.
[277] Lan, K., Wang, D. T., Fong, S., Liu, L. S., Wong, K. K., & Dey, N. (2018). A survey of data mining and deep learning in bioinformatics. *Journal of medical systems*, *42*, 1-20.
[278] Chollet, F. (2017). The limitations of deep learning. *Deep learning with Python*.
[279] de Santana Correia, A., & Colombini, E. L. (2022). Attention, please! A survey of neural attention models in deep learning. *Artificial Intelligence Review*, *55*(8), 6037-6124.
[280] Ding, K., Xu, Z., Tong, H., & Liu, H. (2022). Data augmentation for deep graph learning: A survey. *ACM SIGKDD Explorations Newsletter*, *24*(2), 61-77.
[281] Zhang, H., Liang, W., Li, C., Xiong, Q., Shi, H., Hu, L., & Li, G. (2022). DCML: deep contrastive mutual learning for COVID-19 recognition. *Biomedical Signal Processing and Control*, *77*, 103770.
[282] Ahmed, I., Jeon, G., & Piccialli, F. (2022). From artificial intelligence to explainable artificial intelligence in industry 4.0: a survey on what, how, and where. *IEEE Transactions on Industrial Informatics*, *18*(8), 5031-5042.
[283] Novakovsky, G., Dexter, N., Libbrecht, M. W., Wasserman, W. W., & Mostafavi, S. (2023). Obtaining genetics insights from deep learning via explainable artificial intelligence. *Nature Reviews Genetics*, *24*(2), 125-137.
[284] Loh, H. W., Ooi, C. P., Seoni, S., Barua, P. D., Molinari, F., & Acharya, U. R. (2022). Application of explainable artificial intelligence for healthcare: A systematic review of the last decade (2011–2022). *Computer Methods and Programs in Biomedicine*, 107161.
[285] Van der Velden, B. H., Kuijf, H. J., Gilhuijs, K. G., & Viergever, M. A. (2022). Explainable artificial intelligence (XAI) in deep learning-based medical image analysis. *Medical Image Analysis*, 102470.
[286] Lauriola, I., Lavelli, A., & Aiolli, F. (2022). An introduction to deep learning in natural language processing: Models, techniques, and tools. *Neurocomputing*, *470*, 443-456.
[287] Gunny, A., Rankin, D., Krupa, J., Saleem, M., Nguyen, T., Coughlin, M., ... & Holzman, B. (2022). Hardware-accelerated inference for real-time gravitational-wave astronomy. *Nature Astronomy*, *6*(5), 529-536.
[288] Boukhennoufa, I., Zhai, X., Utti, V., Jackson, J., & McDonald-Maier, K. D. (2022). Wearable sensors and machine learning in post-stroke rehabilitation assessment: A systematic review. *Biomedical Signal Processing and Control*, *71*, 103197.
[289] Luvizutto, G. J., Silva, G. F., Nascimento, M. R., Sousa Santos, K. C., Appelt, P. A., de Moura Neto, E., ... & Bazan, R. (2022). Use of artificial intelligence as an instrument of evaluation after stroke: a scoping review

based on international classification of functioning, disability and health concept: AI applications for stroke evaluation. *Topics in Stroke Rehabilitation*, *29*(5), 331-346.
[290] Kohli, V., Tripathi, U., Chamola, V., Rout, B. K., & Kanhere, S. S. (2022). A review on Virtual Reality and Augmented Reality use-cases of Brain Computer Interface based applications for smart cities. *Microprocessors and Microsystems*, *88*, 104392.
[291] Aggarwal, S., & Chugh, N. (2022). Review of machine learning techniques for EEG based brain computer interface. *Archives of Computational Methods in Engineering*, 1-20.
[292] Ciccarelli, G., Federico, G., Mele, G., Di Cecca, A., Migliaccio, M., Ilardi, C. R., ... & Cavaliere, C. (2023). Simultaneous real-time EEG-fMRI neurofeedback: A systematic review. *Frontiers in Human Neuroscience*, *17*, 132.
[293] Vos, G., Trinh, K., Sarnyai, Z., & Azghadi, M. R. (2023). Generalizable Machine Learning for Stress Monitoring from Wearable Devices: A Systematic Literature Review. *International Journal of Medical Informatics*, 105026.
[294] Al-Turjman, F., & Baali, I. (2022). Machine learning for wearable IoT-based applications: A survey. *Transactions on Emerging Telecommunications Technologies*, *33*(8), e3635.
[295] Francese, R., & Yang, X. (2022). Supporting autism spectrum disorder screening and intervention with machine learning and wearables: a systematic literature review. *Complex & Intelligent Systems*, *8*(5), 3659-3674.
[296] Sujith, A. V. L. N., Sajja, G. S., Mahalakshmi, V., Nuhmani, S., & Prasanalakshmi, B. (2022). Systematic review of smart health monitoring using deep learning and Artificial intelligence. *Neuroscience Informatics*, *2*(3), 100028.
[297] Houssein, E. H., Hammad, A., & Ali, A. A. (2022). Human emotion recognition from EEG-based brain–computer interface using machine learning: a comprehensive review. *Neural Computing and Applications*, *34*(15), 12527-12557.
[298] Noble, W. S. (2006). What is a support vector machine?. *Nature biotechnology*, *24*(12), 1565-1567.
[299] Zhu, D., Lu, S., Wang, M., Lin, J., & Wang, Z. (2020). Efficient precision-adjustable architecture for softmax function in deep learning. *IEEE Transactions on Circuits and Systems II: Express Briefs*, *67*(12), 3382-3386.
[300] Gallagher III, B. J., Jones, B. J., McFalls Jr, J. A., & Pisa, A. M. (2006). Social class and type of schizophrenia. *European Psychiatry*, *21*(4), 233-237.
[301] Alonso, J. F., Romero, S., Ballester, M. R., Antonijoan, R. M., & Mañanas, M. A. (2015). Stress assessment based on EEG univariate features and functional connectivity measures. *Physiological measurement*, *36*(7), 1351.
[302] Moon, S. E., Chen, C. J., Hsieh, C. J., Wang, J. L., & Lee, J. S. (2020). Emotional EEG classification using connectivity features and convolutional neural networks. *Neural Networks*, *132*, 96-107.
[303] Lee, Y. Y., & Hsieh, S. (2014). Classifying different emotional states by means of EEG-based functional connectivity patterns. *PloS one*, *9*(4), e95415.
[304] Mirzaei, S., & Ghasemi, P. (2021). EEG motor imagery classification using dynamic connectivity patterns and convolutional autoencoder. *Biomedical Signal Processing and Control*, *68*, 102584.
[305] Tafreshi, T. F., Daliri, M. R., & Ghodousi, M. (2019). Functional and effective connectivity based features of EEG signals for object recognition. *Cognitive neurodynamics*, *13*, 555-566.
[306] Tan, S. (2006). An effective refinement strategy for KNN text classifier. *Expert Systems with Applications*, *30*(2), 290-298.
[307] Pal, M. (2005). Random forest classifier for remote sensing classification. *International journal of remote sensing*, *26*(1), 217-222.
[308] Jang, J. S. (1993). ANFIS: adaptive-network-based fuzzy inference system. *IEEE transactions on systems, man, and cybernetics*, *23*(3), 665-685.
[309] Quinlan, J. R. (1996). Learning decision tree classifiers. *ACM Computing Surveys (CSUR)*, *28*(1), 71-72.
[310] Arunkumar, N., Kumar, K. R., & Venkataraman, V. (2018). Entropy features for focal EEG and non focal EEG. *Journal of computational science*, *27*, 440-444.
[311] Vafaeikia, P., Namdar, K., & Khalvati, F. (2020). A brief review of deep multi-task learning and auxiliary task learning. *arXiv preprint arXiv:2007.01126*.
[312] Yang, Q., Liu, Y., Chen, T., & Tong, Y. (2019). Federated machine learning: Concept and applications. *ACM Transactions on Intelligent Systems and Technology (TIST)*, *10*(2), 1-19.
[313] Abdar, M., Pourpanah, F., Hussain, S., Rezazadegan, D., Liu, L., Ghavamzadeh, M., ... & Nahavandi, S. (2021). A review of uncertainty quantification in deep learning: Techniques, applications and challenges. *Information Fusion*, *76*, 243-297.

[314] Brauwers, G., & Frasincar, F. (2021). A general survey on attention mechanisms in deep learning. *IEEE Transactions on Knowledge and Data Engineering*.
[315] Hafiz, A. M., Parah, S. A., & Bhat, R. U. A. (2021). Attention mechanisms and deep learning for machine vision: A survey of the state of the art. *arXiv preprint arXiv:2106.07550*.
[316] Liu, H., Chatterjee, I., Zhou, M., Lu, X. S., & Abusorrah, A. (2020). Aspect-based sentiment analysis: A survey of deep learning methods. *IEEE Transactions on Computational Social Systems*, *7*(6), 1358-1375.
[317] Lin, T., Wang, Y., Liu, X., & Qiu, X. (2022). A survey of transformers. *AI Open*.
[318] Guo, D., Shao, Y., Cui, Y., Wang, Z., Zhang, L., & Shen, C. (2021). Graph attention tracking. In *Proceedings of the IEEE/CVF conference on computer vision and pattern recognition* (pp. 9543-9552).
[319] Chen, T., Li, X., Yin, H., & Zhang, J. (2018). Call attention to rumors: Deep attention based recurrent neural networks for early rumor detection. In *Trends and Applications in Knowledge Discovery and Data Mining: PAKDD 2018 Workshops, BDASC, BDM, ML4Cyber, PAISI, DaMEMO, Melbourne, VIC, Australia, June 3, 2018, Revised Selected Papers 22* (pp. 40-52). Springer International Publishing.
[320] Xue, G., Liu, S., & Ma, Y. (2020). A hybrid deep learning-based fruit classification using attention model and convolution autoencoder. *Complex & Intelligent Systems*, 1-11.
[321] Wu, Z., Pan, S., Long, G., Jiang, J., & Zhang, C. (2019). Graph wavenet for deep spatial-temporal graph modeling. *arXiv preprint arXiv:1906.00121*.
[322] Georgousis, S., Kenning, M. P., & Xie, X. (2021). Graph deep learning: State of the art and challenges. *IEEE Access*, *9*, 22106-22140.
[323] Prabhu, A., Varma, G., & Namboodiri, A. (2018). Deep expander networks: Efficient deep networks from graph theory. In *Proceedings of the European Conference on Computer Vision (ECCV)* (pp. 20-35).
[324] Wang, M., El-Fiqi, H., Hu, J., & Abbass, H. A. (2019). Convolutional neural networks using dynamic functional connectivity for EEG-based person identification in diverse human states. *IEEE Transactions on Information Forensics and Security*, *14*(12), 3259-3272.
[325] Zhang, G., Yu, M., Liu, Y. J., Zhao, G., Zhang, D., & Zheng, W. (2021). SparseDGCNN: Recognizing emotion from multichannel EEG signals. *IEEE Transactions on Affective Computing*.
[326] Song, T., Zheng, W., Song, P., & Cui, Z. (2018). EEG emotion recognition using dynamical graph convolutional neural networks. *IEEE Transactions on Affective Computing*, *11*(3), 532-541.
[327] Pan, S., Hu, R., Long, G., Jiang, J., Yao, L., & Zhang, C. (2018). Adversarially regularized graph autoencoder for graph embedding. *arXiv preprint arXiv:1802.04407*.
[328] Guo, K., Hu, Y., Qian, Z., Liu, H., Zhang, K., Sun, Y., ... & Yin, B. (2020). Optimized graph convolution recurrent neural network for traffic prediction. *IEEE Transactions on Intelligent Transportation Systems*, *22*(2), 1138-1149.
[329] Liu, P., Qiu, X., & Huang, X. (2016). Deep multi-task learning with shared memory. *arXiv preprint arXiv:1609.07222*.
[330] Huang, H., Yang, G., Zhang, W., Xu, X., Yang, W., Jiang, W., & Lai, X. (2021). A deep multi-task learning framework for brain tumor segmentation. *Frontiers in Oncology*, *11*, 690244.
[331] Zeng, N., Li, H., & Peng, Y. (2021). A new deep belief network-based multi-task learning for diagnosis of Alzheimer's disease. *Neural Computing and Applications*, 1-12.
[332] Roy, A. G., Siddiqui, S., Pölsterl, S., Navab, N., & Wachinger, C. (2019). Braintorrent: A peer-to-peer environment for decentralized federated learning. *arXiv preprint arXiv:1905.06731*.
[333] Yuan, B., Ge, S., & Xing, W. (2020). A federated learning framework for healthcare iot devices. *arXiv preprint arXiv:2005.05083*.
[334] Połap, D., Srivastava, G., & Yu, K. (2021). Agent architecture of an intelligent medical system based on federated learning and blockchain technology. *Journal of Information Security and Applications*, *58*, 102748.
[335] Zhang, H., Liang, W., Li, C., Xiong, Q., Shi, H., Hu, L., & Li, G. (2022). DCML: deep contrastive mutual learning for COVID-19 recognition. *Biomedical Signal Processing and Control*, *77*, 103770.
[336] Li, G., Wu, G., Xu, G., Li, C., Zhu, Z., Ye, Y., & Zhang, H. (2023). Pathological image classification via embedded fusion mutual learning. *Biomedical Signal Processing and Control*, *79*, 104181.
[337] Yang, Q., Geng, C., Chen, R., Pang, C., Han, R., Lyu, L., & Zhang, Y. (2022). DMU-Net: Dual-route mirroring U-Net with mutual learning for malignant thyroid nodule segmentation. *Biomedical Signal Processing and Control*, *77*, 103805.
[338] Caldeira, J., & Nord, B. (2020). Deeply uncertain: comparing methods of uncertainty quantification in deep learning algorithms. *Machine Learning: Science and Technology*, *2*(1), 015002.

[339] Zhu, Y., Zabaras, N., Koutsourelakis, P. S., & Perdikaris, P. (2019). Physics-constrained deep learning for high-dimensional surrogate modeling and uncertainty quantification without labeled data. *Journal of Computational Physics*, *394*, 56-81.
[340] Hu, R., Fang, F., Pain, C. C., & Navon, I. M. (2019). Rapid spatio-temporal flood prediction and uncertainty quantification using a deep learning method. *Journal of Hydrology*, *575*, 911-920.
[341] Shawahna, A., Sait, S. M., & El-Maleh, A. (2018). FPGA-based accelerators of deep learning networks for learning and classification: A review. *ieee Access*, *7*, 7823-7859.
[342] Wang, C., Gong, L., Yu, Q., Li, X., Xie, Y., & Zhou, X. (2016). DLAU: A scalable deep learning accelerator unit on FPGA. *IEEE Transactions on Computer-Aided Design of Integrated Circuits and Systems*, *36*(3), 513-517.
[343] Manasi, S. D., & Sapatnekar, S. S. (2021, January). DeepOpt: Optimized scheduling of CNN workloads for ASIC-based systolic deep learning accelerators. In *Proceedings of the 26th Asia and South Pacific Design Automation Conference* (pp. 235-241).
[344] Guo, Y. (2018). A survey on methods and theories of quantized neural networks. *arXiv preprint arXiv:1808.04752*.
[345] Gupta, M., & Agrawal, P. (2022). Compression of deep learning models for text: A survey. *ACM Transactions on Knowledge Discovery from Data (TKDD)*, *16*(4), 1-55.
[346] Alzu'bi, A., Amira, A., & Ramzan, N. (2017). Content-based image retrieval with compact deep convolutional features. *Neurocomputing*, *249*, 95-105.
[347] Eren, L., Ince, T., & Kiranyaz, S. (2019). A generic intelligent bearing fault diagnosis system using compact adaptive 1D CNN classifier. *Journal of Signal Processing Systems*, *91*, 179-189.
[348] Agarwal, M., & Singhal, A. (2023). Fusion of pattern-based and statistical features for Schizophrenia detection from EEG signals. *Medical Engineering & Physics*, 103949.
[349] Baygin, M., Barua, P. D., Chakraborty, S., Tuncer, I., Dogan, S., Palmer, E. E., ... & Acharya, U. R. (2023). CCPNet136: automated detection of schizophrenia using carbon chain pattern and iterative TQWT technique with EEG signals. *Physiological Measurement*.
[350] Gosala, B., Kapgate, P. D., Jain, P., Chaurasia, R. N., & Gupta, M. (2023). Wavelet transforms for feature engineering in EEG data processing: An application on Schizophrenia. *Biomedical Signal Processing and Control*, *85*, 104811.
[351] Shanarova, N., Pronina, M., Lipkovich, M., Ponomarev, V., Müller, A., & Kropotov, J. (2023). Application of Machine Learning to Diagnostics of Schizophrenia Patients Based on Event-Related Potentials. *Diagnostics*, *13*(3), 509.
[352] Siuly, S., Guo, Y., Alcin, O. F., Li, Y., Wen, P., & Wang, H. (2023). Exploring deep residual network based features for automatic schizophrenia detection from EEG. *Physical and Engineering Sciences in Medicine*, 1-14.
[353] Li, B., Wang, J., Guo, Z., & Li, Y. (2023). Automatic detection of schizophrenia based on spatial–temporal feature mapping and LeViT with EEG signals. *Expert Systems with Applications*, 119969.
[354] Grover, N., Chharia, A., Upadhyay, R., & Longo, L. (2023). Schizo-Net: A novel Schizophrenia Diagnosis framework using late fusion multimodal deep learning on Electroencephalogram-based Brain connectivity indices. *IEEE Transactions on Neural Systems and Rehabilitation Engineering*.
[355] Göker, H. (2023). 1D-convolutional neural network approach and feature extraction methods for automatic detection of schizophrenia. *Signal, Image and Video Processing*, 1-10.